\RequirePackage{fix-cm}		
\documentclass[%
	paper=a4,				
	BCOR=10mm,				
	DIV=13,					
	headinclude,			
	twoside,				
    cleardoublepage=empty,	
	titlepage,				
	numbers=noenddot,		
	parskip=half,			
	chapterprefix=false,	
	headsepline,			
	headings=normal,		
    bibliography=totoc,     
	listof=totoc,				
	index=totoc,				
	fontsize=11pt,			
	pagesize=pdftex,		
]{scrbook}	

\usepackage[linesnumbered,ruled,vlined]{algorithm2e}

\usepackage{helvet}						
\usepackage{CJKutf8}
\usepackage[utf8]{inputenc}				
\usepackage[T1]{fontenc}				
\usepackage[T1]{url}					
\usepackage{lmodern}					
\usepackage{microtype}					
\usepackage{setspace}					
\usepackage{multicol}					
\usepackage{ellipsis}					

\usepackage[english]{babel}	            
\usepackage[english=american]{csquotes}	

\usepackage{enumerate}					
\usepackage[table]{xcolor}              
\usepackage{longtable}					
\usepackage{arydshln}   				
\usepackage{booktabs}   				

\usepackage[printonlyused]{acronym}		

\usepackage{amsmath} 
\usepackage{multirow}
\usepackage{mathtools}
\usepackage{amsfonts} 
\usepackage{amssymb}
\usepackage{amsthm}
\usepackage{mathrsfs}
\usepackage{bbm}
\usepackage{scrhack}
\usepackage{dsfont}
               
\usepackage{color}						
\usepackage{xcolor}						
\usepackage{graphicx}					
\usepackage{flafter}					
\usepackage{tikz}						
\usepackage{pdfpages}
\usetikzlibrary{
	positioning,
	shapes,
	chains,
	decorations,
	fit,
	matrix
}

\usepackage{fmtcount}                    
\usepackage{datetime}                    
\let\originaltoday\today                 
\newdateformat{mydate}{\ordinaldate{\THEDAY}~\monthname[\THEMONTH], \THEYEAR}
\renewcommand{\today}{\mydate\originaltoday} 

\usepackage{styles/customlistings}

\usepackage{array}                      
\usepackage{adjustbox}                  
\usepackage{caption}
\usepackage{subcaption}                 
\usepackage{pdflscape}                  
\usepackage{changepage} 

\definecolor{darkblue_table}{RGB}{72, 103, 169}     
\definecolor{midblue_table}{RGB}{91, 145, 207}      
\definecolor{lightblue_table}{RGB}{153, 189, 229}   
\definecolor{yellow_table}{RGB}{255, 204, 0}        
\definecolor{gray_table}{RGB}{204, 204, 204}        
\definecolor{orange_table}{RGB}{255, 153, 51}       

\DeclareCaptionLabelFormat{table}{Table #2:}
\usepackage{float}
\DeclareCaptionLabelFormat{figure}{Figure #2:}
\newenvironment{shiftedchapter}{
  \clearpage
  \setlength{\oddsidemargin}{-0.54in} 
  \setlength{\evensidemargin}{-0.54in} 
  \setlength{\textwidth}{7.5in} 
  \centering
}{\clearpage
  \setlength{\oddsidemargin}{0.54in} 
  \setlength{\evensidemargin}{-0.54in} 
  \setlength{\textwidth}{5.5in}
}

\newenvironment{shiftedappendixB}{
  \clearpage
  \setlength{\oddsidemargin}{-0.64in} 
  \setlength{\evensidemargin}{-0.64in} 
  \setlength{\textwidth}{7.5in} 
  \centering
}{\clearpage
  \setlength{\oddsidemargin}{0.64in} 
  \setlength{\evensidemargin}{-0.64in} 
  \setlength{\textwidth}{5.5in}
}

\usepackage{enumitem,amssymb}
\usepackage{pifont}

\RequirePackage[%
	colorinlistoftodos,
	prependcaption,
	textsize=tiny
]{todonotes}

\usepackage[%
	pdftitle={Document title},	            			%
	pdfsubject={Document subject},	        		    %
	pdfauthor={Author Name},            				%
	pdfkeywords={some keywords},						%
    pdfcreator={LaTeX with hyperref and KOMA-Script},   %
	colorlinks,								            %
	urlcolor=black,							            %
	citecolor=black,							        %
	linkcolor=black,							        %
	breaklinks=true,						            %
	pdfpagemode=UseOutlines,				            %
	plainpages=false,						            %
	pdfpagelabels,							            %
	bookmarksnumbered,						            %
	pdfstartview=FitV,						            %
	pdfpagelayout=SinglePage,				            %
	pdfdisplaydoctitle=true					            %
]{hyperref}
\usepackage{cleveref} 

\usepackage{csquotes}
\usepackage{doi}
\usepackage[
	backend=biber,				
	hyperref,					
	sortcites,					
	date=short,					
	bibencoding=inputenc,		
	maxnames=5,					
	minnames=2,					
	giveninits=true,			
    sorting=none,				
	bibstyle=numeric,			
	citestyle=numeric-comp,		
	bibencoding=utf8,			
	bibwarn=true,				
	doi=true,					
	isbn=false,					
	url=true					
]{biblatex}
\DeclareFieldFormat*{title}{#1}
\DeclareFieldFormat*{titlecase}{%
	\ifdef{\currentfield}{\ifcurrentfield{title}{\usefield{\textit}{\currentfield}}{#1}}{#1}
}

\AtEveryBibitem{%
	\clearlist{language}%
}
\graphicspath{{figures/}}
\begin{document}
	\pagenumbering{alph}	

\begin{titlepage}

    \begin{minipage}{0.5\textwidth}
        \flushleft
        \includegraphics[scale=0.42]{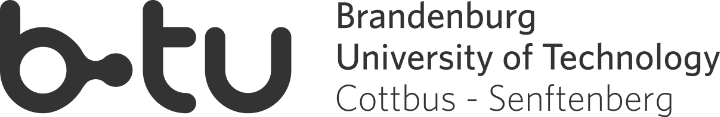}
    \end{minipage}%
    \begin{minipage}{0.5\textwidth}
        \flushright
        \includegraphics[width=0.25\textwidth]{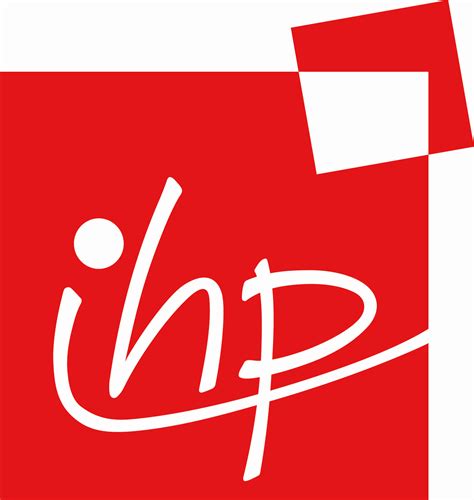}
    \end{minipage}

	\vspace{4em}
	
	\begin{center}
		\Huge{\textsf{\textbf{Investigation Of The Distinguishability Of Giraud-Verneuil Atomic Blocks}}}
		
		\vspace{2.5em}
		\large{Master Thesis}
		
		\vspace{0.2cm}
		\textbf{\large{Philip Laryea Doku}}
		
		\vspace{1.2cm}
		\small{Date of Submission}\\[0.2cm]
        23\textsuperscript{rd} May, 2025

		\vspace{1.5cm}
		This report is submitted to the\\
		\vspace{0.2cm}
		\large{\textbf{Chair of Wireless Systems}}\\[0.2cm]
		\textbf{Brandenburg University of Technology Cottbus-Senftenberg}\\
            \vspace{0.2cm}
            For the degree of Master of Science in Cyber Security\\

            \vspace{1cm}
            \textbf{DOI:} \href{https://doi.org/10.26127/BTUOpen-7140}{https://doi.org/10.26127/BTUOpen-7140} \\

	\end{center}
	
	\vfill
	\vspace{1cm}
	\begin{center}
		\textbf{Supervisor:}\\
		Hon. Prof. Dr. -Ing. Zoya Dyka\\

            \vspace{1.5em}
            \textbf{Co-Supervisor:}\\
            Dr. -Ing. Levgen Kabin\\
	\end{center}
\end{titlepage} 

\newpage
\thispagestyle{empty}
\cleardoublepage
    \cleardoublepage
\thispagestyle{empty}

\begin{center}
\Huge{\textsf{\textbf{Acknowledgements}}}		
\end{center}

\textbf{“Trust in the Lord with all your heart and lean not on your own understanding; in all your ways submit to him, and he will make your paths straight.”} - Proverbs 3:5-6. This verse has been a source of strength and guidance for me during my academic journey. It reminds me to rely on faith and trust in a higher wisdom, especially during times of uncertainty.

I would like to sincerely thank my supervisors, Hon. Prof. Dr.-Ing. Zoya Dyka and Dr.-Ing. Ievgen Kabin, for their invaluable guidance, helpful feedback, and constant support throughout this thesis. Their knowledge and encouragement have been essential to the development and success of this work.

I am also very grateful to Prof. Dr. rer. nat. Peter Langendörfer, Prof. Dr.-Ing. Andriy Panchenko, and Prof. Dr. rer. nat. Oliver Hohlfeld for their mentorship and support during my studies. Their expertise and thoughtful advice have had a strong influence on the direction and quality of this research.

I would also like to thank my colleagues and friends for their support and the many insightful conversations that helped me see different viewpoints and improved my thinking.

Finally, I am deeply thankful to my family for their love, patience, and constant encouragement throughout this academic journey.

	\cleardoublepage
\thispagestyle{empty}

\begin{center}
\Huge{\textsf{\textbf{Abstract}}}		
\end{center}

In this work, we investigate the security of \acf{ECC} implementations against \acf{SCA}. \acs{ECC} is well known for both its efficiency and strong security. However, it is also vulnerable to \acs{SCA}. \acs{SCA} exploits physical information leaked during the execution of the scalar multiplication (\( k\mathbf{P} \)) operations. Countermeasures like regularity and atomicity have been developed against such attacks. This thesis, however, focuses on atomicity as a countermeasure.

In this work, we focused on the atomic pattern proposed by Giraud and Verneuil used in the scalar multiplication operation. We implemented their atomic pattern using the right-to-left scalar multiplication algorithm on the P-256 elliptic curve recommended by NIST. Our implementation was done using the FLECC cryptographic library, which supports constant-time operations, and it was executed on the \acf{TI} LAUNCHXL-F28379D development board. We measured and collected the \acf{EM} emissions generated during the execution of the \( k\mathbf{P} \) operation using measurement tools such as the Lecroy WavePro 604HD Oscilloscope, the Langer ICS 105 Integrated Circuit Scanner, and the Langer MFA-R 0.2-75 Near Field Probe.

Our analysis focused on whether the Giraud and Verneuil atomic blocks could be distinguished based on their electromagnetic trace. We identified that when additionally inserted clock cycle processes were present in the execution, the atomic blocks could be visually distinguished from each other. However, after removing these processes, the atomic blocks became more synchronised and harder to distinguish, which reduced the risk of a successful \acs{SCA} attack.

These findings suggest that, although the atomic pattern was implemented correctly with dummy operations, its resistance to \acs{SCA} can still be affected by the additional processes that are inserted on the hardware or software level. This means that atomicity alone as a countermeasure might not be enough to fully protect \acs{ECC} implementations against \acs{SCA} attacks. More research is required to investigate the reasons for the additionally inserted clock cycle processes that caused the distinguishability in the atomic blocks and also how intermediate operations are addressed in memory registers. This will help to better understand the processes that lead to the insertion of these additional clock cycles.

This thesis is the first to experimentally implement and investigate Giraud and Verneuil's atomic pattern on hardware and offer useful results that can be used to improve countermeasures against \acs{SCA}.

	\frontmatter								
	\pagenumbering{roman}						
	\pdfbookmark[1]{\contentsname}{front-toc-anc}
	\tableofcontents
    \listoffigures
    \listoftables
    \listofalgorithms
    \addcontentsline{toc}{chapter}{List of Algorithms}

%
%
\cleardoublepage
\chapter*{List of Abbreviations and Acronyms}
\markboth{List of Abbreviations and Acronyms}{List of Abbreviations and Acronyms}
\addcontentsline{toc}{chapter}{List of Abbreviations and Acronyms}  

\begin{acronym}[MCU]
    \renewcommand{\baselinestretch}{0.9} 
    \setlength{\parskip}{0pt}
    \setlength{\itemsep}{5pt} 

    \acro{A}{Addition}

    \acro{ASIC}{Application Specific Integrated Circuit}
    \acrodef{asic}{application specific integrated circuit}
    
    \acro{CCS}{Code Composer Studio}
    \acrodef{ccs}{code composer studio}
    
    \acro{DES}{Data Encryption Standard}
    \acrodef{des}{data encryption standard}
    
    \acro{DPA}{Differential Power Analysis}
    \acrodef{dpa}{differential power analysis}
    
    \acro{DSCA}{Differential Side-Channel Analysis}
    \acrodef{dsca}{differential side-channel analysis}

    \acro{DSP}{Digital Signal Processing}
    \acrodef{dsp}{digital signal processing}

    \acro{EM}{Electromagnetic}
    \acrodef{em}{electromagnetic}
    
    \acro{EC}{Elliptic Curve}
    \acrodef{ec}{elliptic curve}
    
    \acro{ECC}{Elliptic Curve Cryptosystem}
    \acrodef{ecc}{elliptic curve cryptosystem}
    
    \acro{ECDLP}{Elliptic Curve Discrete Logarithm Problem}
    \acrodef{ecdlp}{elliptic curve discrete logarithm problem}
    
    \acro{ECDSA}{Elliptic Curve Digital Signature Algorithm}
    \acrodef{ecdsa}{elliptic curve digital signature algorithm}
    
    \acro{ECSM}{Elliptic Curve Scalar Multiplication}
    \acrodef{ecsm}{elliptic curve scalar multiplication}

    \acro{HCCA}{Horizontal Collision Correlation Attack}
    
    \acro{IoT}{Internet of Things}
    \acrodef{iot}{internet of things}

    \acro{M}{Multiplication}
    
    \acro{MCU}{Microcontroller Unit}
    \acrodef{mcu}{microcontroller unit}

    \acro{NIST}{National Institute of Standards and Technology}
    
    \acro{NAF}{Non-Adjacent Form}
    \acrodef{naf}{non-adjacent form}
    
    \acro{NOP}{No Operation}
    \acrodef{nop}{no operation}
    
    \acro{PA}{Point Addition}
    \acrodef{pa}{point addition}
    
    \acro{PD}{Point Doubling}
    \acrodef{pd}{point doubling}
    
    \acro{PKC}{Public Key Cryptography}
    \acrodef{pkc}{public key cryptography}
    
    \acro{RSA}{Rivest-Shamir-Adleman}
    \acrodef{rsa}{rivest-shamir-adleman}
    
    \acro{SCA}{Side-Channel Analysis}
    \acrodef{sca}{side-channel analysis}
    
    \acro{SMA}{Scalar Multiplication Algorithm}
    \acrodef{sma}{scalar multiplication algorithm}
    
    \acro{SPA}{Simple Power Analysis}
    \acrodef{spa}{simple power analysis}

    \acro{SSCA}{Simple Side-Channel Analysis}
    \acrodef{ssca}{simple side-channel analysis}
    
    \acro{STA}{Simple Timing Analysis}
    \acrodef{sta}{simple timing analysis}

    \acro{S}{Subtraction}
    \acro{Sq}{Squaring}

    \acro{TI}{Texas Instruments}

\end{acronym}

	\mainmatter									
	\pagenumbering{arabic}						
	\chapter{Introduction}
\label{chap:introduction}

\section{Motivation}
\acf{ECC} is a public-key cryptosystem, and it is well-known for its strong security properties. The advantage of \ac{ECC} in comparison to \ac{RSA} \cite{rivest_method_1978} is the ability to maintain the same level of security as \ac{RSA} but making use of significantly shorter key \cite{technology_digital_2023}. Because of this property, it is especially useful in environments with limited resources, such as wireless sensor networks and the \ac{IoT}.

The main operation in \ac{ECC} is \acf{EC} point multiplication (\( k\mathbf{P} \)), which is done by multiplying a scalar value by a point on an \ac{EC}. Digital signatures, key generation, and secure communication all rely on this operation. However, \ac{ECC} implementations are vulnerable to \acf{SCA}. \ac{SCA} exploits physical leakages such as power consumption or \acf{EM} emissions during the execution of the \( k\mathbf{P} \) operation to extract sensitive information.

A known countermeasure against \ac{SCA} is the atomicity principle introduced by Chevallier-Manes et al. \cite{chevallier-mames_low-cost_2004}. It ensures that point operations on different key bits operate using the same sequence of field operations and their measured traces should theoretically produce indistinguishable patterns. This prevents attackers from being able to differentiate between point operations performed on various key bits. Nonetheless, horizontal (single-trace) attacks can exploit address-bit vulnerabilities in scalar multiplication algorithms, as demonstrated by Kabin et al. in \cite{kabin_horizontal_2023} and \cite{kabin_atomicity_2022}. Key-dependent addressing of registers is an important property of both regular and atomic-pattern-based \( k\mathbf{P} \) algorithms, raising concerns about whether atomicity alone is a sufficient countermeasure against \ac{SCA}.

While many studies have theoretically analysed the efficiency and security of atomic pattern \( k\mathbf{P} \) algorithms, there are very few experimental implementations and evaluations of these countermeasures on embedded systems. No prior experimental research has been conducted on the side-channel resistance of Giraud and Verneuil's atomic pattern \cite{giraud_atomicity_2010}. This thesis fills that research gap by implementing and analysing the resistance of the atomic pattern to horizontal \ac{SCA} attacks.

Previous research by Sigourou et al. \cite{sigourou_successful_2023} and Li et al. \cite{li_practical_2024} investigated other atomic patterns and reached different conclusions about their resistance to \ac{SCA}. This highlights the need for further experimental analysis to determine the resistance of atomic patterns in hardware implementations.
\section{Contributions of the thesis}
This thesis has three main goals:

\begin{enumerate}
    \item To determine whether Giraud and Verneuil's proposed atomic pattern \cite{giraud_atomicity_2010} has been experimentally investigated by examining the existing literature that references it.
    \item Implement the Giraud-Verneuil atomic pattern \cite{giraud_atomicity_2010} for scalar multiplication on an embedded device and measure the \ac{EM} emanations.
    \item To analyse the \ac{EM} trace produced during the execution of the atomic pattern \( k\mathbf{P} \) algorithm and investigate the distinguishability of the atomic blocks while performing a horizontal \ac{SCA} attack.
\end{enumerate}

The thesis employs the right-to-left \( k\mathbf{P} \) algorithm, which is based on the atomic pattern by Giraud and Verneuil \cite{giraud_atomicity_2010}, and is implemented with the FLECC \cite{noauthor_iaikflecc_in_c_2022} open-source cryptographic library, to achieve these objectives. The \ac{EM} emanations were measured from the TMS320F28379D microcontroller and analysed using simple \ac{EM} \ac{SCA} techniques.

To the best of our knowledge, this work is the first to experimentally implement the Giraud-Verneuil atomic pattern on an embedded system. 
\section{Structure of this thesis}
The remaining chapters of this thesis are structured as follows:

\textbf{Chapter 2:} Introduces some basics of \ac{ECC} and \ac{SCA} that are needed to understand the rest of the work.

\textbf{Chapter 3:} Reviews the state-of-the-art literature in relation to the work by Giraud and Verneuil, and we identify the gap this thesis aims to address.

\textbf{Chapter 4:} Describes our experimental setup used for the measurements of the \ac{EM} emanations. It also describes our approach for measuring the \ac{EM} trace during the \( k\mathbf{P} \) operation.

\textbf{Chapter 5:} Focuses on the criteria for the selection of the open-source cryptographic library used for the implementation of the Giraud-Verneuil atomic pattern \( k\mathbf{P} \) algorithm used in this work.

\textbf{Chapter 6:} Describes how the Giraud-Verneuil atomic pattern was implemented using the FLECC cryptographic library.

\textbf{Chapter 7:} Describes the analysis and investigations performed on the measured \ac{EM} trace. It also describes the results and findings from our investigations.

The thesis concludes with a list of references and an appendix that contains additional information, such as a detailed evaluation of the state-of-the-art literature review, an analysis of the execution times of the \( k\mathbf{P} \) operations, the domain parameters of the implemented \ac{EC} and finally details about the samples from the measured \ac{EM} trace.

	\chapter{Backgrounds: ECC and SCA}
\label{chap:background}

In this chapter, we will describe theoretical basics that is required to understand this thesis. We will start by explaining the basics of \ac{ECC}, including its mathematical foundations and cryptographic protocols for \ac{EC} over prime finite fields. We will also discuss \ac{SCA} attacks, with a focus on horizontal, or single-trace analysis attacks and discuss the vulnerabilities we aim to address in this work. We will finally present the Giraud-Verneuil atomic pattern \cite{giraud_atomicity_2010} and explain how it helps defend against \ac{SCA}.

\section{Introduction to Prime Finite Fields}

A finite field, also known as a Galois field (\(GF\)), is a field that consists of a finite number of elements. The order of a finite field, denoted by \(n\), refers to the total number of elements in that field. When the order of a finite field is a prime number \(p\), then the field is termed a prime finite field, which is denoted here as \(GF(p)\) and this finite field will have exactly \(p\) elements in it.

All arithmetic operations in \(GF(p)\) are performed modulo \(p\) and its elements are integers represented as \( \{ 0, 1, 2, \dots, p-1 \} \). Also, for any integer \(a\), the expression \( a \mod p \) denotes the unique integer remainder \(r\), where \( 0 \leq r \leq p-1 \), obtained when dividing \(a\) by \(p\). This operation is termed reduction modulo \(p\) and it is there to ensure that results of the arithmetic operations are always within the field.

A prime finite field satisfies some key mathematical properties including closure, associativity, commutativity, and distributivity. Also, each element in \(GF(p)\) has an additive inverse, and every nonzero element has a multiplicative inverse.
\section{Elliptic Curves over GF(p)}
\label{sec:ecvovergfp}

An elliptic curve \(E\) over \(GF(p)\) is a set of points \( \boldsymbol{(x, y)} \) that satisfy the general equation:
\[
y^2 \mod p = x^3 + ax + b \mod p \tag{1} \label{eq:elliptic_curve}
\]

where coefficients \(a, b\) as well as the coordinates \( \boldsymbol{x, y} \) are elements of \(GF(p)\). The coefficients \(a, b\) satisfy the condition \(4a^3 + 27b^2 \neq 0 \mod p\). In addition to all points \( \boldsymbol{(x, y)} \) , the \ac{EC} over \(GF(p)\) includes a special element known as the point at infinity, (denoted here as point \( \mathcal{O} \)), which serves as the identity element (zero-element) in \ac{EC} point arithmetic. The complete set of all points on the \ac{EC}, including the point \( \mathcal{O} \) forms the \acl{EC} over the finite field. The \ac{EC} over \(GF(p)\) is denoted as \(E(GF(p))\).
\section{Elliptic Curve Point Arithmetic}
\label{sec:ecparithmetic}

\ac{EC} point arithmetic is the set of operations that make it possible to combine points on an elliptic curve in a clear way. There are 2 main point operations; \ac{PA}, denoted as \(P+Q\), which is the process of adding two distinct points \(P\) and \(Q\) to make a third point on the curve and \ac{PD}, denoted as \(2P\), which is the process of adding a point \(P\) to itself to make another point. 

On an elliptic curve, these operations show how to perform point addition and point doubling. It is important to remember that the \ac{EC} point operations are performed in a finite field and computed using modular arithmetic. In particular, all calculations are done modulo a prime number \(p\) when working over a prime finite field \(GF(p)\). This prime number \(p\) defines the prime finite field over which the elliptic curve operates.

\subsection{Point Addition (P + Q)}

Given two distinct points \( \boldsymbol{P} = (x_1, y_1) \in E(GF(p)) \) and \( \boldsymbol{Q} = (x_2, y_2) \in E(GF(p)) \) on the elliptic curve, their sum results in a third point \( \boldsymbol{R} = (x_3, y_3) \in E(GF(p)) \) that also lies on the curve. For the addition to be valid, \(P\) and \(Q\) must satisfy the curve equation.

If \( P \neq Q \) and \( P \neq -Q \), the sum \( R = P + Q \) is given by:

\begin{itemize}
    \item x-coordinate of the result (\( x_3 \)): \(x_3 = \left( \frac{y_2 - y_1}{x_2 - x_1} \right)^2 - x_1 - x_2 \mod p\)
    \item y-coordinate of the result (\( y_3 \)): \(y_3 = \left( \frac{y_2 - y_1}{x_2 - x_1} \right)(x_1 - x_3) - y_1 \mod p\)
\end{itemize}

If \( \boldsymbol{P} = (x, y) \in E(GF(p)) \), then \( P + (-P) = \mathcal{O} \) (point at infinity), where \( -P \) is known as the negative of \( P \) and is represented as \( -P = (x, -y) \). This means that \( P + \mathcal{O} = \mathcal{O} + P = P \) for all \( P \in E(GF(p)) \).

\subsection{Point Doubling (2P)}

\acl{PD} is the process of adding a point \( \boldsymbol{P} = (x_1, y_1) \in E(GF(p)) \) to itself to produce another point \( \boldsymbol{R} = (x_3, y_3) \in E(GF(p)) \) on the curve.

If \( P \neq -P \), then the resulting point \( R = 2P \) is given by:

\begin{itemize}
    \item x-coordinate of the result (\( x_3 \)): \(x_3 = \left( \frac{3x_1^2 + a}{2y_1} \right)^2 - 2x_1 \mod p\)
    \item y-coordinate of the result (\( y_3 \)): \(y_3 = \left( \frac{3x_1^2 + a}{2y_1} \right)(x_1 - x_3) - y_1 \mod p\)
\end{itemize}

If \( y_1 = 0 \), the result is the point at infinity \( \mathcal{O} \), meaning that \( 2P = \mathcal{O} \).
\section{Coordinate Systems for EC over GF(p)}

Different coordinate systems can be used to perform \ac{EC} computations over \(GF(p)\). Points on an \ac{EC} are represented using the coordinate systems.

\subsection{Affine Coordinate System}

In the Affine Coordinate System, points on the \ac{EC} are represented as \( \boldsymbol{P} = (x, y) \in E(GF(p)) \) and satisfy \hyperref[eq:elliptic_curve]{equation~(\ref*{eq:elliptic_curve})}. This is the simplest and most common way to represent points on an \ac{EC}. In \hyperref[sec:ecvovergfp]{section~\ref*{sec:ecvovergfp}} and \hyperref[sec:ecparithmetic]{section~\ref*{sec:ecparithmetic}}, all the formulas use the affine coordinate representation, which means that the \(\boldsymbol{x}\) and \(\boldsymbol{y}\) values are used directly in the computations. 

The main disadvantage with affine coordinates is that they need modular inversion when performing \ac{PA} and \ac{PD} operations \cite{joye_fast_2008}, which is a computationally expensive operation to perform in finite fields. Because modular inversion is significantly more complex and thus slower than multiplication and addition, affine coordinates are frequently unsuitable for implementations in resource-constrained devices.

\subsection{Projective Coordinates}
\label{sec:projectivecoordinate}

The Projective Coordinate System provides an alternative representation that eliminates the need for modular inversion. It introduces a third point \(Z\), to represent points as \( \boldsymbol{P} = (X, Y, Z) \) where \( \boldsymbol{X, Y, Z}\) are elements of \( E(GF(p)) \) and \( \boldsymbol {Z} \neq 0 \). This point corresponds to the affine point representation as:
\[
x = \frac{X}{Z}, \quad y = \frac{Y}{Z}.
\]

The elliptic curve equation in projective form is given by:

\[
Y^2Z = X^3 + aX Z^2 + bZ^3 \mod p \tag{2}
\]

The point at infinity \( \mathcal{O} \) is represented as \( \boldsymbol {(0 : 1 : 0)} \), while the negative of a point \( \boldsymbol{P} = (X : Y : Z) \) is \( -\boldsymbol{P} = (X : -Y : Z) \). Since modular inversion is replaced with a sequence of additional multiplications, projective coordinates improve the efficiency of point addition and point doubling. However, after computations, converting back to affine coordinates requires a single modular inversion.

More details about affine and projective coordinate system point representation can be found in \cite{noauthor_guide_2004}.

\subsection{Jacobian Coordinates}

The Jacobian Coordinate System is a variant of the projective coordinate system that is intended to further optimise computations by reducing the number of required modular multiplications and additions when compared to the projective coordinates described in \hyperref[sec:projectivecoordinate]{sub-section~\ref*{sec:projectivecoordinate}}. Points are represented as \( \boldsymbol{P} = (X, Y, Z) \) with \( \boldsymbol {Z} \neq 0 \), and they correspond to the affine point as:
\[
x = \frac{X}{Z^2}, \quad y = \frac{Y}{Z^3}.
\]

The equation for the \ac{EC} in this representation is given by: 
\[
Y^2 = X^3 + aXZ^4 + bZ^6 \tag{3}.
\] 

The point at infinity is also represented as \( \boldsymbol {(1 : 1 : 0)} \), whereas the negative of a point \( \boldsymbol{P} = (X : Y : Z) \) remains \( -\boldsymbol{P} = (X : -Y : Z) \). Jacobian coordinates are more efficient than projective coordinates because they do not require modular inversion during \acl{PA} and \acl{PD}. It uses specialised formulas to reduce the number of modular multiplications that are required. These formulas make use of the representations \( Z^2 \) and \( Z^3 \), which can be reused in multiple computations.
\section{Scalar Multiplication Operation and Algorithms}
\label{sec:SMA-operations}

A basic operation in \ac{ECC}, \ac{EC} point multiplication with a scalar is denoted as \(Q = k\boldsymbol{P}\) and results from adding a point \( \boldsymbol{P} = (x, y) \) on an \ac{EC}, to itself \(k\) times. Here, \(k\) is a scalar (an integer), and the goal is to compute \(Q = k\boldsymbol{P}\) by means of a combination of repeated point addition and point doubling operations, that is, \(k\boldsymbol{P} = \boldsymbol{P} + \boldsymbol{P} + \dots + \boldsymbol{P} \quad (k \, \text{times})\). In an \ac{ECC}, this operation, which is also known as scalar multiplication or \( k\mathbf{P} \) operation, is essential for the generation of key pairs, encryption/decryption, and digital signatures/verification, as well as authentication protocols. In binary \( k\mathbf{P} \) algorithms, the scalar \(k\) is represented in its binary form and the bits are processed sequentially to perform the required \ac{EC} point arithmetic operation.

\ac{ECC} performs \( k\mathbf{P} \) operation using Left-to-Right double-and-add Scalar Multiplication or Right-to-Left double-and-add Scalar Multiplication among several other binary \( k\mathbf{P} \) algorithms.

\subsection{Left-to-Right Scalar Multiplication (Double-and-Add)}
\label{sec:left-to-right-SMA}

Given the binary representation of the scalar \(k\), \( Q = k\mathbf{P} \) can be computed by iterating through the binary representation of \(k\) bit-by-bit from the second most significant bit (left) to the least significant bit (right). The algorithm is given below:

\begin{samepage}
\textbf{Algorithm:}
\begin{enumerate}
    \item Set \( Q = P \).
    \item For each bit \( k_i \) of \(k\), from \( k_{n-2} \) (the second most significant bit) to \( k_0 \) (the least significant bit):
    \begin{enumerate}
        \item Point Doubling: \( Q = 2Q \),
        \item Point Addition (only if \( k_i = 1 \)): \( Q = Q + P \).
    \end{enumerate}
    \item After processing all bits, \(Q = k\boldsymbol{P}\).
\end{enumerate}
\end{samepage}

The algorithm can be expressed as:
\[
Q =
\begin{cases} 
    2Q, \\
    Q + P, & \text{if } k_i = 1.
\end{cases}
\]

\begin{algorithm}[H]
\DontPrintSemicolon     
\caption{Left-to-right double-and-add scalar multiplication algorithm}
\label{alg:l2r-binary-kp}

    \KwIn{$\boldsymbol{k} = (\boldsymbol{k}_{n-1} \boldsymbol{k}_{n-2} \ldots \boldsymbol{k}_1 \boldsymbol{k}_0)_2$, an EC point \(\boldsymbol{P} \in \text{GF}(\boldsymbol{p})\)}
    \KwOut{$\boldsymbol{Q} = \boldsymbol{k}\boldsymbol{P}$}

    $\boldsymbol{Q} \leftarrow \boldsymbol{P}$ \\
    \For{$i = n - 1$ \textbf{downto} $0$ }{
        $\boldsymbol{Q} \leftarrow 2\boldsymbol{Q}$ \tcc*{\textcolor{blue}{Point Doubling}}
        \If{$\boldsymbol{k}_i = 1$}{
            $\boldsymbol{Q} \leftarrow \boldsymbol{Q} + \boldsymbol{P}$ \tcc*{\textcolor{blue}{Point Addition}}
        }
    }
\Return $\boldsymbol{Q} = \boldsymbol{k}\boldsymbol{P}$
\end{algorithm}

\subsection{Right-to-Left Scalar Multiplication}
\label{sec:right-to-left-SMA}

Given the binary representation of the scalar \(k\), \(Q = k\boldsymbol{P}\) can be computed by iterating through the binary representation of \(k\) bit-by-bit from the least significant bit (rightmost) to the most significant bit (leftmost). The algorithm is given below:

\textbf{Algorithm:}
\begin{enumerate}
    \item Initialize \( Q = \mathcal{O} \) (the point at infinity) and \( R = P \).
    \item For each bit \( k_i \) in \( k \), from \( k_0 \) (the least significant bit) to \( k_{n-1} \) (the most significant bit):
    \begin{itemize}
        \item Point Addition (only if \( k_i = 1 \)): \( Q = Q + R \).
        \item Point Doubling: \( R = 2R \).
    \end{itemize}
    \item After processing all bits, \(Q = k\boldsymbol{P}\).
\end{enumerate}

The algorithm can be expressed as:
\[
Q =
\begin{cases} 
    Q + R, & \text{if } k_i = 1, \\
    2R. & 
\end{cases}
\]

\begin{algorithm}[H]
\DontPrintSemicolon     
\caption{Right-to-Left double-and-add scalar multiplication algorithm}
\label{alg:r2l-binary-kp}

    \KwIn{$\boldsymbol{k} = (\boldsymbol{k}_{n-1} \boldsymbol{k}_{n-2} \dots \boldsymbol{k}_1 \boldsymbol{k}_0)_2$, an EC point \(\boldsymbol{P} \in \text{GF}(\boldsymbol{p})\)}
    \KwOut{$\boldsymbol{Q} = \boldsymbol{k}\boldsymbol{P}$}
    
    $\boldsymbol{Q} \leftarrow \mathcal{O}$ \tcc*{\textcolor{blue}{Point of infinity}}
    $\boldsymbol{R} \leftarrow \boldsymbol{P}$ \;
    
    \For{$i = 0$ \textbf{to} $n - 1$ }{
        \If{$\boldsymbol{k}_i = 1$}{
            $\boldsymbol{Q} \leftarrow \boldsymbol{Q} + \boldsymbol{R}$ \tcc*{\textcolor{blue}{Point Addition}}
        }
        $\boldsymbol{R} \leftarrow 2\boldsymbol{R}$ \tcc*{\textcolor{blue}{Point Doubling}}
    }
    
\Return $\boldsymbol{Q} = \boldsymbol{k}\boldsymbol{P}$
\end{algorithm}
\section{Elliptic Curve Cryptosystem (ECC)}
\label{sec:ECC}

A cryptosystem is a framework or set of protocols to guarantee information security including confidentiality, integrity and authenticity. A cryptosystem consists of algorithms that are used for generating key pairs, encryption and decryption that work together to protect and maintain the security of data. There are two main classifications of cryptosystems. These are symmetric-key cryptosystems and asymmetric-key cryptosystems. In a symmetric-key cryptosystem, the same secret key is used for both encryption and decryption while in an asymmetric-key cryptosystem (also known as a public-key cryptosystem), two pairs of keys are used, that is, a public key for encryption and a private key for decryption. In a public-key cryptosystem, the security is based on hard mathematical problems that makes it impossible to decrypt a message if the private key is unknown \cite{noauthor_guide_2004}. Some examples of public-key cryptosystems include \ac{RSA}, Diffie-Hellman, and \ac{ECC}.

N. Koblitz \cite{koblitz_elliptic_1987} and V. Miller \cite{miller_use_1986} first introduced \ac{ECC} in 1987 and 1985, respectively. It is a public-key cryptosystem based on \ac{EC}'s defined over finite fields. In \ac{ECC}, security is based on the difficulty of solving the \ac{ECDLP}. \ac{ECC} provides security with substantially smaller key sizes than other public-key cryptosystems, including Diffie-Hellman and \ac{RSA} and Diffie-Hellman \cite{noauthor_guide_2004}, \cite{technology_digital_2023}. For example, a 160-bit \ac{ECC} key provides the same level of security as a 1024-bit \ac{RSA} key, whereas a 256-bit \ac{ECC} key is equivalent to a 3072-bit \ac{RSA} key. Because smaller keys are used, encryption, decryption, and digital signature verification take less time and cost less computing power. This makes \ac{ECC} a good choice for devices with limited resources.

\subsection{Elliptic Curve Discrete Logarithm Problem (ECDLP)}

Given an elliptic curve \(E\) defined over a finite field \(GF(p)\), a point \( \boldsymbol{P} \in E(GF(p)) \) of order \( n \), and a point \( Q \in P \), it is computationally infeasible to find the integer \( d \in [0, n - 1] \) such that \(Q = dP\) \cite{noauthor_guide_2004}.

The difficulty of solving this problem is the basis for the security of \ac{ECC} because no efficient algorithms that exist to compute \(d\) for large values. The computational complexity of \ac{ECDLP} increases exponentially with the size of \(d\). The best-known naive approach is an exhaustive search as mentioned in \cite{noauthor_guide_2004}, where one computes successive multiples of \( P \) (i.e., \( P, 2P, 3P, 4P, \ldots \)) until reaching \(Q\). This method requires approximately \(d\) steps in the worst case and \( d/2 \) steps on average, making it impractical when \(d\) is large. \ac{EC} parameters must be carefully selected to resist attacks on the \ac{ECDLP} in order to ensure security. The curve must be defined over a prime field \(GF(p)\) where the order \(n\) is a sufficiently large prime number. At a minimum, it is recommended that \(n > 2^{256}\) to ensure security against exhaustive search attacks.

\subsection{Elliptic Curve Domain Parameters}

\ac{EC} domain parameters are a set of information that allows communicating parties to identify a particular elliptic curve group for cryptographic purposes \cite{noauthor_bsi_nodate}. These parameters comprise the finite field \(GF(p)\), the coefficients \(a\) and \(b\) in the \ac{EC} equation represented in \hyperref[eq:elliptic_curve]{equation~(\ref*{eq:elliptic_curve})}, a base point \(G\), its order \(n\), and the cofactor \(h\). In summary, the domain parameters are:

\begin{itemize}
    \item \( p \): A prime number that defines the finite field \( GF(p) \).
    \item \( a, b \): The coefficients that define the equation of the elliptic curve: \(y^2 = x^3 + ax + b \mod p.\)
    \item \( G (x, y) \): A base point in \( E(GF(p)) \) with the \(\boldsymbol{x}\) and \(\boldsymbol{y}\) coordinates of the base point \( G \).
    \item \( n \): The prime order of the base point \( G \) in \( E(GF(p)) \).
    \item \( h \): The cofactor of \( G \) in \( E(GF(p)) \), which is the ratio of the total number of points on the elliptic curve to the order of the base point and mathematically denoted as \(h = \frac{\# E(GF(p))}{n}.\)
\end{itemize}

Several standardised elliptic curves are recommended for secure cryptographic implementations to ensure that elliptic curve parameters are carefully selected for both security and efficiency. The \acp{EC} defined by the \ac{NIST} in the FIPS 186-5 standard \cite{technology_digital_2023} specify elliptic curves over five prime fields with moduli. These five \ac{NIST} recommended elliptic curves are randomly chosen over prime fields, where the primes \( p \) were specifically selected to enable very fast reduction of integers modulo \( p \). These include:

\begin{itemize}
    \item P-192: Defined over a 192-bit prime field, providing equivalent strength to a 96-bit symmetric key. This level of security is considered weak for modern security standards and the curve is designated for legacy-use only, such as decryption and verification of already protected information.
    
    \item P-224: Defined over a 224-bit prime field and offers a comparable security strength to a 112-bit symmetric key. This means, breaking it would require approximately \( 2^{112} \) operations.
    
    \item P-256 (secp256r1): A widely used 256-bit prime field that offers a comparable security strength to a 128-bit symmetric key, meaning that breaking its \ac{ECDLP} requires approximately \( 2^{128} \) operations.
    
    \item P-384: A 384-bit prime field that offers comparable security strength to a 192-bit symmetric key.
    
    \item P-521: A 521-bit prime field that provides comparable security strength to a 256-bit symmetric key.
\end{itemize}

Other elliptic curves have been developed to offer both security and efficiency in addition to \ac{NIST} curves. One of such curves is the Montgomery Curve25519 which it is defined over a 256-bit key size and offers a security level that is comparable to that of a 128-bit symmetric key. It is commonly used in cryptographic protocols such as X25519 for key exchange and has been standardized by the Internet Engineering Task Force as RFC 7748 \footnote{Source: \url{https://datatracker.ietf.org/doc/rfc7748/}}.

In this work, the \ac{NIST} EC P-256 was defined as secp256r1 and the domain parameters of this curve is detailed in \hyperref[appendixC:DomainParams]{Appendix~\ref*{appendixC:DomainParams}}.

\subsection{Key Pair Generation}
\ac{ECC} key pair generation is based on scalar multiplication. The private key, often denoted as \(d\) \cite{noauthor_guide_2004} is a randomly chosen integer between \(1\) and \(n-1\), where \(n\) is the order of the base point. The private key must remain secret and is used for cryptographic operations such as digital signature generation and key exchange. 

The public key denoted as \(Q\) \cite{noauthor_guide_2004} is derived from the private key \(d\) and the publicly known base point \(P\) on the \ac{EC}. The public key is computed as \(Q = d \cdot G\), where "\(\cdot\)" represents \ac{EC} point multiplication. The public key \(Q\) is made known to the public and used for encryption, key exchange and digital signature verification.

\subsection{EC Encryption and Decryption Scheme}

The \ac{EC} Encryption and Decryption Scheme is part of some set of \ac{ECC} protocols that rely on scalar \( k \) and the computation of \ac{EC} point multiplication to perform different cryptographic operations such as authentication and generation of digital signatures. As mentioned earlier, these protocols rely on the difficulty of the \ac{ECDLP}. The primary goal of an attacker is to determine \( k \) from \( k\mathbf{P} \). We present here the \ac{EC} encryption and decryption procedure according to the ElGamal encryption scheme as presented in \cite{noauthor_guide_2004}.

In this encryption scheme, a plaintext message \( m \) is first represented as a point \( M \) on the elliptic curve. The sender then encrypts a message by adding \( M \) to \( kQ \), where \( k \) is a randomly selected integer and \( Q \) is the recipient's public key. The sender then transmits the points \( C_1 = kP \) and \( C_2 = M + kQ \) as the encrypted message. The algorithm for encryption is as follows:

\begin{enumerate}
    \item Represent the message \( m \) as a point \( M \) in \( E(GF(p)) \).
    \item Select a random integer \( k \) from the range \( [1, n-1] \), where \( n \) is the order of the base point.
    \item Compute \( C_1 = kP \).
    \item Compute \( C_2 = M + kQ \), where \( Q \) is the recipient's public key.
    \item Transmit the ciphertext pair \( (C_1, C_2) \) to the recipient.
\end{enumerate}

To decrypt the message, the recipient uses their private key \( d \) to recover \( M \) by computing:

\begin{enumerate}
    \item Since \( C_2 = M + kQ \), the recipient can compute \( M = C_2 - kQ \).
    \item Extract \( m \) from \( M \).
\end{enumerate}

The security of this encryption scheme is based on the difficulty of computing \( kQ \) from the known parameters \( Q, C_1 = kP \), and the elliptic curve's domain parameters. In addition to encryption and decryption, \ac{ECC} is used in other cryptographic protocols that depend on \ac{EC} point multiplication. Some of these protocols are:

\begin{itemize}
    \item \ac{ECDSA} \cite{johnson_elliptic_2001}: Used for digital signature generation and verification.
    \item Elliptic Curve Diffie-Hellman (ECDH) \cite{miller_use_1986}: A key exchange protocol that enables two parties to establish a shared secret.
    \item Elliptic Curve Integrated Encryption Scheme (ECIES) \cite{noauthor_guide_2004}: A hybrid encryption scheme combining \ac{ECC} with symmetric encryption for secure message transmission.
\end{itemize}

Further details can be found here: \cite{noauthor_guide_2004},~\cite{noauthor_bsi_nodate}.
\section{Side-Channel Analysis}
\label{sec:SCA}

\acl{SCA} is a security exploit which is used to analyse and extract sensitive information from physical parameters (such as timing variations, power consumption, electromagnetic emissions, and acoustic signals) of cryptographic systems by exploiting unintended data leakage during the execution of cryptographic operations \cite{standaert_introduction_2010} such as the \( k\mathbf{P} \) operation. \ac{SCA} infers secret information by focussing on the physical implementation of cryptographic algorithms and their operating environment, rather than directly targeting the mathematical properties of these algorithms. As a result, \ac{SCA} can break cryptographic implementations and allow attackers to recover sensitive information.

A trace consists of side-channel measurements gathered during a cryptographic operation; in our case it is the \ac{EC} \( k\mathbf{P} \) operation which is the main part of different \ac{EC}-based cryptographic operations. The trace represents a sequence of measurements, such as power consumption or electromagnetic emanations, recorded over time as the \( k\mathbf{P} \) operation is executed.

\subsection{Classification of SCA}

Depending on the number of traces one needs for the analysis, \ac{SCA} can be classified into Horizontal \ac{SCA} and Vertical \ac{SCA}. Horizontal \ac{SCA} extracts sensitive information from a single execution of the \( k\mathbf{P} \) algorithm by analysing patterns within a single trace. Vertical \ac{SCA}, on the other hand, requires multiple traces \footnote{At least two traces required} collected from different \( k\mathbf{P} \) executions to perform statistical analysis and identify correlations. This thesis focuses on horizontal \ac{SCA}, i.e. on single-trace analysis attacks.

\ac{SCA} can also be categorised based on the analysis method used. \acf{SSCA} examines a single trace or a small number of traces to identify patterns in side-channel information. Unlike more complex methods, \ac{SSCA} does not require advanced statistical tools. For example, Simple Timing Analysis \cite{kocher_timing_1996} measures execution time variations to infer secret information.

In contrast, \acf{DSCA} is a more advanced technique that compares multiple traces, i.e. at least two traces but often hundreds or thousands of traces, captured under different conditions or processing different inputs to extract sensitive information. Attackers use statistical techniques, such as analysis of differences in the traces (known as \acf{DPA} \cite{kocher_differential_1999}), to find observable patterns between the side-channel information and the secret/private key. It is important to note, however, that it is also possible to apply statistical and machine learning methods to analyse a single \( k\mathbf{P} \) trace, as demonstrated in \cite{bauer_horizontal_2015}, \cite{kabin_vulnerability_2023}.

\subsection{Vulnerability of kP Implementations to SCA}
Due to the differences between \(2P\) and \(P+Q\) point operations, \( k\mathbf{P} \) implementations in \ac{ECC} are vulnerable to \ac{SCA} \cite{coron_resistance_1999} \cite{kabin_ec_2021}.The sequence of mathematical operations that are necessary for \(2P\) is different from those that are necessary for \(P+Q\), which results in side-channel profiles that are distinct and easily distinguishable.

The shape of a measured \( k\mathbf{P} \) trace is determined by several factors, including the value of the scalar key \(k\) that is processed, the \( k\mathbf{P} \) algorithm that is implemented, the operating frequency of the hardware equipment, and the hardware characteristics of the device under attack. The differences in the shapes of a \( k\mathbf{P} \) trace can be used by attackers to determine whether the processed key bit corresponds to \(0\) or \(1\), which allows the attacker to reconstruct the scalar \(k\) bit by bit. Since \ac{EC} point multiplication iterates through all bits of \(k\), this may eventually result in the full recovery of the secret key, which is an attacker’s goal.

The main challenge in securing \ac{ECC} implementations is ensuring that side-channel traces do not reveal whether the observed operation corresponds to \(2P\) or \(P+Q\), that is, the shapes of the trace while processing a key bit value of \(0\) should be indistinguishable from that of \(1\). In other words, an \ac{SCA}-resistant algorithm should produce similar power or electromagnetic profiles for both '\(0\)' and '\(1\)' or completely randomised patterns that do not reveal any correlation with the key bits.

\section{Countermeasures for EC Scalar Multiplication}
Several methods have been proposed to prevent unintentional side-channel information leakages \cite{brier_weierstras_2002}, \cite{edwards_normal_2007}, \cite{coron_resistance_1999}, \cite{chevallier-mames_low-cost_2004}, \cite{joye_montgomery_2003}, \cite{itoh_address-bit_2003}, \cite{izumi_improved_2010} as countermeasures against simple \ac{SCA}. Among these, two well-known and widely studied countermeasures are “Regularity” and “Atomicity”. The double-and-add-always algorithm, proposed by Coron in 1999 \cite{coron_resistance_1999} is an example of a regular \( k\mathbf{P} \) algorithm. This algorithm processes each bit of the scalar by performing both a point doubling and a point addition operation, regardless of the value of the processed key bit. However, it is known to be inefficient due the inefficient use of dummy operations \cite{noauthor_guide_2004}. Other examples of regular algorithms include \cite{joye_montgomery_2003}, \cite{joye_highly_2009}.

Atomic pattern algorithms, introduced in \cite{chevallier-mames_low-cost_2004}, are a faster and a more energy-efficient type of \( k\mathbf{P} \) algorithm countermeasure because they use fewer dummy operations. Other algorithmic methods can also be used to make \( k\mathbf{P} \) implementations more resistant to \ac{SCA}, including key masking or blinding, which hides the value of the key being processed and randomising certain steps in the \( k\mathbf{P} \) algorithm.

This thesis focuses only on atomicity as a countermeasure to \ac{SSCA} for implementations in embedded devices. The goal is to investigate if the atomic pattern introduced in \cite{giraud_atomicity_2010} is vulnerable to horizontal \ac{SCA}.

\subsection{Atomicity As a Countermeasure Against Simple SCA}
Atomicity as a countermeasure was first presented in 2004 by Chevallier-Mames et al. \cite{chevallier-mames_low-cost_2004}, although Catherine Gebotys and Robert Gebotys \cite{gebotys_secure_2003} had presented an approach earlier in 2003 to modify software designs by inserting redundant operations into the different point operations, thereby reducing the distinguishability in the power profile. The approach by Catherine Gebotys and Robert Gebotys was similar to the atomicity principle introduced by Chevallier-Mames et al. \cite{chevallier-mames_low-cost_2004} but they did not term it Atomicity.

In the atomicity principle, as explained by Chevallier-Mames et al. \cite{chevallier-mames_low-cost_2004}, each \ac{EC} point operation is treated as a process (or atomic block) that follows the same sequence of instructions. Two processes are considered side-channel equivalent if they are indistinguishable through \acl{sca}. A series of these atomic blocks is repeated consistently to form the atomic pattern for an \ac{EC} point operation. In atomic patterns, \ac{EC} point operations such as \(2P\) and \(P+Q\) are structured as atomic blocks with the same sequences of instructions in a regular sequence. This method theoretically effectively eliminates the variations that are often exploited in \ac{SCA}. This means, and attacker using \ac{SCA} cannot theoretically distinguish between different point operations to recover the secret keys or identify the starting and ending points of each point operation. Several algorithms for atomic pattern implementations have been presented in various studies \cite{giraud_atomicity_2010}, \cite{longa_accelerating_2008}, \cite{rondepierre_revisiting_2014}.

The atomicity principle relies on the assumption that reading to and writing data from registers are indistinguishable operations and, hence, do not create any distinguishable patterns through side-channel information leakages. This approach is similar to the security model used in the Montgomery ladder, which is known for its resistance to \ac{SSCA} attacks \cite{joye_montgomery_2003}.

\subsection{Vulnerabilities Of Atomic Patterns To SCA}
\label{sec:key-dependent-operations}
Key-dependent operations, such as data-bit processing and addressing of registers, also introduce vulnerabilities in \ac{SCA}-protected algorithms that attackers can exploit using \ac{SCA}.

\subsubsection{Key-dependent Data Processing}
Key-dependent data processing (data-bit vulnerability) refers to operations or computations that vary based on whether the scalar \(k\) bit being processed is \(0\) or \(1\). These variations are usually small and depend on the Hamming distance of the processed key bit. Using various statistical analysis methods, such as the correlation analysis, it is possible to analyse and exploit these variations in a trace to recover the secret key. This is often referred to as exploiting the data-dependent vulnerabilities through data-bit \ac{SCA}. The \ac{HCCA}, introduced in the work by Bauer et al. \cite{bauer_horizontal_2015} is an example of a successful horizontal \ac{SCA} attack against \( k\mathbf{P} \) implementations that exploited data-dependent vulnerabilities using a single \( k\mathbf{P} \) execution trace. The assumption is that the adversary is able to distinguish when two field multiplications have a common operand. The \ac{HCCA} attack allows the attacker to detect distinguishable patterns of operand sharing between point doubling and point addition operations. Bauer et al. \cite{bauer_horizontal_2015} showed that implementations combining atomicity and randomisation techniques are, in fact, also vulnerable to \ac{HCCA}. They demonstrated through experiments that the atomic patterns introduced by Chevallier-Mames et al. \cite{chevallier-mames_low-cost_2004}, Longa \cite{longa_accelerating_2008} and Christophe Giraud and Vincent Verneuil \cite{giraud_atomicity_2010} are all theoretically vulnerable to \ac{HCCA}.

\subsubsection{Key-dependent Addressing of Registers}
\label{sec:key-dep-add-of-reg}
Key-dependent addressing of registers refers to selection of specific registers or memory locations for storing intermediate results based on key bits. For example, separate registers may be used for point doubling and point addition operations, depending on the value of the bit being processed. An attacker can analyse the power or electromagnetic trace of a \( k\mathbf{P} \) execution associated with the key-dependent addressing register to identify distinguishable patterns that could help recover the secret keys. This vulnerability arises because accessing different memory registers causes different levels of bus activity, especially in hardware implementations. Messerges et al. \cite{messerges_investigations_nodate} in 1999 first introduced the address-bit \acl{DPA}. In their work, they showed the correlation between the secret data and how it is related to the addressing of registers. They highlighted the importance of robust countermeasures to protect against such attacks. Their work, however, was thought to be of no effect if the intermediate data is randomised. Itoh et al. \cite{itoh_address-bit_2003} further extended this work and proposed address-bit \ac{DPA} against \ac{EC}-based cryptosystems. They showed that randomising the data was not effective for their analysis and that multiple-trace attack is also successful when a small number of registers are used. Kabin et al. \cite{kabin_horizontal_2017} also showed the successful horizontal, i.e., single-trace, address-bit \ac{DPA} attack against Montgomery \( k\mathbf{P} \) implementation in an \ac{ECDSA} implementation. They mentioned that the success of their horizontal \ac{SCA} attack was due to the addressing of register operation, and since this process is not controlled by the designer, it cannot be prevented using means such as the atomicity principle or randomisation. Kabin et al. again described in \cite{kabin_horizontal_2018} the successful horizontal (single trace) address-bit Differential \acl{EM} Attack against a hardware implementation of the Montgomery \( k\mathbf{P} \) algorithm. They again showed that the key-dependent bus activity and addressing of registers was the reason for the success of the attack.

\subsection{Proposal From Christophe Giraud and Vincent Verneuil}
Giraud and Vincent Verneuil’s \cite{giraud_atomicity_2010} proposed an atomic pattern for scalar multiplication on \ac{EC} over \(GF(p)\). They claimed their atomic pattern allows for the implementation of \ac{EC} point multiplication in embedded devices in a more efficient way compared to any of the existing atomic patterns such as those from Chevallier-Mames et al. \cite{chevallier-mames_low-cost_2004}, and Longa \cite{longa_accelerating_2008}. Their approach was based on two main strategies:

\begin{enumerate}
    \item \textbf{Maximising the Use of Squaring Operations:} The Jacobian point addition and modified Jacobian point doubling operations were rewritten in order to take advantage of field squarings, since field squarings are faster than field multiplications.
    
    \item \textbf{Minimising the Use of Field Additions and Negations:} In this step, they reduced the number of dummy field operations. They mentioned that classical assumptions taken by designers to neglect the cost of field additions and subtractions in \(GF(p)\) in the context of embedded devices does not hold true because these operations actually introduced significant costs.
\end{enumerate}

Compared to earlier atomic patterns, their method provided:

\begin{enumerate}
    \item \( 18.3\% \) faster execution compared to the right-to-left scalar multiplication algorithm from Joye \cite{joye_fast_2008} protected using Chevallier-Mames' atomic pattern \cite{chevallier-mames_low-cost_2004}.
    \item \( 10.6\% \) faster execution compared to the left-to-right scalar multiplication algorithm protected using Longa's atomic pattern \cite{longa_accelerating_2008}.
\end{enumerate}

Although their study was mostly focused on the right-to-left scalar multiplication, they made it clear that their atomic pattern improvement method can be used to speed up other atomically protected algorithms.

They denoted dummy operations by "\(*\)" to indicate additional field operations required to complete the atomic pattern. These improvements resulted in the development of two optimised atomic blocks for point addition (Addition 1 and Addition 2) and one for point doubling. The details of their proposed atomic pattern are summarised in the \hyperref[tab:extended-atomic-pattern]{Table~\ref*{tab:extended-atomic-pattern}} below:

\begin{table}[H]
\renewcommand{\arraystretch}{1.5} 
\centering
\begin{tabular}{|l|llll|ll|}
\hline
\rowcolor{orange!20} 
\textbf{Operations} & \multicolumn{4}{c|}{\textbf{\begin{tabular}[c]{@{}c@{}}EC Point Addition\\ $P = (X_1: Y_1: Z_1)$\\ $Q = (X_2: Y_2: Z_2)$\\ $P + Q = (X_3: Y_3: Z_3)$\end{tabular}}} & \multicolumn{2}{c|}{\textbf{\begin{tabular}[c]{@{}c@{}}EC Point Doubling\\ $P = (X_1: Y_1: Z_1: W_1)$\\ $2P = (X_2: Y_2: Z_2: W_2)$\end{tabular}}} \\ \hline

OP1 & \multicolumn{1}{r|}{\multirow{18}{*}{\textbf{Add. 1}}} & \multicolumn{1}{l|}{$R_1 \leftarrow Z_2^2$} & \multicolumn{1}{r|}{\multirow{18}{*}{\textbf{Add. 2}}} & $R_1 \leftarrow R_6^2$ & \multicolumn{1}{r|}{\multirow{18}{*}{\textbf{Dbl.}}} & $R_1 \leftarrow X_1^2$ \\ \cline{1-1} \cline{3-3} \cline{5-5} \cline{7-7} 

OP2 & \multicolumn{1}{r|}{} & \multicolumn{1}{l|}{*} & \multicolumn{1}{r|}{} & * & \multicolumn{1}{r|}{} & $R_2 \leftarrow Y_1 + Y_1$ \\ \cline{1-1} \cline{3-3} \cline{5-5} \cline{7-7} 

OP3 & \multicolumn{1}{r|}{} & \multicolumn{1}{l|}{$R_2 \leftarrow Y_1 \times Z_2$} & \multicolumn{1}{r|}{} & $R_4 \leftarrow R_5 \times R_1$ & \multicolumn{1}{r|}{} & $Z_2 \leftarrow R_2 \times Z_1$ \\ \cline{1-1} \cline{3-3} \cline{5-5} \cline{7-7} 

OP4 & \multicolumn{1}{r|}{} & \multicolumn{1}{l|}{*} & \multicolumn{1}{r|}{} & * & \multicolumn{1}{r|}{} & $R_4 \leftarrow R_1 + R_1$ \\ \cline{1-1} \cline{3-3} \cline{5-5} \cline{7-7} 

OP5 & \multicolumn{1}{r|}{} & \multicolumn{1}{l|}{$R_5 \leftarrow Y_2 \times Z_1$} & \multicolumn{1}{r|}{} & $R_5 \leftarrow R_1 \times R_6$ & \multicolumn{1}{r|}{} & $R_3 \leftarrow R_2 \times Y_1$ \\ \cline{1-1} \cline{3-3} \cline{5-5} \cline{7-7} 

OP6 & \multicolumn{1}{r|}{} & \multicolumn{1}{l|}{*} & \multicolumn{1}{r|}{} & * & \multicolumn{1}{r|}{} & $R_6 \leftarrow R_3 + R_3$ \\ \cline{1-1} \cline{3-3} \cline{5-5} \cline{7-7} 

OP7 & \multicolumn{1}{r|}{} & \multicolumn{1}{l|}{$R_3 \leftarrow R_1 \times R_2$} & \multicolumn{1}{r|}{} & $R_1 \leftarrow Z_1 \times R_6$ & \multicolumn{1}{r|}{} & $R_2 \leftarrow R_6 \times R_3$ \\ \cline{1-1} \cline{3-3} \cline{5-5} \cline{7-7} 

OP8 & \multicolumn{1}{r|}{} & \multicolumn{1}{l|}{*} & \multicolumn{1}{r|}{} & * & \multicolumn{1}{r|}{} & $R_1 \leftarrow R_4 + R_1$ \\ \cline{1-1} \cline{3-3} \cline{5-5} \cline{7-7} 

OP9 & \multicolumn{1}{r|}{} & \multicolumn{1}{l|}{*} & \multicolumn{1}{r|}{} & * & \multicolumn{1}{r|}{} & $R_1 \leftarrow R_1 + W_1$ \\ \cline{1-1} \cline{3-3} \cline{5-5} \cline{7-7} 

OP10 & \multicolumn{1}{r|}{} & \multicolumn{1}{l|}{$R_4 \leftarrow Z_1^2$} & \multicolumn{1}{r|}{} & $R_6 \leftarrow R_2^2$ & \multicolumn{1}{r|}{} & $R_3 \leftarrow R_1^2$ \\ \cline{1-1} \cline{3-3} \cline{5-5} \cline{7-7} 

OP11 & \multicolumn{1}{r|}{} & \multicolumn{1}{l|}{$R_2 \leftarrow R_5 \times R_4$} & \multicolumn{1}{r|}{} & $Z_3 \leftarrow R_1 \times Z_2$ & \multicolumn{1}{r|}{} & $R_4 \leftarrow R_6 \times X_1$ \\ \cline{1-1} \cline{3-3} \cline{5-5} \cline{7-7} 

OP12 & \multicolumn{1}{r|}{} & \multicolumn{1}{l|}{*} & \multicolumn{1}{r|}{} & $R_1 \leftarrow R_4 + R_4$ & \multicolumn{1}{r|}{} & $R_5 \leftarrow W_1 + W_1$ \\ \cline{1-1} \cline{3-3} \cline{5-5} \cline{7-7} 

OP14 & \multicolumn{1}{r|}{} & \multicolumn{1}{l|}{$R_5 \leftarrow R_1 \times X_1$} & \multicolumn{1}{r|}{} & $R_1 \leftarrow R_5 \times R_3$ & \multicolumn{1}{r|}{} & $W_2 \leftarrow R_2 \times R_5$ \\ \cline{1-1} \cline{3-3} \cline{5-5} \cline{7-7} 

OP15 & \multicolumn{1}{r|}{} & \multicolumn{1}{l|}{*} & \multicolumn{1}{r|}{} & $X_3 \leftarrow R_6 - R_5$ & \multicolumn{1}{r|}{} & $X_2 \leftarrow R_3 - R_4$ \\ \cline{1-1} \cline{3-3} \cline{5-5} \cline{7-7} 

OP16 & \multicolumn{1}{r|}{} & \multicolumn{1}{l|}{*} & \multicolumn{1}{r|}{} & $R_4 \leftarrow R_4 - X_3$ & \multicolumn{1}{r|}{} & $R_6 \leftarrow R_4 - X_2$ \\ \cline{1-1} \cline{3-3} \cline{5-5} \cline{7-7} 

OP17 & \multicolumn{1}{r|}{} & \multicolumn{1}{l|}{$R_6 \leftarrow X_2 \times R_4$} & \multicolumn{1}{r|}{} & $R_3 \leftarrow R_4 \times R_2$ & \multicolumn{1}{r|}{} & $R_4 \leftarrow R_6 \times R_1$ \\ \cline{1-1} \cline{3-3} \cline{5-5} \cline{7-7} 

OP18 & \multicolumn{1}{r|}{} & \multicolumn{1}{l|}{$R_6 \leftarrow R_6 - R_5$} & \multicolumn{1}{r|}{} & $Y_3 \leftarrow R_3 - R_1$ & \multicolumn{1}{r|}{} & $Y_2 \leftarrow R_4 - R_2$ \\ \hline

\end{tabular}
\caption{Improved atomic pattern for point addition and point doubling operations proposed by Christophe Giraud and Vincent Verneuil.}
\label{tab:extended-atomic-pattern}
\end{table}

With this final optimisation, they reduced the number of field operations, saving six (6) field additions and removing eight (8) field negations each time the pattern was used. In the optimised pattern, the modified Jacobian point doubling operation (atomic block for point doubling) no longer required any dummy operations. The final atomic pattern consists of 8 field multiplications and 10 field additions or subtractions. Based on these optimisations, they concluded that the atomic pattern presented in \hyperref[tab:atomic-pattern]{Table~\ref*{tab:atomic-pattern}} below is the most efficient configuration for this operation. The field operations in this pattern include \ac{Sq}, \ac{M}, \ac{A} and \ac{S}

\begin{table}[H]
\renewcommand{\arraystretch}{1.5} 
\centering
\begin{tabular}{|c|l|}
\hline
\rowcolor{orange!20} 
\textbf{Field Operation} & \textbf{Instruction} \\ \hline
\ac{Sq} & $R_1 \leftarrow R_2^2$ \\ \hline
\ac{A} & $R_3 \leftarrow R_4 + R_5$ \\ \hline
\ac{M} & $R_6 \leftarrow R_7 \cdot R_8$ \\ \hline
\ac{A} & $R_9 \leftarrow R_{10} + R_{11}$ \\ \hline
\ac{M} & $R_{12} \leftarrow R_{13} \cdot R_{14}$ \\ \hline
\ac{A} & $R_{15} \leftarrow R_{16} + R_{17}$ \\ \hline
\ac{M} & $R_{18} \leftarrow R_{19} \cdot R_{20}$ \\ \hline
\ac{A} & $R_{21} \leftarrow R_{22} + R_{23}$ \\ \hline
\ac{A} & $R_{24} \leftarrow R_{25} + R_{26}$ \\ \hline
\ac{Sq} & $R_{27} \leftarrow R_{28}^2$ \\ \hline
\ac{M} & $R_{29} \leftarrow R_{30} \cdot R_{31}$ \\ \hline
\ac{A} & $R_{32} \leftarrow R_{33} + R_{34}$ \\ \hline
\ac{S} & $R_{35} \leftarrow R_{36} - R_{37}$ \\ \hline
\ac{M} & $R_{38} \leftarrow R_{39} \cdot R_{40}$ \\ \hline
\ac{S} & $R_{41} \leftarrow R_{42} - R_{43}$ \\ \hline
\ac{S} & $R_{44} \leftarrow R_{45} - R_{46}$ \\ \hline
\ac{M} & $R_{47} \leftarrow R_{48} \cdot R_{49}$ \\ \hline
\ac{S} & $R_{50} \leftarrow R_{51} - R_{52}$ \\ \hline
\end{tabular}
\caption{Resulting atomic pattern proposed by Giraud and Verneuil}
\label{tab:atomic-pattern}
\end{table}

	\chapter{State-of-The-Art Literature Review}
\label{chap:state-of-art}

The paper “Atomicity Improvement for Elliptic Curve Scalar Multiplication” by Christophe Giraud and Vincent Verneuil \cite{giraud_atomicity_2010} has significantly contributed to the field, counting 58 citations on Google Scholar \cite{noauthor_google_nodate}. It was important to review all papers citing Christophe Giraud and Vincent Verneuil's work \cite{giraud_atomicity_2010}  for several reasons. To begin with, it allowed us to understand the influence and impact of their work within the academic community. The goal of this literature review was to organise, analyse, and categorise all studies that mentioned Giraud and Verneuil's work within the broader context of improving atomicity in scalar multiplication with a particular focus on those studies involving embedded devices. By exploring these citations, we gained insight into how their proposed atomic pattern has been received, critiqued, and extended by other researchers. Additionally, it also helped to pinpoint the direction of research that has been influenced by Giraud and Verneuil's work, thereby highlighting the ongoing discourse and areas of interest within the field.

Google Scholar \cite{noauthor_google_nodate} was used as an academic search engine to collect a comprehensive set of research works that cited Giraud and Verneuil's work \cite{giraud_atomicity_2010}. By simply entering the title of their work in Google Scholar \cite{noauthor_atomicity_nodate}, I accessed a list of 58 publications citing it \cite{noauthor_giraud_nodate}. Through this search, I identified a total of 58 cited papers that referenced Giraud and Verneuil's work. I analysed each citation to assess the extent to which the work has been reviewed, compared with related works, implemented or improved upon in subsequent studies and referenced as a state-of-the-art reference in the context of mitigating \acl{SCA}.

Of the 58 identified papers, 13 could not be read for various reasons:
\begin{itemize}
    \item Four (4) of these papers were in languages other than English, making them impossible to review in detail. To determine their relevance to the thesis, I translated only the abstract and conclusion of these papers using DeepL \cite{noauthor_deepl_nodate}.
    \item Two (2) papers were inaccessible, as they were merely citations without accompanying full texts. 
    \item Seven (7) instances of duplicate works contributed to the subset of papers that could not be included in the comprehensive analysis. 
\end{itemize}

Despite these limitations, the papers that were accessible and could be reviewed in detail formed the basis for the findings provided in the subsequent sections of this chapter. In total, 45 of the papers were thoroughly reviewed.

\hyperref[appendixA:LiteratureReview]{Appendix~\ref*{appendixA:LiteratureReview}} of this work provides a detailed summary of my findings, presented as a table. It includes a short overview of the 45 papers, plus the additional four (4) papers that were in other languages other than English, detailing the main contributions and limitations of each. The summary focuses particularly on horizontal (single-trace) attacks, a category which includes simple SCA, among other methods that use various statistical and correlation analysis techniques. As discussed in \hyperref[sec:key-dependent-operations]{Section~\ref*{sec:key-dependent-operations}}, horizontal \ac{SCA} attacks can exploit data-bit vulnerabilities or address-bit vulnerabilities and since these vulnerabilities are important to consider in this work, the table indicates whether these points were investigated in the literature.

The literature review has been categorised into five overarching aspects for the reasons outlined below:
\begin{enumerate}
    \item \textbf{Literature on \acfp{SMA}:} In this subsection, we will discuss research works that focused on binary \( k\mathbf{P} \) algorithms to more advanced scalar multiplication algorithms like the Montgomery Ladder and the \ac{NAF}. The goal here is to better understand the research works that have reviewed the efficiency and security improvements of these algorithms, particularly in the context of embedded systems. This investigation was important because Giraud and Verneuil proposed implementing these advanced algorithms using atomic patterns for \ac{EC} point doubling and \ac{EC} point addition.
    
    \item \textbf{Literature on \acf{SCA}:} This subsection begins with defining and classifying \ac{SCA} attacks, focusing on evaluating the resistance of \( k\mathbf{P} \) algorithms to horizontal \ac{SCA} attacks. Understanding \ac{SCA} is also important because the investigated atomic pattern proposed by Giraud and Verneuil claim to countermeasure simple \ac{SCA}, and so investigating their resistance to \ac{SCA} requires that we also understand these attacks.

    \item \textbf{Literature on Data-Bit Dependency and Address-Bit Key Dependency:} This subsection explores studies on data-bit and address-bit dependency attacks, which influence the susceptibility of scalar multiplication to \ac{SCA}. We focused on such research papers because Kabin et al. \cite{kabin_horizontal_2017} highlighted that the Montgomery ladder and \( k\mathbf{P} \) algorithms using atomic patterns are inherently vulnerable to horizontal address-bit \ac{SCA} attacks. Given that Giraud’s proposal is also an atomic pattern, it was important to review how these vulnerabilities might impact its effectiveness.
    
    \item \textbf{Literature on \ac{SCA} Countermeasures:} In this subsection, we will review research papers that proposed and evaluated countermeasures such as regularity and atomicity. Atomic patterns are recognised as effective countermeasures against \ac{SCA}. These countermeasures are investigated for their potential to mitigate the impact of \ac{SCA} on scalar multiplication.
    
    \item \textbf{Categorisation of Literature on Giraud's Atomic Pattern:} In this subsection, we will explore literature claiming Giraud's work as innovative and investigates any implementations or improvements of their proposed atomic pattern.
    
\end{enumerate}

\section{Literature on Scalar Multiplication Algorithms}

In \hyperref[sec:SMA-operations]{Section~\ref*{sec:SMA-operations}}, we discussed the Left-to-Right and Right-to-Left double-and-add scalar multiplication algorithms in detail. Also in \hyperref[sec:ECC]{Section~\ref*{sec:ECC}}, we covered \ac{ECC} and its main concepts. Here, we will review research that has been done to improve these scalar multiplication algorithms with the focus on simply reviewing their efficiency or improving them.

Several methods have been proposed to enhance the efficiency of scalar multiplication algorithms especially in the context of embedded systems. One popular method is the Montgomery ladder based on the paper investigation of Peter Montgomery \cite{montgomery_speeding_1987} and J. Lopez, et al. \cite{oliveira_montgomery_2018}, which reduces memory requirements in resource-constrained devices by allowing one to omit the y-coordinate of the involved \ac{EC} points. Another improvement was proposed by Meloni \cite{meloni_new_2007} in 2007, which demonstrated that points on an \ac{EC} can be efficiently added when they share a common coordinate, like the projective Z-coordinate. This method, applied to specific Euclid addition chains, not only speeds up \ac{ECC} implementations but also reduces memory requirements by one coordinate, as proven by Lee and Verbauwhede \cite{lee_compact_2007} in binary fields. Baldwin et al. \cite{baldwin_co-ecc_2012} further extended Meloni's idea by providing formulas over prime fields, which can be applied to classical binary scalar multiplication methods.  They introduced a new operation called conjugate co-Z addition, which enables fast computations with points sharing the same Z-coordinate (co-Z arithmetic) when combined with Meloni's addition formula. They further analysed co-Z-based versions of the Montgomery ladder and Joye’s double-add algorithm \cite{joye_fast_2008}, noting their inherent resistance to \ac{SPA} attacks and comparing their performance across software, hardware, and co-design implementations.

Over the years, numerous algorithms have further been proposed to optimise scalar multiplication, especially in embedded systems. These algorithms focus on improving computational efficiency and reducing resource consumption. Key algorithms, including the double-and-add by Marc Joye \cite{joye_fast_2008} and windowed methods, present specific trade-offs in terms of speed and complexity. The literature extensively evaluates these \aclp{SMA}, focusing on their performance, speed, efficiency, and security. 

Significant contributions in this field include the work of Hutter et al. \cite{nitaj_memory-constrained_2011}, who presented new formulas for efficient point addition and point doubling using the Montgomery ladder. They emphasised memory usage reduction in low-resource environments by employing a common projective Z-coordinate representation for all computations. Duquesne et al. \cite{duquesne_choosing_2018} focused on generating optimal parameters for pairing-based cryptography, addressing pairing algorithms, tower field construction, \textit{Fp} arithmetic, and system coordinates. Furthermore, Abdulrahman and Reyhani-Masoleh \cite{abdulrahman_new_2013} introduced an innovative method for evaluating \ac{ECSM} that can be used on any abelian group, processing three bits of the scalar uniformly with five arithmetic operations (3 DBLs and 2 ADDs) for all eight-bit combinations without resorting to dummy operations or pre-computed point lookup tables. Other notable research works in this field include but are not limited to the studies by \cite{zulberti_script-based_2022}, \cite{guerrini_randomized_2017}, \cite{kim_speeding_2017}, \cite{al-somani_method_2014} and \cite{hasan_abdulrahman_efficient_2013}.
\section{Literature on Side-Channel Analysis (SCA)}

We earlier discussed \ac{SCA} in \hyperref[sec:SCA]{Section~\ref*{sec:SCA}}, its classification, and some vulnerabilities of scalar multiplication implementations to \ac{SCA}. As mentioned in that section, \ac{SCA} techniques can be broadly categorised into simple and differential methods. In this section, we will review existing research on the various \ac{SCA} methods used in these categories.

\ac{ECC} is widely used in smart cards and microcontrollers, so it faces a broader range of security threats, including physical attacks, which must be considered when evaluating its security in \ac{ECC} implementations. In 2010, a new classification of \acl{sca} was proposed by Clavier et al., classifying them into vertical and horizontal (single-trace) attacks \cite{clavier_horizontal_2010}. Before the classification of the attacks into horizontal and vertical by Clavier in 2010, single-trace attacks were often classified as simple \ac{SCA} even if statistical methods were applied for the analysis. 

\acp{SCA} have become more well-known in recent years because of how well they can be used to compromise sensitive data and circumvent conventional secure systems. For instance Lo’ai et al. in \cite{loai_review_2017} reviewed recent \ac{SCA} attacks on \ac{ECC}, \ac{RSA}, and AES cryptosystems and discussed effective countermeasures to protect these systems from cyber threats. Similarly, Danger et al. in \cite{danger_synthesis_2013} provided a comprehensive overview of various \ac{SCA} attacks targeting \ac{ECC} in embedded systems, particularly smart cards. For these reasons, it is important to study and understand the vulnerabilities that are exploited by \ac{SCA} techniques and develop effective countermeasures to mitigate these vulnerabilities.

\subsection{Horizontal Side-Channel Analysis}

Horizontal \ac{SCA} attacks exploit a single execution trace to recover secret information. The concept of \acl{sca} was introduced in 1996 by Paul Kocher and his colleagues Joshua Jaffe, and Benjamin Jun. Their seminal work highlighted the vulnerability of \( k\mathbf{P} \) implementations to \ac{SCA}. According to Kocher’s research \cite{kocher_timing_1996}, the physical implementation of an algorithm may leak information through side channel, even if it is theoretically secure. This vulnerability can be exploited to recover secret keys. 

Building on this foundational work, Feix et al. in \cite{meier_side-channel_2014} presented a new attack vector against protected \acl{ECSM} algorithms, revealing that even state-of-the-art countermeasures like regular algorithms and scalar blinding can be compromised using a combination of vertical and horizontal \ac{SCA}. Järvinen and Balasch in \cite{lemke-rust_single-trace_2017} investigated the susceptibility of \acl{ECSM} algorithms to single-trace \acl{SCA}, even when precomputations are used as a countermeasure. They demonstrated that attackers could correlate measurements with precomputed values or use clustering techniques to reveal secret scalar values without prior knowledge of these values. Similarly, Kabin et al. in \cite{kabin_ec_2021} also implemented the atomic pattern from Rondepierre \cite{rondepierre_revisiting_2014} and explored its resistance to horizontal \ac{SCA} attacks. They highlighted a vulnerability where key-dependent addressing of registers/blocks can be exploited to reveal the secret key thereby challenging the assumption that different registers/blocks are indistinguishable in operations. They also mentioned that all atomic patterns may be vulnerable to horizontal address-bit \ac{SCA} at least when implemented in hardware. Danger et al. also in \cite{ryan_improving_2016} improved the Big Mac attack, originally introduced by C. D. Walter \cite{walter_sliding_2001} for \ac{RSA} implementations and later adapted by Bauer et al. \cite{bauer_horizontal_2015} to allow for comparing multiple multiplications in elliptic curve implementations, leading to a higher success rate. Other research works have also focused on both simple and horizontal \ac{SCA} techniques including studies by \cite{diffie_new_1976}, \cite{rivest_method_1978}, \cite{elgamal_public_1985}, \cite{yan_cryptanalytic_2007}, \cite{sako_side-channel_2016}.

\section{Literature on Data-Bit and Address-Bit Key Dependency}
In \hyperref[sec:key-dependent-operations]{Section~\ref*{sec:key-dependent-operations}}, we discussed the vulnerabilities that can be caused by key-dependent operations including how differences in key-dependent operations can lead to the leakage of information through \ac{SCA}. Here, we will summarise these concepts and review research studies that have explored this topic in more detail.

As discussed earlier, research has challenged the assumption that reading and writing intermediate results into registers or memory during the execution of \( k\mathbf{P} \) operations is an indistinguishable operation in atomic patterns. Studies have demonstrated that subtle variations in power or \ac{EM} consumption during read and write operations may indeed be detected in certain situations. 

The key-dependent addressing of registers was identified as a vulnerability by Itoh et al. in \cite{itoh_address-bit_2003} that allows distinguishing which key bit was processed. They presented a successful vertical, address-bit \acf{DPA} attack against \ac{ECC} in their work in 2002. Although Messerges et al. \cite{messerges_investigations_nodate} originally investigated this analysis, it was initially considered ineffective if the intermediate results were randomised. However, later studies confirmed that reading data from different memory addresses can affect the power consumption or \ac{EM} emanations due to differences in the Hamming weight of addresses. This discovery has significant security implications, as it exposes a potential vulnerability that could be exploited by an attacker through \ac{SCA}. Address-bit attacks, as documented in \cite{kabin_randomized_2023}, \cite{kabin_ec_2021} and \cite{kabin_ec_2021-2} represent some examples of these vulnerabilities.

A work by Kabin et al. \cite{kabin_vulnerability_2023} studied the \( k\mathbf{P} \) algorithm based on atomic patterns proposed by Rondepierre \cite{rondepierre_revisiting_2014}, which was implemented for an \ac{ASIC} in the 250 nm IHP \cite{noauthor_ihp_nodate} technology hardware. A simple \ac{SCA} attack was successful by analysing a simulated power trace, exploiting key-dependent addressing of registers. Kabin et al. \cite{kabin_randomized_2023} further investigated the breaking of well-known randomised addressing countermeasures by M. Izumi et al. \cite{izumi_improved_2010} and K. Itoh et al. \cite{itoh_practical_2003} with a single trace attack and concluded that the success of their attacks was due to exploiting key-dependent addressing of registers. Again in \cite{kabin_horizontal_2018} Kabin et al. presented findings indicating that the Montgomery ladder, which employs Lopez-Dahab projective coordinates for \ac{EC} over \(GF(2n)\), is susceptible to horizontal (single-trace) address-bit \ac{DPA}. Additionally, Sigourou et al. \cite{sigourou_successful_2023} investigated the atomic patterns \( k\mathbf{P} \) algorithm proposed by Rondepierre \cite{rondepierre_revisiting_2014} for \ac{EC} over \(GF(p)\) using a horizontal \ac{SCA} attack. They revealed that the scalar \(k\) processed during the \( k\mathbf{P} \) execution could be uncovered by analysing the measured \ac{EM} trace.
\section{Literature on SCA Countermeasures}
As \ac{SCA} becomes more common and sophisticated, finding effective ways to protect cryptographic systems from these threats has become important. Researchers have explored various countermeasures to mitigate the risks posed by these attacks, focusing on both vertical and horizontal attack vectors. Effective countermeasures against vertical attacks include \ac{EC} point blinding or key randomisation as proposed by Coron \cite{coron_resistance_1999}, different masking techniques \cite{goubin_and_1999}, \cite{messerges_securing_2001} and randomization techniques such as randomising point calculations within the main loop of \ac{EC} point multiplication algorithms \cite{mentens_power_2008} and randomisation of the steps of cryptographic algorithms \cite{stottinger_procedures_2010}, \cite{oswald_randomized_2001} and \cite{madlener_novel_2009}. However, many of these countermeasures are less effective against horizontal \ac{SCA} attacks, particularly those exploiting key-dependent addressing in \ac{EC} point multiplication hardware accelerators \cite{kabin_breaking_2020} and \cite{kabin_horizontal_2017}.

When performing \ac{EC} point multiplication on \acp{EC} over \(GF(p)\), particularly in the context of embedded devices like smart cards, efficiency and resistance to \ac{SCA} are critical concerns. To address these issues, various algorithms have been developed to enhance resistance to simple \ac{SCA}. Since 1996, many proposals have aimed to protect \ac{EC} point multiplication algorithms from these attacks, as evidenced by works cited in \cite{coron_resistance_1999}, \cite{joye_protections_2001}, \cite{brier_weierstras_2002}, \cite{baldwin_co-ecc_2012}, \cite{nitaj_memory-constrained_2011}, \cite{abdulrahman_new_2013}, \cite{loai_review_2017}, \cite{abarzua_survey_2019} and \cite{tawalbeh_towards_2016}.

In this section, we will explore research works that address the two main countermeasure principles: regularity and atomicity, both widely recognised and used as effective countermeasures against simple \ac{SCA}.

\subsection{Regularity}
The designer's approach to mitigate the effectiveness of \ac{SCA} attacks involves ensuring that the power or electromagnetic trace(s) measured during the processing of all key bits are identical, regardless of the specific values of the processed key bit. The “regularity” strategy is achieved when designers implement a consistent sequence of operations for processing each scalar bit, that is, either \(0\) or \(1\). This strategy helps protect against certain types of attacks. For example, in \ac{RSA}, the “\textit{square-and-multiply-always}” method, as mentioned in \cite{messerges_power_1999} is an example of the regularity principle. This method is designed to counter simple \ac{SCA} attacks by ensuring that the same sequence of operations is executed, no matter which key bit is processed. Similarly, in \ac{ECC}, the Montgomery \ac{EC} point multiplication algorithm \cite{montgomery_speeding_1987} and the “\textit{double-and-add-always}” algorithm \cite{coron_resistance_1999} are examples of regular algorithms. The Montgomery ladder is specifically claimed to be resistant to simple \ac{SCA} attacks, as mentioned by Joye in \cite{joye_highly_2009} due to its regularity.

Numerous studies have been published on efficient and regular scalar multiplication algorithms. Rivain \cite{rivain_fast_2011} surveyed efficient and regular scalar multiplication algorithms for \acp{EC} over large prime fields, including the Montgomery ladder algorithm, and proposed a new algorithm to improve its efficiency. Abdulrahman and Reyhani \cite{abdulrahman_new_2013} also proposed a new regular algorithm for computing \ac{EC} point multiplication that is resistant to simple \ac{SCA} and safe-error fault attacks. Chabrier \cite{chabrier_arithmetic_2013} also proposed different methods to achieve regularity by using scheduling sequences independent of the recoded scalar. This was aimed to protect \ac{ECC} against simple \ac{SCA}.

\subsection{Atomicity}

The atomicity principle addresses the main problem of the regularity principle, which is that each processed key bit requires the execution of both point doubling and point addition operations which then increases the execution time and energy consumption of the algorithm. Since the atomicity principle was introduced, it has been extensively studied, with recent improvements proposed by Longa for certain scalar multiplication algorithms [5]. The main idea by Chevallier-Mames et al. in \cite{chevallier-mames_low-cost_2004} and Longa \cite{longa_accelerating_2008} is to split the process of point doubling and point addition into a sequence repeating atomic blocks. These blocks are designed to be identical and to minimise the leakage of any side-channel information. 

Over the years, several methods for protecting algorithms against simple \ac{SCA} using the atomicity principle have been developed. The well-known ones are the atomic patterns introduced by Chevallier-Mames et al. \cite{chevallier-mames_low-cost_2004}, Longa \cite{longa_accelerating_2008}, Giraud and Verneuil \cite{giraud_atomicity_2010} and Rondepierre \cite{rondepierre_revisiting_2014}. Additionally, many researchers have studied the applications and performance of these atomic pattern algorithms, primarily focusing on evaluating their efficiency rather than proposing new algorithms. Such studies include \cite{bauer_horizontal_2015}, \cite{abarzua_survey_2019}, \cite{abarzua_survey_2021} and \cite{kabin_ec_2021}.
\section{Categorisation of Literature on Giraud and Verneuil’s Atomic Pattern}

Christophe Giraud and Vincent Verneuil’s proposed atomic pattern \cite{giraud_atomicity_2010} is well-known for its theoretical approach to improving the security of \ac{EC} point multiplication operations against \ac{SCA}. Although its resistance has not been extensively investigated experimentally, their atomic pattern aims to reduce the vulnerability of \ac{EC} point multiplication to \ac{SCA} by ensuring that the same sequence of field operations is executed for each atomic block. This way, theoretically, an attacker finds it harder to gain useful information from physical leakages.

\begin{figure}[H]
    \centering
    \includegraphics[width=0.75\linewidth]{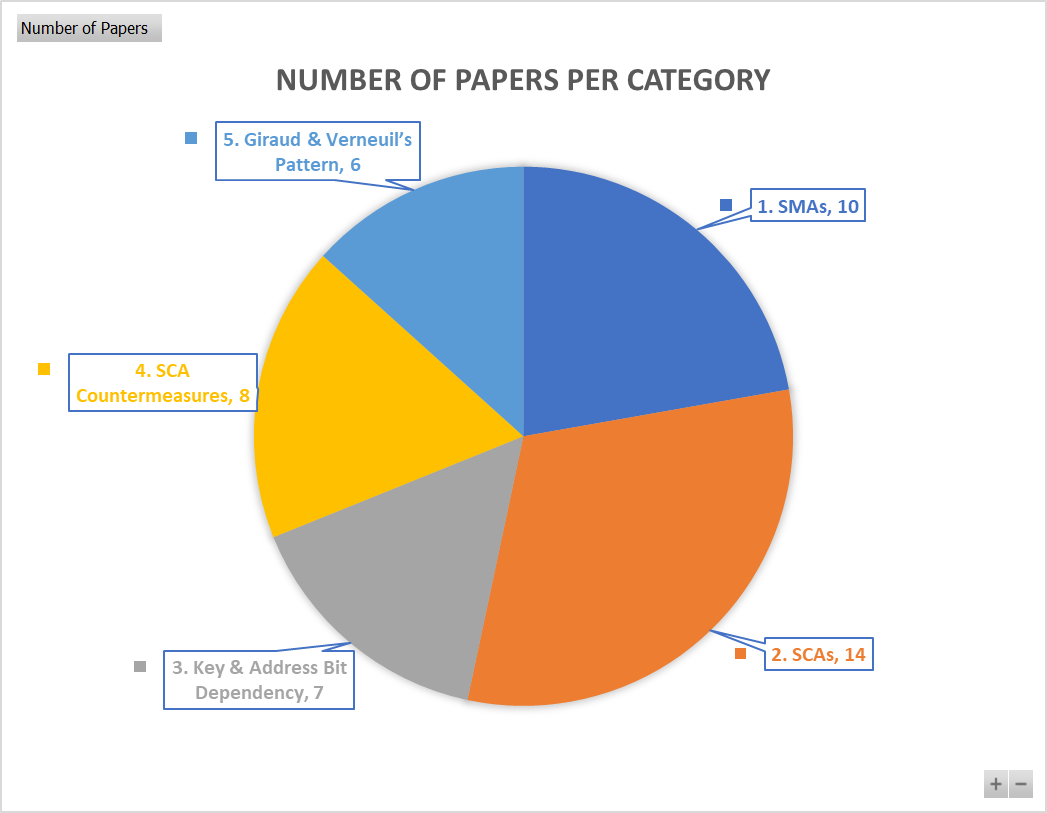}
    \caption{Categorisation of research works showing the number of papers that have cited Giraud and Verneuil's work based on focus areas.}
    \label{fig:Categorization}
\end{figure}

To understand the impact and application of their work, I refer to \hyperref[fig:Categorization]{Figure~\ref*{fig:Categorization}} presented below and the detailed table in the \hyperref[appendix:A]{Appendix~\ref*{appendix:A}} of this work. These resources detail the various research works that have cited their work and the focus areas of the research.


\hyperref[fig:Categorization]{Figure~\ref*{fig:Categorization}} shows the categorisation of the 45 papers we reviewed based on their focus areas and the results. These papers were categorised into five main focus areas, that is, research works that focused on:

\begin{enumerate}
    \item Scalar Multiplication Algorithms: \cite{baldwin_co-ecc_2012}, \cite{nitaj_memory-constrained_2011}, \cite{duquesne_choosing_2018}, \cite{abdulrahman_new_2013}, \cite{zulberti_script-based_2022}, \cite{guerrini_randomized_2017}, \cite{kim_speeding_2017}, \cite{al-somani_method_2014}, \cite{hasan_abdulrahman_efficient_2013} and \cite{baldwin_hardware_2013}.
    
    \item Side-Channel Analysis: \cite{loai_review_2017}, \cite{danger_synthesis_2013}, \cite{meier_side-channel_2014}, \cite{lemke-rust_single-trace_2017}, \cite{ryan_improving_2016}, \cite{lee_single-trace_2022}, \cite{russon_exploiting_2020}, \cite{sako_side-channel_2016}, \cite{tawalbeh_towards_2016}, \cite{jin_enhancing_2023}, \cite{murdica_physical_2014}, \cite{danger_dynamic_2013}, \cite{murdica_securite_2014} and \cite{el_mrabet_contributions_2017}.

    \item Data-bit and address-bit key dependency: \cite{kabin_horizontal_2017}, \cite{kabin_ec_2021}, \cite{sigourou_successful_2023}, \cite{kabin_vulnerability_2023}, \cite{kabin_randomized_2023}, \cite{kabin_ec_2021-2} and \cite{kabin_horizontal_2023}.
    
    \item \ac{SCA} Countermeasures: \cite{al-somani_system_2014}, \cite{houssain_elliptic_2012}, \cite{abarzua_survey_2019}, \cite{rivain_fast_2011}, \cite{chabrier_arithmetic_2013}, \cite{abarzua_survey_2021}, \cite{lu_general_2013} and \cite{nascimento_comparison_2014}.
    
    \item Review and improvement of Giraud and Verneuil's proposed atomic pattern: \cite{rondepierre_revisiting_2014}, \cite{bauer_horizontal_2015}, \cite{das_improved_2016}, \cite{ryan_safe-errors_2016}, \cite{abarzua_complete_2012} and \cite{abarzua_method_2015}.
\end{enumerate}

\begin{figure}[H]
    \centering
    \includegraphics[width=1\linewidth]{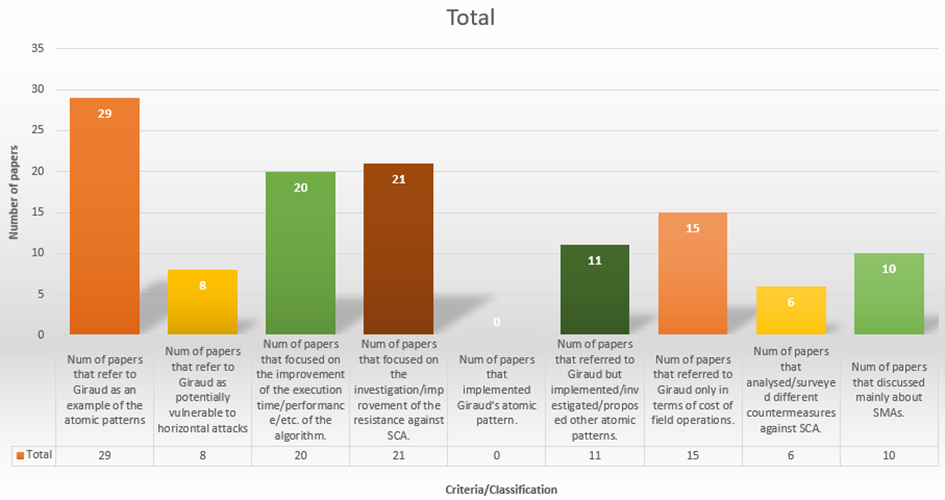}
    \caption{Classification of research works based on relevant questions}
    \label{fig:Classification}
\end{figure}

We further categorised these papers based on specific criteria, where we assessed them on specific questions. These criteria are as follows, and the corresponding results can be seen graphically in \hyperref[fig:Classification]{Figure~\ref*{fig:Classification}}.

\begin{enumerate}
    \item 29 papers refer to Giraud and Verneuil’s work as an example of an atomic pattern: \cite{rondepierre_revisiting_2014}, \cite{bauer_horizontal_2015}, \cite{kabin_horizontal_2017}, \cite{guerrini_randomized_2017}, \cite{hasan_abdulrahman_efficient_2013}, \cite{loai_review_2017}, \cite{meier_side-channel_2014}, \cite{lemke-rust_single-trace_2017}, \cite{kabin_ec_2021}, \cite{ryan_improving_2016}, \cite{lee_single-trace_2022}, \cite{sigourou_successful_2023}, \cite{kabin_vulnerability_2023}, \cite{das_improved_2016}, \cite{al-somani_system_2014}, \cite{houssain_elliptic_2012}, \cite{abarzua_survey_2019}, \cite{tawalbeh_towards_2016}, \cite{chabrier_arithmetic_2013}, \cite{abarzua_survey_2021}, \cite{jin_enhancing_2023}, \cite{murdica_physical_2014}, \cite{murdica_securite_2014}, \cite{lu_general_2013}, \cite{nascimento_comparison_2014}, \cite{ryan_safe-errors_2016}, \cite{abarzua_complete_2012}, \cite{kabin_ec_2021-2} and \cite{kabin_horizontal_2023}.
    
    \item 8 papers refer to Giraud and Verneuil’s work as potentially vulnerable to horizontal \ac{SCA} attacks: \cite{bauer_horizontal_2015}, \cite{kabin_horizontal_2017}, \cite{abdulrahman_new_2013}, \cite{kabin_ec_2021}, \cite{kabin_vulnerability_2023}, \cite{kabin_randomized_2023}, \cite{kabin_ec_2021-2} and \cite{kabin_horizontal_2023}.
    
    \item 20 papers focused on improving the execution time/performance/etc. of the atomic pattern algorithm: \cite{rondepierre_revisiting_2014}, \cite{baldwin_co-ecc_2012}, \cite{nitaj_memory-constrained_2011}, \cite{duquesne_choosing_2018}, \cite{abdulrahman_new_2013}, \cite{zulberti_script-based_2022}, \cite{guerrini_randomized_2017}, \cite{kim_speeding_2017}, \cite{al-somani_method_2014}, \cite{hasan_abdulrahman_efficient_2013}, \cite{al-somani_system_2014}, \cite{houssain_elliptic_2012}, \cite{rivain_fast_2011}, \cite{chabrier_arithmetic_2013}, \cite{baldwin_hardware_2013}, \cite{el_mrabet_contributions_2017}, \cite{lu_general_2013}, \cite{ryan_safe-errors_2016}, \cite{abarzua_complete_2012}, \cite{abarzua_method_2015}.
    
    \item 21 papers focused on the investigation/improvement of the atomic pattern's resistance against \ac{SCA}: \cite{bauer_horizontal_2015}, \cite{kabin_horizontal_2017}, \cite{kim_speeding_2017}, \cite{hasan_abdulrahman_efficient_2013}, \cite{ryan_improving_2016}, \cite{sigourou_successful_2023}, \cite{kabin_vulnerability_2023}, \cite{das_improved_2016}, \cite{sako_side-channel_2016}, \cite{al-somani_system_2014}, \cite{kabin_randomized_2023}, \cite{houssain_elliptic_2012}, \cite{rivain_fast_2011}, \cite{chabrier_arithmetic_2013}, \cite{murdica_physical_2014}, \cite{danger_dynamic_2013}, \cite{murdica_securite_2014}, \cite{ryan_safe-errors_2016}, \cite{abarzua_complete_2012}, \cite{abarzua_method_2015} and \cite{kabin_horizontal_2023}.
    
    \item No paper experimentally implemented Giraud and Verneuil’s atomic pattern: None
    
    \item 11 papers referred to Giraud and Verneuil’s work but implemented/investigated/proposed other atomic patterns: \cite{rondepierre_revisiting_2014}, \cite{kabin_ec_2021}, \cite{sigourou_successful_2023}, \cite{kabin_vulnerability_2023}, \cite{sako_side-channel_2016}, \cite{kabin_randomized_2023}, \cite{lu_general_2013}, \cite{abarzua_complete_2012}, \cite{abarzua_method_2015}, \cite{kabin_ec_2021-2} and \cite{kabin_horizontal_2023}.
    
    \item 15 papers referred to Giraud Verneuil’s work only in terms of the cost of field operations: \cite{baldwin_co-ecc_2012}, \cite{nitaj_memory-constrained_2011}, \cite{duquesne_choosing_2018}, \cite{abdulrahman_new_2013}, \cite{kim_speeding_2017}, \cite{hasan_abdulrahman_efficient_2013}, \cite{loai_review_2017}, \cite{lemke-rust_single-trace_2017}, \cite{abarzua_survey_2019}, \cite{tawalbeh_towards_2016}, \cite{rivain_fast_2011}, \cite{abarzua_survey_2021}, \cite{baldwin_hardware_2013}, \cite{danger_dynamic_2013} and \cite{el_mrabet_contributions_2017}.
    
    \item 6 papers analysed or surveyed different countermeasures against \ac{SCA}: \cite{danger_synthesis_2013}, \cite{meier_side-channel_2014}, \cite{abarzua_survey_2019}, \cite{abarzua_survey_2021}, \cite{nascimento_comparison_2014} and \cite{ryan_safe-errors_2016}.
    
    \item 10 papers discussed mainly \acp{SMA}: \cite{baldwin_co-ecc_2012}, \cite{nitaj_memory-constrained_2011}, \cite{duquesne_choosing_2018}, \cite{abdulrahman_new_2013}, \cite{zulberti_script-based_2022}, \cite{guerrini_randomized_2017}, \cite{kim_speeding_2017}, \cite{al-somani_method_2014}, \cite{hasan_abdulrahman_efficient_2013} and \cite{baldwin_hardware_2013}.
\end{enumerate}

Matthieu Rivain \cite{rivain_fast_2011}, for instance, addressed the computational costs of field operations crucial for cryptographic implementations, citing Giraud and Verneuil’s work. Other significant studies that have discussed the relevance of computational costs of field operations in relation to Giraud’s and Verneuil’s atomic pattern include \cite{danger_synthesis_2013}, \cite{duquesne_choosing_2018}, \cite{lemke-rust_single-trace_2017}, \cite{baldwin_co-ecc_2012}, \cite{hasan_abdulrahman_efficient_2013}, \cite{baldwin_hardware_2013} and \cite{danger_dynamic_2013}.

Some studies also evaluated the effectiveness of Giraud and Verneuil's atomic pattern in mitigating side-channel vulnerabilities across various attack scenarios and its compatibility with different \ac{EC} point multiplication algorithms to evaluate its impact on overall system security. For instance, Franck Rondepierre \cite{rondepierre_revisiting_2014} used the atomic pattern to propose secure \ac{EC} point multiplication formulas optimised for double scalar multiplications and resistant to simple \ac{SCA}. R. Abarzúa et al. \cite{abarzua_survey_2019} surveyed several algorithmic countermeasures to prevent passive \ac{SCA} on \ac{ECC} defined over prime fields, focusing on \acp{SMA} without precomputation.

Other notable contributions include Bauer et al. \cite{bauer_horizontal_2015}, who detailed an attack (Horizontal Collision Correlation Attack) on \ac{EC} atomic implementations with input randomisation and showed how it was successfully applied on Giraud and Verneuil’s atomic pattern. Fouque et al. \cite{ryan_safe-errors_2016} also studied the vulnerability of side-channel atomicity countermeasures in \ac{ECC} to C safe-error attacks, which exploit dummy operations in atomic patterns to recover bits of ephemeral nonces used in \aclp{ECDSA} and also show how it can be applied to Giraud and Verneuil’s atomic pattern. Abarzúa and Thériault \cite{abarzua_complete_2012} proposed new atomic blocks to enhance safety against \ac{SCA} and C-safe fault attacks during scalar multiplication for \acp{EC} over prime fields, offering both compactness and enhanced protection for improved performance on some atomic blocks including that of Giraud and Verneuil. H. Houssain \cite{houssain_elliptic_2012}, presented architectures for \ac{EC} cryptoprocessors resilient to power analysis attacks on resource-constrained devices like wireless sensor networks and Das et al. \cite{das_improved_2016} also discussed the Big Mac attack, which exploits operand sharing in \ac{RSA} and \ac{ECC} implementations despite conventional countermeasures, referencing Giraud and Verneuil's atomic pattern.

\vspace{2em}
\hspace{1em} While numerous research works have explored and analysed Giraud and Verneuil's atomic pattern in various contexts, to the best of our knowledge, there has been a notable gap in the literature regarding its practical implementation and the investigation of the indistinguishability of the atomic blocks when used for implementation of a binary \( k\mathbf{P} \) algorithm, particularly in embedded systems. In this thesis work, our primary objective is to address the aforementioned gap in the literature by implementing the atomic pattern proposed by Giraud and Verneuil in hardware on an embedded system. For this, we will measure and analyse the \ac{EM} trace generated during the execution of the \( k\mathbf{P} \) operation to investigate its resistance to \ac{SCA}.

        \chapter{Experimental Setup}
\label{chap:board}

This chapter describes the software and hardware platforms used for experiments and the implementation of the atomic pattern \( k\mathbf{P} \) algorithm investigated in this thesis.

\section{Board Selected For Experiments}
\begin{figure}[H]
    \centering
    \includegraphics[width=0.57\linewidth]{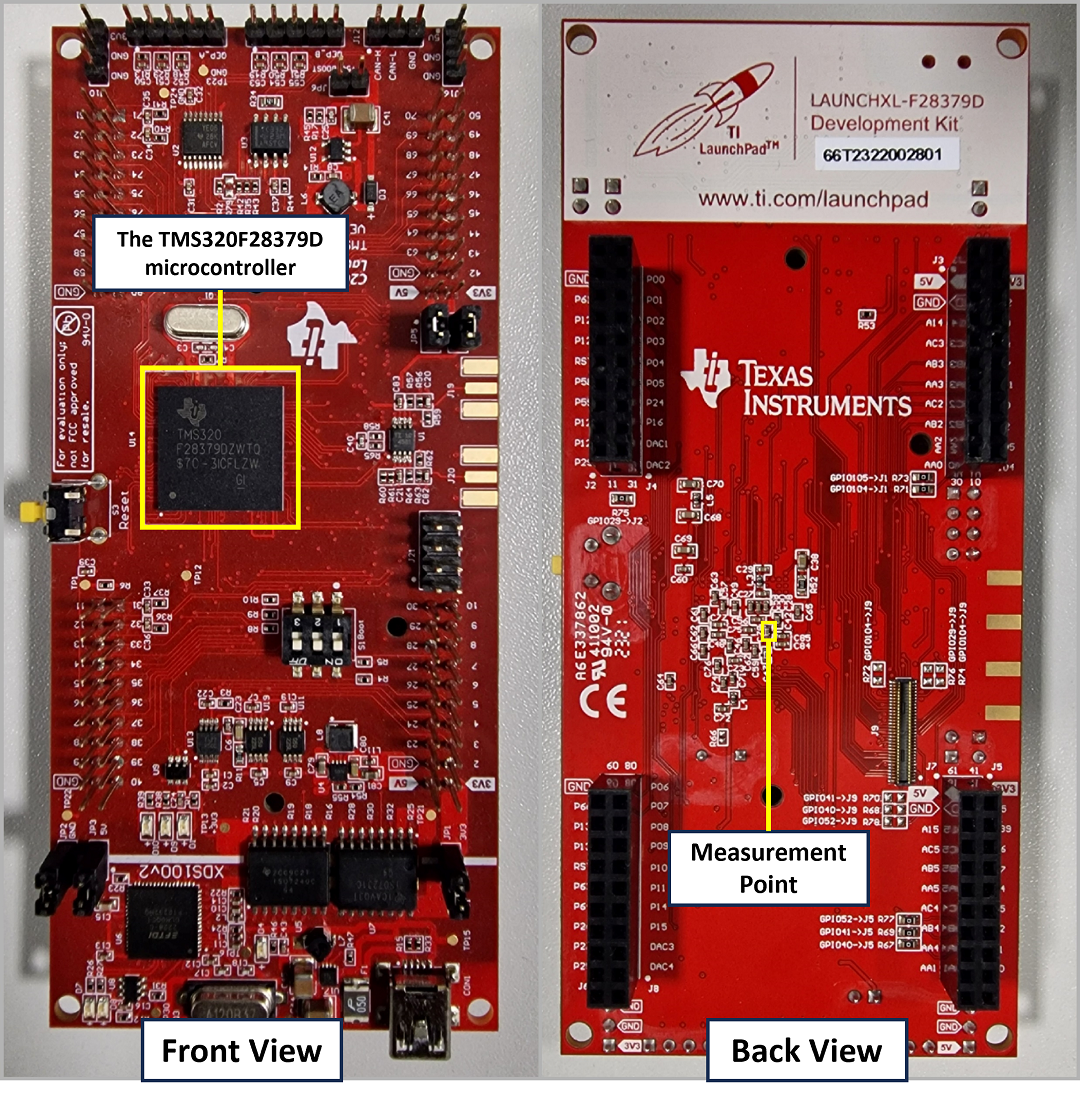}
    \caption{The front and back of the LAUNCHXL-F28379D development board}
    \label{fig:TMS320F28379D-mc}
\end{figure}

The experiments were conducted using the \acf{TI} LAUNCHXL-F28379D development board/kit \cite{noauthor_launchxl-f28379d_nodate}, an evaluation and development tool for the TMS320F28379D \acf{MCU}. This board is part of the \ac{TI} \ac{MCU} LaunchPad development kit family and is designed for prototyping and development. \hyperref[fig:TMS320F28379D-mc]{Figure~\ref*{fig:TMS320F28379D-mc}} shows the front and the back sides of the development board.

The TMS320F28379D microcontroller \cite{noauthor_tms320f28377dpdf_nodate} (\hyperref[fig:TMS320F28379D-mc]{Figure~\ref*{fig:TMS320F28379D-mc}}), is the core component of this development kit. This microcontroller was our target device for executing the atomic pattern \( k\mathbf{P} \) algorithm. It is a 32-bit \ac{MCU} from the C2000™ Delfino™ series of products. The \ac{MCU} operates real-time maximum at 200 MHz with dual 32-bit CPU cores. It also has 1 MB flash memory and 204 KB RAM making it suitable for various applications, including real-time control and digital signal processing.

\section{Suitability Of The Board For Thesis Objectives}
The \ac{TI} LAUNCHXL-F28379D development kit, was selected for this work because of its suitability for executing the atomic pattern \( k\mathbf{P} \) algorithm and conducting our experiments. 

The TMS320F28379D \ac{MCU} allowed for the execution of the \( k\mathbf{P} \) algorithm without running parallel processes. This was especially important for minimising noise in the measured \ac{EM} trace and ensuring accurate measurement results.

Additionally, the same board was used for similar investigations by Li et al. \cite{li_practical_2024} experimenting with Longa’s atomic pattern \cite{longa_accelerating_2008}. Moreover, a similar evaluation board \cite{noauthor_launchxl-f280025c_nodate} was applied for investigation of the resistance of Rondepierr’s atomic pattern \cite{rondepierre_revisiting_2014} by Sigourou et al. \cite{sigourou_successful_2023}. The use of the same microcontroller and measurement equipment as in this work enabled direct comparison of the results of this work with results published by Li et al. in \cite{li_practical_2024} and Sigourou et al. in \cite{sigourou_successful_2023}.
\section{Measurement Equipment}
\label{sec:other-hardware-tools}

\begin{enumerate}
    \item \textbf{LeCroy  WavePro 604HD Oscilloscope \cite{noauthor_teledyne_nodate}:} The LeCroy WavePro 604HD oscilloscope was used for capturing and measuring the \ac{EM} trace during \( k\mathbf{P} \) operations. It is a 4-channel high-definition oscilloscope featuring 6 GHz bandwidth, 20 GS/s sample rate, and 12-bit vertical resolution.
    
            \begin{figure}[H]
                \centering
                \includegraphics[width=0.5\linewidth]{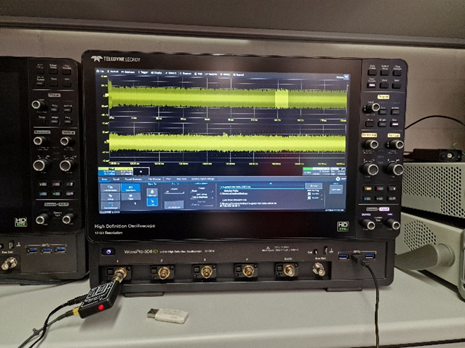}
                \caption{The LeCroy WavePro 604HD Oscilloscope}
                \label{fig:oscilloscope}
            \end{figure}

    \item \textbf{Langer ICS 105 Integrated Circuit Scanner \cite{noauthor_langer_nodate-1}:} The Langer ICS 105 ICS is a 4-axis positioning system designed for the movement and rotation of near-field probes over integrated circuits on printed circuit boards. It has a maximum positioning distance of (50 × 50 × 50) mm with \(\pm180^\circ\) rotational capability and a minimum positioning distance of (10 × 10 × 10) \(\mu\)m with \(1^\circ\) rotational increments. The positioning speed is specified at 2 mm/s\footnote{\url{https://www.langer-emv.de/fileadmin/2020.04.28\%20ICS\%20105\%20IC\%20Scanner\%20user\%20manual.pdf}}, with rotational speeds up to \(45^\circ/\text{s}\) or \(90^\circ/\text{s}\) depending on operation. It facilitates high-resolution measurements of high-frequency near fields. The system is compatible with various probes from Langer EMV-Technik GmbH, including the MFA-R 0.2-75 near-field probe used in this work.
    
            \begin{figure}[H]
                \centering
                \includegraphics[width=0.4\linewidth]{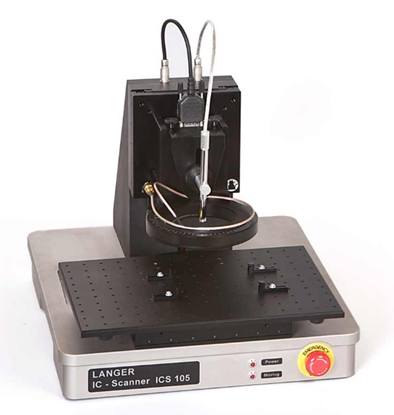}
                \caption{The Langer ICS 105 Integrated Circuit Scanner.\protect \footnotemark}
                \label{fig:integrated-circuit-scanner}
            \end{figure}

            \footnotetext{Source: \url{https://www.langer-emv.de/en/product/langer-scanner/41/ics-105-set-ic-scanner-4-axis-positioning-system/144} \cite{noauthor_langer_nodate-1}}

    \item \textbf{Langer MFA-R 0.2-75 Near Field Probe} \cite{noauthor_langer_nodate}: The Langer MFA-R 0.2-75 near-field probe was used to measure the \ac{EM} emissions produced during the execution of the \( k\mathbf{P} \) algorithm. It is an active near-field magnetic microprobe made for measuring high-resolution radio frequency magnetic fields ranging from 1 MHz to 1 GHz. It has a small and high-resolution probe head with a resolution of 300 \(\mu\)m, that is suitable for measurements near IC pins, fine conducting paths, or small surface-mount device components. The probe uses the BT 706 bias tee for operation, which supplies power to the integrated amplifier stage (9 V, 100 mA) and has an impedance of 50 \(\Omega\). The probe is electrically shielded to reduce interference and it also includes correction curves to convert output voltage into corresponding magnetic field strength or current measurements.

            \begin{figure}[H]
                \centering
                \includegraphics[width=0.5\linewidth]{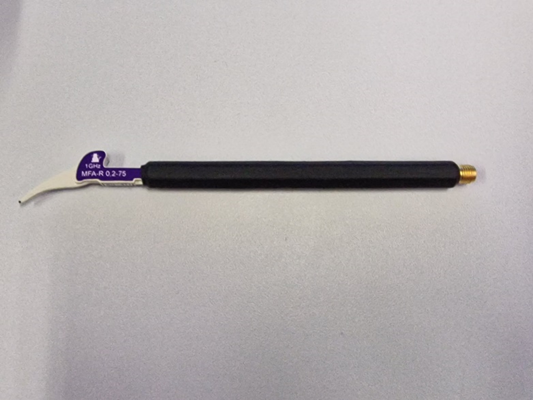}
                \caption{Langer MFA-R 0.2-75 Near Field Probe}
                \label{fig:near-field-probe}
            \end{figure}
            
\end{enumerate}
\section{Software Used In This Work}
\begin{enumerate}
    \item \textbf{\acf{CCS} \cite{noauthor_ccstudio_nodate}:} \ac{CCS} is an integrated development environment built by \acl{TI} for the specific purpose of supporting their \ac{MCU} and embedded processors. \ac{CCS} (version 11.2.0) was used to transfer the implementation of the atomic pattern \( k\mathbf{P} \) algorithm to the TMS320F28379D \ac{MCU}. It is compatible with Windows, Linux, and macOS platforms and offers a collection of tools for building, debugging, analysing, and optimising embedded applications. It supports the C/C++ programming languages and provides optimisation tools to improve the performance of code. For this work, \ac{CCS} helped to connect to the TMS320F28379D \ac{MCU}, which allowed us to execute the atomic pattern \( k\mathbf{P} \) algorithm. The debugging environment also helped a lot with finding and fixing possible problems with code.
    
    \item \textbf{C Programming:} C programming language was chosen for implementing the atomic pattern on the \ac{MCU} because the cryptographic library used for the atomic pattern \( k\mathbf{P} \) algorithm is written in C which makes it the most suitable option for this implementation. Additionally, using C allowed us to maintain the same development environment to that of Li et al. \cite{li_practical_2024} and Sigourou et al. \cite{sigourou_successful_2023} who both used the same cryptographic library in their work. This was again necessary to enable direct comparison of results. In addition, C is efficient, provides low-level control over hardware, and is widely used in embedded systems.
    
    \item \textbf{Python Programming \cite{noauthor_welcome_2024}:} The analysis of the measured \ac{EM} trace was done using Python. Its libraries, including Pandas for data handling and Matplotlib for data visualisation, was used for the processing of the \ac{EM} trace collected during the measurements and understanding the results.
\end{enumerate}
\section{Measurement Setup And Collection Of EM Trace}

\begin{figure}[H]
            \centering
            \includegraphics[width=0.3\linewidth]{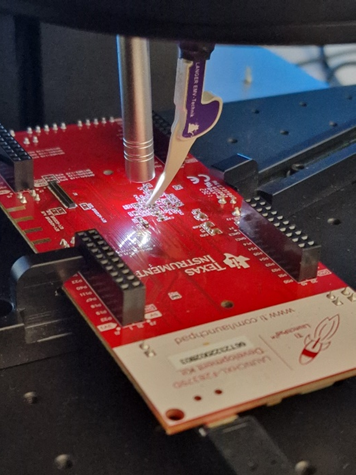}
            \caption{The Near-Field Probe being positioned on the capacitor with the help of the ICS}
            \label{fig:equipment-setup}
        \end{figure}

The TMS320F28379D microcontroller was connected to the PC via a USB interface on the development board, allowing interaction with \ac{CCS} for code execution. The Integrated Circuit Scanner was then used to precisely position the development board and the near-field probe. \hyperref[fig:equipment-setup]{Figure~\ref*{fig:equipment-setup}} shows the position of the \ac{EM} probe over the development board during measurements.

\begin{figure}[H]
            \centering
            \includegraphics[width=0.4\linewidth]{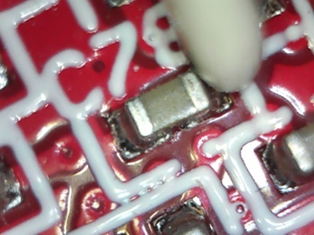}
            \caption{Positioning of the near-field probe near capacitor C78 on the \ac{MCU}}
            \label{fig:capacitor78}
        \end{figure}

To determine the optimal position of the probe, we reviewed the schematics of the LAUNCHXL-F28379D development board \cite{noauthor_sprui77cpdf_nodate}. We focused on capacitors in the power line (C42, C46-49, and C75-78) because these are most likely to produce strong \ac{EM} signals. After testing various probe placements while executing the atomic pattern \( k\mathbf{P} \) algorithm in RAM, we found that positioning the probe near C78 provided the strongest signal-to-noise ratio. This position was selected for all subsequent \ac{EM} trace measurements. \hyperref[fig:capacitor78]{Figure~\ref*{fig:capacitor78}} shows the near-field probe positioned near the C78 power line. 

For our \( k\mathbf{P} \) algorithm implementation, with a development board operating at a reduced frequency of 100 MHz, executing a \acl{PD} operation in RAM takes an average execution time of \( 278,826 \) clock cycles, while the first \acl{PA} atomic block takes an average of \( 278,822 \) clock cycles and the second \acl{PA} atom takes an average of \( 278,817 \) clock cycles (implementation details are described in \hyperref[sec:em-data-processing]{Section~\ref*{sec:em-data-processing}}). When performing \( k\mathbf{P} \) with a 256-bit key, a "dummy" scalar \(k\) with the last significant bit set to \(1\) and all other bits set to \(0\) \((k = 000 \dots 1)\) would require an execution time of about \((278,826 + 278,822 + 278,817) + (278,826 \cdot 255) = 71,937,095\) clock cycles, while a scalar \(k\) with all bits set to \(1\) \((k = 111 \dots 1)\) would require \((278,826 + 278,822 + 278,817) \cdot 256 = 214,135,040\) clock cycles. When using a sufficiently high sampling rate, for example \( 10 \, \text{GS/s} \), this would result in 100 samples per clock cycle, generating a large amount of data which would require a large amount of storage space. The storage space needed to store this data would be large, and processing and analysing it would require significant time and computational resources, even if the oscilloscope could technically handle the requirements. Despite using modern equipment, these limitations made it infeasible to perform \( k\mathbf{P} \) using the full 256-bit key space specified by the \ac{NIST} standard \cite{technology_digital_2023}. The \ac{NIST} standard defines the key space up to 256 bits to ensure cryptographic strength, however in our experiments, we decided to use a 10-bit long scalar key. In this work, we selected the 10-bit long scalar \(k = 1111111111\) to make it easier to identify and analyse the point operations. The selected scalar value ensured an equal number of \acl{PD} and \acl{PA} operations, which provided a balanced dataset for our analysis. 

As the input point for the executed \( k\mathbf{P} \) operation we selected the base point \( \boldsymbol{G} \) of the \ac{EC} P-256 as given below:

        \begin{table}[H]
        \renewcommand{\arraystretch}{1.5} 
        \centering
        \Large %
        \resizebox{\textwidth}{!}{%
        \begin{tabular}{|c|l|}
        \hline
        \rowcolor{orange!20} 
        \multicolumn{2}{|l|}{\textbf{Base Point \( \boldsymbol{G = (x, y)} \)}} \\ \hline
        \textbf{\( x \)} & 6B17D1F2E12C4247F8BCE6E563A440F277037D812DEB33A0F4A13945D898C296 \\ \hline
        \textbf{\( y \)} & 4FE342E2FE1A7F9B8EE7EB4A7C0F9E162BCE33576B315ECECBB6406837BF51F5 \\ \hline
        \end{tabular}
        }
        \caption{Base Point \( G \) Coordinates of the \ac{EC} P-256 Curve}
        \label{tab:base_point_g}
        \end{table}

In real-world scenarios, where a 256-bit key space is standard, attackers would face significant challenges in measuring such a lengthy key due to the high memory and expensive resources required to handle these operations. We believe that these constraints are the primary reasons why similar experimental investigations are limited in this area. Our experiment, focused on identifying potential distinguishability in \acl{PD} and \acl{PA} operations, assumes that a 10-bit key suffices to meet this objective.

The \ac{EM} emanations were measured with an oscilloscope sampling rate of \( 5 \, \text{GS/s} \) during the \( k\mathbf{P} \) execution. The signal frequency of the \ac{MCU} board was reduced to 100 MHz to improve the balance of signal-to-noise ratio and reduce noise in signal capture. Lowering the frequency minimised high-frequency noise in the signals, resulting in cleaner and more analysable \ac{EM} traces. This configuration resulted in 50 samples per clock cycle. In total, the measured \ac{EM} trace consisted of 500 million samples and was approximately \( 15.2 \, \text{GB} \) in size.

The \( k\mathbf{P} \) execution with the 10-bit scalar key generated ten (10) atomic patterns for the point doubling operations and twenty (20) atomic patterns for point addition (10 \acp{PA} for Addition 1 atomic block and another 10 \acp{PA} for Addition 2 atomic block). See further implementation details in \hyperref[chap:implementation]{Chapter~\ref*{chap:implementation}}. 

\hyperref[tab:data_collection_config]{Table~\ref*{tab:data_collection_config}} summarises the measurement settings:

        \begin{table}[H]
        \renewcommand{\arraystretch}{1.5} 
        \centering
        \begin{tabular}{|p{7cm}|p{5cm}|}
        \hline
        \rowcolor{orange!20} 
        \textbf{Setting} & \textbf{Value} \\ \hline
        Microcontroller Clock Frequency & 100 MHz \\ \hline
        Sampling Rate of oscilloscope & 5 GS/s \\ \hline
        Samples per clock cycle & 50 \\ \hline
        Number of Samples & 500 million \\ \hline
        Total File Size & \textasciitilde 15.2 GB \\ \hline
        Scalar Key-Bit Length & 10-bit \((k = 1111111111)\) \\ \hline
        Number of Atomic Patterns for \ac{EC} \acl{PD} & 10 \\ \hline
        Number of Atomic Patterns for \ac{EC} \acl{PA} & 20 (10 for each \ac{PA} atomic block) \\ \hline
        \end{tabular}
        \caption{Configurations applied in our setup for the measurement of the \ac{EM} trace.}
        \label{tab:data_collection_config} 
        \end{table}
        
In this work, we did not capture \ac{EM} trace from Flash memory because it was not practical for this type of analysis. Flash memory generally has slower read/write speed and fewer \ac{EM} signal transitions compared to RAM. This means that the information provided is less useful for \ac{SCA}. Variability in the amplitude signals in an \ac{EM} trace is important for extracting useful information for \ac{SCA}. To obtain meaningful data from Flash memory, a much higher sampling rate would be required to capture subtle variations in the \ac{EM} emanations. This would generate a significantly larger trace file, requiring substantial storage and computational resources to process, which was beyond the scope of this study.

Details about the processing and analysis of the data will further be described in \hyperref[chap:analysis]{Chapter~\ref*{chap:analysis}}.

        \chapter{Selection of an Open-Source Cryptographic Library}
\label{chap:crypto-library}

In this chapter, we will discuss the selection of the open-source \ac{ECC} cryptographic library used in the implementation of the atomic pattern \( k\mathbf{P} \) algorithm.  

To prevent \ac{SCA} attacks, the library that is chosen must support \ac{ECC}, let you perform modular arithmetic operations in finite prime fields, and support constant-time functions. The objectives of this thesis are closely aligned with the work by Sigourou et al. \cite{sigourou_successful_2023} who implemented and evaluated the resistance of the atomic pattern \( k\mathbf{P} \) algorithm introduced by Rondepierre \cite{rondepierre_revisiting_2014}. Their study evaluated five cryptographic libraries, including OpenSSL \cite{noauthor_openssl_nodate} [3], MIRACL \cite{noauthor_miraclmiracl_2024}, Cryptlib \cite{noauthor_cryptlib_nodate} [5], Crypto ++ \cite{weidai11_weidai11cryptopp_2024}, and \texttt{FLECC\_IN\_C} (FLECC) \cite{noauthor_iaikflecc_in_c_2022} to see which would best suit \ac{ECC} implementation on resource-constrained devices. They found that FLECC would be the most suitable cryptographic library while considering factors such as portability of their implemented \( k\mathbf{P} \) algorithm to the device under attack, memory usage, and the library's access to constant-time functions, although some functions were not fully constant-time at the hardware level.

In this work, we build on the findings of Sigourou et al. \cite{sigourou_successful_2023}, to implement a binary \( k\mathbf{P} \) algorithm using the atomic pattern for \ac{EC} point doubling and addition proposed by Giraud and Verneuil in \cite{giraud_atomicity_2010}. As mentioned earlier, an important requirement for this implementation is that the cryptographic library can be ported to the device under test. Since Sigourou et al. \cite{giraud_atomicity_2010} concluded that FLECC \cite{noauthor_iaikflecc_in_c_2022} fulfilled all the requirements for the implementation of the atomic pattern \( k\mathbf{P} \) algorithm on a resource-constrained device, we examined FLECC to determine whether it would support our \( k\mathbf{P} \) implementation in our specific environment and on our target device.

We also reviewed two other libraries, i.e. mbed TLS \cite{noauthor_mbed-tlsmbedtls_2024} and Micro-ECC \cite{mackay_kmackaymicro-ecc_2024} that were mentioned in the study by Tjerand Silde \cite{silde_comparative_nodate} for their constant-time capabilities. We finally chose FLECC for the implementation of the atomic pattern \( k\mathbf{P} \) algorithm after comparing the suitability of other choices since it most satisfied the requirements of this work.

Another important reason we chose FLECC was to ensure that our development environment stayed the same as that used in previous studies by Li et al. \cite{li_distinguishability_2024} and Sigourou et al. \cite{sigourou_successful_2023}, who both used the same cryptographic library. In order to accurately compare the results of this work with their findings, it was important to ensure that our setup was the same as theirs, which was achieved by selecting FLECC.

\hyperref[tab:library_comparison]{Table~\ref*{tab:library_comparison}} below summarises the results of the assessment.

\begin{table}[H]
\renewcommand{\arraystretch}{2}
\centering
\Large %
\resizebox{\textwidth}{!}{%
\begin{tabular}{|l|l|l|l|l|}
\hline
\rowcolor{orange!20} 
\textbf{Library} & \textbf{\begin{tabular}[c]{@{}c@{}}Suitable for\\ Embedded Devices\end{tabular}} & \textbf{\begin{tabular}[c]{@{}c@{}}Access to\\ Modular Arithmetic\\ Functions\end{tabular}} & \textbf{\begin{tabular}[c]{@{}c@{}}Constant\\ Time Functions\end{tabular}} & \textbf{\begin{tabular}[c]{@{}c@{}}Portable to\\ TMS320F28379D\\ Microcontroller Board\end{tabular}} \\ \hline
Mbed TLS \cite{noauthor_mbed-tlsmbedtls_2024} & Yes & Yes & No & Not confirmed \\ \hline
Micro-ECC \cite{mackay_kmackaymicro-ecc_2024} & Yes & Yes & Yes & Not confirmed \\ \hline
FLECC\_IN\_C \cite{noauthor_iaikflecc_in_c_2022} & Yes & Yes & Yes & Yes \\ \hline
\end{tabular}
}
\caption{Comparison of Cryptographic Libraries}
\label{tab:library_comparison}
\end{table}

\section{Mbed TLS}

Mbed TLS \cite{noauthor_mbed-tlsmbedtls_2024} is cryptographic library that is written in C and supports the SSL/TLS and DTLS protocols as well as the cryptographic primitives. It is well-suited for embedded systems. According to the work by Tjerand Silde \cite{silde_comparative_nodate}, Mbed TLS supports constant-time functions to protect against timing attacks.

Despite this, Mbed TLS was not chosen for this project for multiple reasons. First, it was not clear if it was compatible with the TMS320F28379D microcontroller board used in this work. Because there was no clear evidence that it could be easily used with this specific hardware, Mbed TLS did not meet the portability requirement and was thus excluded from further consideration. Also, none of the Mbed TLS modular arithmetic functions were explicitly defined or documented as constant-time functions.
\section{Micro-ECC}

Micro-ECC \cite{mackay_kmackaymicro-ecc_2024} is a lightweight, open-source cryptographic library designed for the use in embedded systems. Micro-ECC is implemented in C and is known for its small code size. The author of Micro-ECC claims the library is resistant to known \ac{SCA} attacks and it includes modular arithmetic functions that is designed to support constant-time executions to prevent timing attacks. Additionally, it supports the \ac{NIST} curve P-256 used in this work.

Despite these advantages, Micro-ECC was not selected for this project. As documented by Tjerand Silde \cite{silde_comparative_nodate}, the library’s claims of \ac{SCA} resistance lacks substantial support, both in academic literature and the library’s official documentation. Again, Tjerand Silde \cite{silde_comparative_nodate} mentioned that randomisation is inconsistently applied in certain operations, increasing its vulnerability to \ac{SCA} attacks. Due to these concerns, Micro-ECC did not meet our requirements of this work and was therefore excluded from further consideration.
\section{FLECC\_IN\_C (FLECC)}

FLECC (The Flexible Elliptic Curve Cryptography) \cite{noauthor_iaikflecc_in_c_2022} is a lightweight and efficient cryptographic library written in C. It is suitable for resource-constrained devices and it was compatible with our experimental device. It passed all our checks including providing constant-time functions as documented in the library’s description \footnote{\url{https://github.com/IAIK/flecc_in_c/blob/develop/src/gfp/gfp_const_runtime.c}}  and could easily be ported to the device under test. Additionally, although FLECC does not have a single function for constant-time modular multiplication of two numbers (i.e., \( a \cdot b \mod p \)), its constant-time Montgomery modular multiplication can be applied twice to achieve the same result for field multiplications. In a study by Tjerand Silde \cite{silde_comparative_nodate}, FLECC was found to be one of the fastest and most secure \ac{ECC} libraries for embedded devices, with strong countermeasures against \ac{SCA} attacks. These security features, along with research done by Sigourou et al. \cite{sigourou_successful_2023} on FLECC and its strong performance, made FLECC the best choice for implementing the \( k\mathbf{P} \) algorithm in this work.
        \chapter{Implementation Of The Atomic Pattern kP Algorithm Using FLECC Library}
\label{chap:implementation}

In this chapter, we present our implementation of the \ac{EC} point multiplication using the atomic patterns for
\ac{EC} point doubling and point addition proposed by Giraud and Verneuil \cite{giraud_atomicity_2010}. We implemented the right-to-left \( k\mathbf{P} \) algorithm for the \ac{NIST} \ac{EC} P-256 (also known as secp256r1) curve \cite{technology_digital_2023}. The implemented atomic pattern, illustrated in \hyperref[tab:implemented-atomic-pattern]{Table~\ref*{tab:implemented-atomic-pattern}},  use a sequence of operations to perform \ac{EC} point addition and point doubling operations. The \ac{EC} point addition pattern consist of two atomic blocks. The EC point doubling pattern consists of one atomic block. Each atomic block is a sequence of eight (8) field multiplications and ten (10) field additions.

The implemented atomic block for point doubling uses the same point twice as input, denoted here as \(P1 = P2\), where \( \boldsymbol{P1 = (X_1: Y_1: Z_1: W_1)} \). Hence, the double of the point \( \boldsymbol{P = (X_1: Y_1: Z_1: W_1)} \) is the point \( \boldsymbol{2P = (X_2: Y_2: Z_2: W_2)} \), where \(W_1\) is denoted by \( aZ^4 \) (with \( a = -3 \)).

The implemented atomic blocks for point addition takes two distinct points, \(P\) and \(Q\), and computes \(P + Q\). The sum of \( \boldsymbol{P = (X_1: Y_1: Z_1)} \) and \( \boldsymbol{Q = (X_2: Y_2: Z_2)} \), with \(P, Q\) not equal to \( \mathcal{O} \) and \(P\) not equal to \( \pm Q \), is the point \( \boldsymbol{P + Q = (X_3: Y_3: Z_3)} \).

As stated earlier, FLECC emerged as the most suitable cryptographic library for our implementation of the \( k\mathbf{P} \) algorithm. Its built-in constant-time arithmetic functions, including \texttt{gfp\_cr\_add} (for field addition), \texttt{gfp\_cr\_subtract} (for field subtraction), and \texttt{gfp\_cr\_mont\_multiply\_sos} (for field multiplication) were used to perform the \ac{EC} point doubling and \ac{EC} point addition operations and to generate points on the \acl{ec}.

Montgomery modular multiplication operates by transforming values into a special "Montgomery space", where modular multiplications can be performed efficiently. Afterwards, the values are converted back to their original form. In FLECC, this method uses the Separated Operand Scanning (SOS) approach, with predefined parameters like \( R = 2^{(256) (8)} = 2^{2048} \). The constant-time multiplication function independently did not always yield correct results, so the \texttt{gfp\_mult\_two\_mont} function in FLECC, which uses two constant-time Montgomery multiplications, was implemented to ensure the results were correct. This change doubled the number of field multiplications in the pattern, reducing the original design's performance due to the increased computational cost. In summary, the multiplication \( c = a \cdot b \mod p \) is performed in two steps: first as \( c = a \cdot b \cdot R^{-1} \), and then \( c = c \cdot R^2 \cdot R^{-1} \). The final function is shown in \hyperref[fig:modified-montgomery]{Figure~\ref*{fig:modified-montgomery}} below:

\begin{figure}[H]
    \centering
    \includegraphics[width=1\linewidth]{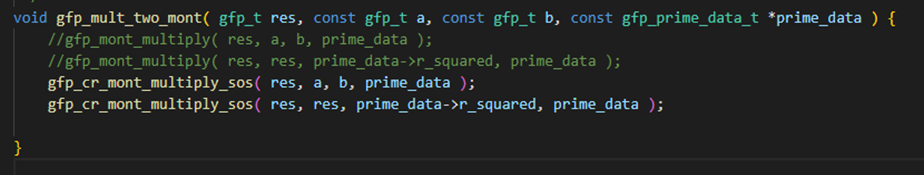}
    \caption{Modified Montgomery Multiplication Function in FLECC to Perform Field Multiplications}
    \label{fig:modified-montgomery}
\end{figure}

The atomic blocks for both point doubling and point addition operations, consist entirely of the same sequence of field operations, with the only difference being the values of the curve point coordinates used at each step of the execution process and the registers used for storing the intermediate operations. Dummy operations were included in both point addition atomic blocks for uniformity. These dummy operations do not alter the results in any way. They were only included to ensure that the atomic block for point doubling and the atomic blocks for point addition executed the same sequence of field operations each time. You can see OP 2, OP 4, OP 6, OP 8, OP 9, OP 12, OP 15, and OP 16 for reference in \hyperref[tab:implemented-atomic-pattern]{Table~\ref*{tab:implemented-atomic-pattern}}. In \cite{noauthor_aromalaryeaanalysis--kp-algorithm_nodate}, you can find the complete source code, demonstrating the implementation of the atomic pattern with the right-to-left scalar multiplication.

\begin{table}[H]
\renewcommand{\arraystretch}{1.5} 
\centering
\resizebox{1.1\textwidth}{\height}{%
\begin{tabular}{|l|llll|ll|}
\hline
\rowcolor{orange!20} 
\textbf{Operations} & \multicolumn{4}{c|}{\textbf{\begin{tabular}[c]{@{}c@{}}EC Point Addition\\ $P = (X_1: Y_1: Z_1)$\\ $Q = (X_2: Y_2: Z_2)$\\ $P + Q = (X_3: Y_3: Z_3)$\end{tabular}}} & \multicolumn{2}{c|}{\textbf{\begin{tabular}[c]{@{}c@{}}EC Point Doubling\\ $P = (X_1: Y_1: Z_1: W_1)$\\ $2P = (X_2: Y_2: Z_2: W_2)$\end{tabular}}} \\ \hline

OP1 & \multicolumn{1}{r|}{\multirow{18}{*}{\textbf{Add. 1}}} & \multicolumn{1}{l|}{$R_1 \leftarrow Z_2^2$} & \multicolumn{1}{r|}{\multirow{18}{*}{\textbf{Add. 2}}} & $R_1 \leftarrow R_6^2$ & \multicolumn{1}{r|}{\multirow{18}{*}{\textbf{Dbl.}}} & $R_1 \leftarrow X_1^2$ \\ \cline{1-1} \cline{3-3} \cline{5-5} \cline{7-7} 

OP2 & \multicolumn{1}{r|}{} & \multicolumn{1}{l|}{\textbf{tmpReg \( \leftarrow R_1 + Z_1 \)}} & \multicolumn{1}{r|}{} & \textbf{tmpReg \( \leftarrow R_2 + R_3 \)} & \multicolumn{1}{r|}{} & \( R_2 \leftarrow Y_1 + Y_1 \) \\ \cline{1-1} \cline{3-3} \cline{5-5} \cline{7-7}

OP3 & \multicolumn{1}{r|}{} & \multicolumn{1}{l|}{$R_2 \leftarrow Y_1 \times Z_2$} & \multicolumn{1}{r|}{} & $R_4 \leftarrow R_5 \times R_1$ & \multicolumn{1}{r|}{} & $Z_2 \leftarrow R_2 \times Z_1$ \\ \cline{1-1} \cline{3-3} \cline{5-5} \cline{7-7} 

OP4 & \multicolumn{1}{r|}{} & \multicolumn{1}{l|}{\textbf{tmpReg \( \leftarrow R_2 + Z_2 \)}} & \multicolumn{1}{r|}{} & {\textbf{tmpReg \( \leftarrow R_4 + R_3 \)}} & \multicolumn{1}{r|}{} & $R_4 \leftarrow R_1 + R_1$ \\ \cline{1-1} \cline{3-3} \cline{5-5} \cline{7-7} 

OP5 & \multicolumn{1}{r|}{} & \multicolumn{1}{l|}{$R_5 \leftarrow Y_2 \times Z_1$} & \multicolumn{1}{r|}{} & $R_5 \leftarrow R_1 \times R_6$ & \multicolumn{1}{r|}{} & $R_3 \leftarrow R_2 \times Y_1$ \\ \cline{1-1} \cline{3-3} \cline{5-5} \cline{7-7} 

OP6 & \multicolumn{1}{r|}{} & \multicolumn{1}{l|}{\textbf{tmpReg \( \leftarrow R_2 + Z_2 \)}} & \multicolumn{1}{r|}{} & {\textbf{tmpReg \( \leftarrow R_5 + R_6 \)}} & \multicolumn{1}{r|}{} & $R_6 \leftarrow R_3 + R_3$ \\ \cline{1-1} \cline{3-3} \cline{5-5} \cline{7-7} 

OP7 & \multicolumn{1}{r|}{} & \multicolumn{1}{l|}{$R_3 \leftarrow R_1 \times R_2$} & \multicolumn{1}{r|}{} & $R_1 \leftarrow Z_1 \times R_6$ & \multicolumn{1}{r|}{} & $R_2 \leftarrow R_6 \times R_3$ \\ \cline{1-1} \cline{3-3} \cline{5-5} \cline{7-7} 

OP8 & \multicolumn{1}{r|}{} & \multicolumn{1}{l|}{\textbf{tmpReg \( \leftarrow R_3 + R_1 \)}} & \multicolumn{1}{r|}{} & {\textbf{tmpReg \( \leftarrow R_1 + Z_1 \)}} & \multicolumn{1}{r|}{} & $R_1 \leftarrow R_4 + R_1$ \\ \cline{1-1} \cline{3-3} \cline{5-5} \cline{7-7} 

OP9 & \multicolumn{1}{r|}{} & \multicolumn{1}{l|}{\textbf{tmpReg \( \leftarrow R_5 + Z_2 \)}} & \multicolumn{1}{r|}{} & {\textbf{tmpReg \( \leftarrow R_2 + R_6 \)}} & \multicolumn{1}{r|}{} & $R_1 \leftarrow R_1 + W_1$ \\ \cline{1-1} \cline{3-3} \cline{5-5} \cline{7-7} 

OP10 & \multicolumn{1}{r|}{} & \multicolumn{1}{l|}{$R_4 \leftarrow Z_1^2$} & \multicolumn{1}{r|}{} & $R_6 \leftarrow R_2^2$ & \multicolumn{1}{r|}{} & $R_3 \leftarrow R_1^2$ \\ \cline{1-1} \cline{3-3} \cline{5-5} \cline{7-7} 

OP11 & \multicolumn{1}{r|}{} & \multicolumn{1}{l|}{$R_2 \leftarrow R_5 \times R_4$} & \multicolumn{1}{r|}{} & $Z_3 \leftarrow R_1 \times Z_2$ & \multicolumn{1}{r|}{} & $R_4 \leftarrow R_6 \times X_1$ \\ \cline{1-1} \cline{3-3} \cline{5-5} \cline{7-7} 

OP12 & \multicolumn{1}{r|}{} & \multicolumn{1}{l|}{\textbf{tmpReg \( \leftarrow R_2 + R_4 \)}} & \multicolumn{1}{r|}{} & $R_1 \leftarrow R_4 + R_4$ & \multicolumn{1}{r|}{} & $R_5 \leftarrow W_1 + W_1$ \\ \cline{1-1} \cline{3-3} \cline{5-5} \cline{7-7} 

OP14 & \multicolumn{1}{r|}{} & \multicolumn{1}{l|}{$R_5 \leftarrow R_1 \times X_1$} & \multicolumn{1}{r|}{} & $R_1 \leftarrow R_5 \times R_3$ & \multicolumn{1}{r|}{} & $W_2 \leftarrow R_2 \times R_5$ \\ \cline{1-1} \cline{3-3} \cline{5-5} \cline{7-7} 

OP15 & \multicolumn{1}{r|}{} & \multicolumn{1}{l|}{\textbf{tmpReg \( \leftarrow R_5 - X_1 \)}} & \multicolumn{1}{r|}{} & $X_3 \leftarrow R_6 - R_5$ & \multicolumn{1}{r|}{} & $X_2 \leftarrow R_3 - R_4$ \\ \cline{1-1} \cline{3-3} \cline{5-5} \cline{7-7} 

OP16 & \multicolumn{1}{r|}{} & \multicolumn{1}{l|}{\textbf{tmpReg \( \leftarrow R_5 - X_2 \)}} & \multicolumn{1}{r|}{} & $R_4 \leftarrow R_4 - X_3$ & \multicolumn{1}{r|}{} & $R_6 \leftarrow R_4 - X_2$ \\ \cline{1-1} \cline{3-3} \cline{5-5} \cline{7-7} 

OP17 & \multicolumn{1}{r|}{} & \multicolumn{1}{l|}{$R_6 \leftarrow X_2 \times R_4$} & \multicolumn{1}{r|}{} & $R_3 \leftarrow R_4 \times R_2$ & \multicolumn{1}{r|}{} & $R_4 \leftarrow R_6 \times R_1$ \\ \cline{1-1} \cline{3-3} \cline{5-5} \cline{7-7} 

OP18 & \multicolumn{1}{r|}{} & \multicolumn{1}{l|}{$R_6 \leftarrow R_6 - R_5$} & \multicolumn{1}{r|}{} & $Y_3 \leftarrow R_3 - R_1$ & \multicolumn{1}{r|}{} & $Y_2 \leftarrow R_4 - R_2$ \\ \hline

\end{tabular}
}
\caption{The implemented atomic pattern using constant-time FLECC functions and dummy operations.}
\label{tab:implemented-atomic-pattern}
\end{table}

        \chapter{Measurement and Analysis of Electromagnetic Trace}
\label{chap:analysis}

\acl{EM} emanations can be measured non-invasively unlike power traces, which require physical changes to the microcontroller, such as soldering or adding components like resistors to detect current. Thus, a probe positioned close to the microcontroller may identify the \ac{EM} signals without changing the device or its circuit so the device can maintain its usual operation. For this reason, during the execution of the \( k\mathbf{P} \) operation, we concentrated on measuring and analysing the \ac{EM} emanations only.

\section{Evaluation of The Execution Time of The FLECC Constant-time Operations}
\label{sec:eval-flecc-cr-ops}

To verify that the operations in the FLECC library were genuinely executing in constant-time, we measured the execution times of the first instance of each field operation i.e. the constant-time Montgomery field multiplication, constant-time field addition, and constant-time field subtraction - used by the first point doubling operation while processing the first key bit only \texttt{(Doubling)\textsubscript{1}}. Using \ac{CCS}, we inserted software breakpoints between the various field operations in the Doubling atomic block and executed the code from RAM to measure the execution time. For consistency, we repeatedly measured the execution time of the first key bit for ten (10) times. In each iteration, we obtained the same execution time for each first field operation in the Doubling atomic block which indicated that the point operations were indeed executed in constant-time as intended. The results of these measurements, provided in \hyperref[tab:flecc_execution_time]{Table~\ref*{tab:flecc_execution_time}}, show the execution time in clock cycles and the equivalent time in milliseconds for the first field multiplication, the first field addition and the first field subtraction, in the atomic block for point doubling. This consistency supports that the FLECC library’s implementation operates with a constant runtime for each point doubling operation while processing the first key-bit.

\begin{table}[H]
\renewcommand{\arraystretch}{1.5} 
\centering
\begin{tabular}{|l|l|l|}
\hline
\rowcolor{yellow!40} 
\multicolumn{3}{|p{\textwidth}|}{\textbf{Execution Time of the FLECC Constant Time operation of \texttt{(Doubling)\textsubscript{1}}, measured 10 Times.}} \\ \hline
\rowcolor{orange!20} 
\textbf{Doubling Atomic Block} & \textbf{Num of Clock Cycles} & \textbf{Time (ms)} \\ \hline
1st Field Multiplication (Measured 10x) & 33153 & 0.3315 \\ \hline
1st Field Addition (Measured 10x) & 1355 & 0.0136 \\ \hline
1st Field Subtraction (Measured 10x) & 1351 & 0.0135 \\ \hline
\end{tabular}
\caption{Execution times of the FLECC constant-time functions for the first field Multiplication, the first field Additon and first field Subtraction, measured 10 times while processing the first key bit - \texttt{(Doubling)\textsubscript{1}}}
\label{tab:flecc_execution_time} 
\end{table}

A variation of 5 clock cycles was observed while analysing the results of the execution times of the first distinct field operation across the atomic blocks. For example, the first field multiplication while processing the first key bit took \(33,153\) clock cycles, while the first field multiplication while processing the seventh key bit took \(33,158\) clock cycles. \hyperref[tab:execution_time_analysis]{Table~\ref*{tab:execution_time_analysis}} presents the execution times of the first field multiplication, the first field addition and the first field subtraction while processing each key bit across the atomic blocks measured in clock cycles. For a comprehensive view, \hyperref[appendix:B]{Appendix~\ref*{appendix:B}} provides a detailed table showing the execution time of each field operation within each atomic block for all ten key bits.

{
\renewcommand{\arraystretch}{1.0} 
\begin{longtable}{|p{5cm}|p{3.5cm}|p{2.5cm}|p{2.5cm}|}
\hline
\rowcolor{orange!20}
\textbf{Addition 1 Atomic Block} & \textbf{1st Field Multiplication} & \textbf{1st Field Addition} & \textbf{1st Field Subtraction} \\ \hline
\endfirsthead
1st Key Bit & 33158 & 1355 & 1351 \\ \hline
2nd Key Bit & 33158 & 1355 & 1356 \\ \hline
3rd Key Bit & 33153 & 1360 & 1356 \\ \hline
4th Key Bit & 33153 & 1360 & 1351 \\ \hline
5th Key Bit & 33153 & 1355 & 1356 \\ \hline
6th Key Bit & 33153 & 1355 & 1356 \\ \hline
7th Key Bit & 33158 & 1360 & 1356 \\ \hline
8th Key Bit & 33153 & 1360 & 1356 \\ \hline
9th Key Bit & 33158 & 1360 & 1356 \\ \hline
10th Key Bit & 33153 & 1355 & 1351 \\ \hline
\textbf{Max} & \textbf{33158} & \textbf{1360} & \textbf{1356} \\ \hline
\textbf{Min} & \textbf{33153} & \textbf{1355} & \textbf{1351} \\ \hline
\textbf{Difference in Clock Cycles} & \textbf{5} & \textbf{5} & \textbf{5} \\ \hline
\rowcolor{orange!20}
\textbf{Addition 2 Atomic Block} & \textbf{1st Field Multiplication} & \textbf{1st Field Addition} & \textbf{1st Field Subtraction} \\ \hline
1st Key Bit & 33151 & 1355 & 1351 \\ \hline
2nd Key Bit & 33151 & 1360 & 1356 \\ \hline
3rd Key Bit & 33151 & 1360 & 1351 \\ \hline
4th Key Bit & 33151 & 1355 & 1351 \\ \hline
5th Key Bit & 33151 & 1360 & 1351 \\ \hline
6th Key Bit & 33156 & 1360 & 1356 \\ \hline
7th Key Bit & 33151 & 1360 & 1351 \\ \hline
8th Key Bit & 33156 & 1360 & 1351 \\ \hline
9th Key Bit & 33156 & 1360 & 1351 \\ \hline
10th Key Bit & 33151 & 1355 & 1356 \\ \hline
\textbf{Max} & \textbf{33156} & \textbf{1360} & \textbf{1356} \\ \hline
\textbf{Min} & \textbf{33151} & \textbf{1355} & \textbf{1351} \\ \hline
\textbf{Difference in Clock Cycles} & \textbf{5} & \textbf{5} & \textbf{5} \\ \hline
\rowcolor{orange!20}
\textbf{Doubling Atomic Block} & \textbf{1st Field Multiplication} & \textbf{1st Field Addition} & \textbf{1st Field Subtraction} \\ \hline
1st Key Bit & 33153 & 1355 & 1351 \\ \hline
2nd Key Bit & 33153 & 1355 & 1356 \\ \hline
3rd Key Bit & 33153 & 1360 & 1356 \\ \hline
4th Key Bit & 33153 & 1355 & 1351 \\ \hline
5th Key Bit & 33153 & 1360 & 1356 \\ \hline
6th Key Bit & 33153 & 1355 & 1351 \\ \hline
7th Key Bit & 33158 & 1360 & 1351 \\ \hline
8th Key Bit & 33158 & 1360 & 1356 \\ \hline
9th Key Bit & 33158 & 1360 & 1351 \\ \hline
10th Key Bit & 33153 & 1355 & 1351 \\ \hline
\textbf{Max} & \textbf{33158} & \textbf{1360} & \textbf{1356} \\ \hline
\textbf{Min} & \textbf{33153} & \textbf{1355} & \textbf{1351} \\ \hline
\textbf{Difference in Clock Cycles} & \textbf{5} & \textbf{5} & \textbf{5} \\ \hline
\caption{Execution times (measured in clock cycles) of the FLECC constant-time functions for the first field multiplication, first field addition and the first field subtraction while processing all 10 key bits across all the atomic blocks.} 
\label{tab:execution_time_analysis}
\end{longtable}
}

The 5-clock cycle variations in the execution times across the various point operations during the execution of each key bit are negligible but, however, could be a cause for \ac{SCA} leakages. The fact that these differences are small and stay the same across point operations supports the idea that they behave in constant time.
\section{Processing of Electromagnetic Trace}
\label{sec:em-data-processing}

We inserted \ac{NOP} instructions in our implementation code between each atomic block before measuring the \ac{EM} emanations to aid distinguish the samples corresponding to the different atomic blocks. These \acp{NOP} execute simple instructions that produce low \ac{EM} emanations, which create distinguishable gaps in the \ac{EM} trace. Specifically, we added:

\begin{enumerate}
    \item A loop of 1,000 \acp{NOP} between all Addition 1 and Addition 2 atomic blocks in each point addition operation.
    \item A loop of 2,000 \acp{NOP} after each Addition 2 atomic block (i.e. before each Doubling atomic block) to separate \ac{EC} point addition from \ac{EC} point doubling operations.
    \item A longer sequence of 3,000 \acp{NOP} following each Doubling atomic block to indicate the end of the processing of each key bit and perhaps the start of the processing of the next key bit.
\end{enumerate}

\begin{figure}[H]
    \centering
    \includegraphics[width=1\linewidth]{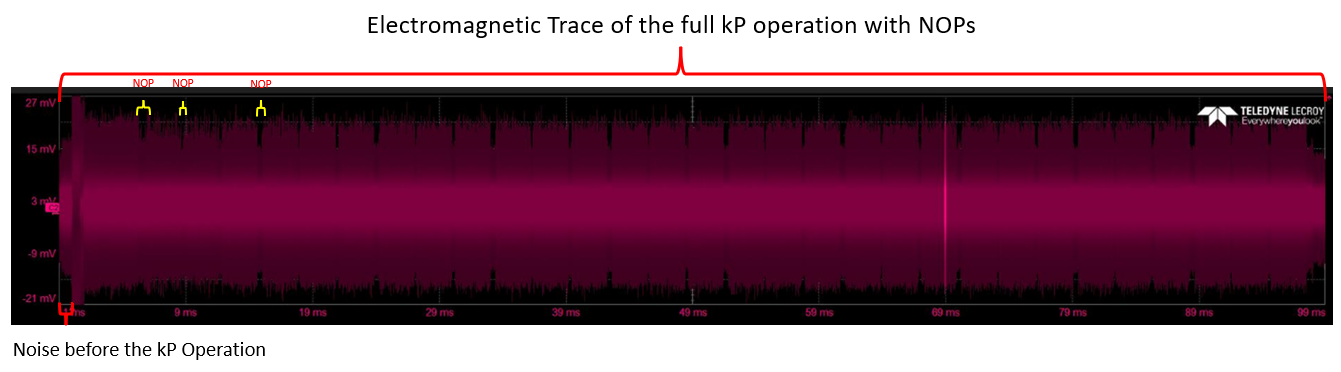}
    \caption{The \acl{EM} trace of the full \( k\mathbf{P} \) execution time with \acp{NOP} Full.}
    \label{fig:full-kP-trace}
\end{figure}

\hyperref[fig:full-kP-trace]{Figure~\ref*{fig:full-kP-trace}} shows the measured \ac{EM} trace, with clear \ac{NOP} sequences. These distinct \ac{NOP} segments clearly separate the atomic blocks for the point operations, which made it easier for us to extract and analyse each atomic block used in the point operations in the next phases.

To measure the execution times of the \ac{NOP} operations, we placed software breakpoints in \ac{CCS} at the start and end of the execution of each \ac{NOP} operation. This allowed us to measure the exact duration (in clock cycles) of each \ac{NOP} operation between the different atomic blocks during the execution of each key bit. We performed this measurement ten (10) times, and the results were identical each time, further supporting our assumption about FLECC’s constant-time functions. The measured durations were as follows:

\begin{enumerate}
    \item The sequence of 1,000 \acp{NOP} inserted between all Addition 1 and Addition 2 atomic blocks took 14,019 clock cycles.
    \item The sequence of 2,000 \acp{NOP} after each Addition 2 atomic block and the Doubling atomic block required 27,878 clock cycles.
\end{enumerate}

However, some variations were observed in the execution times for the 3,000 \ac{NOP} sequence inserted after each Doubling atomic block. Specifically, the sequence of 3,000 \acp{NOP}, which is used to differentiate the execution of the different key bits, always took 42,044 clock cycles. However, the first and last key bits took 42,019 and 42,028 clock cycles, respectively. Most likely, these differences are caused by memory wait states, which can make it hard to keep execution times for some operations consistent. The right-to-left \( k\mathbf{P} \) algorithm also processes the first key bit differently than the others, which changed how long it took to perform the initial point doubling operation. The first key bit is the only one with this small difference and this does not show up in the subsequent key bits.

Microcontrollers often experience non-zero wait states when the CPU’s clock speed exceeds that of the memory, requiring the CPU to wait for certain clock cycles to complete memory read or write operations. While modern RAM uses various techniques to minimise these wait states, the duration can still vary depending on factors like the type of operation (read/write) and access patterns (sequential or random). As such, these differences in operational times can affect the performance of \ac{SCA}.

\begin{figure}[H]
    \centering
    \includegraphics[width=1\linewidth]{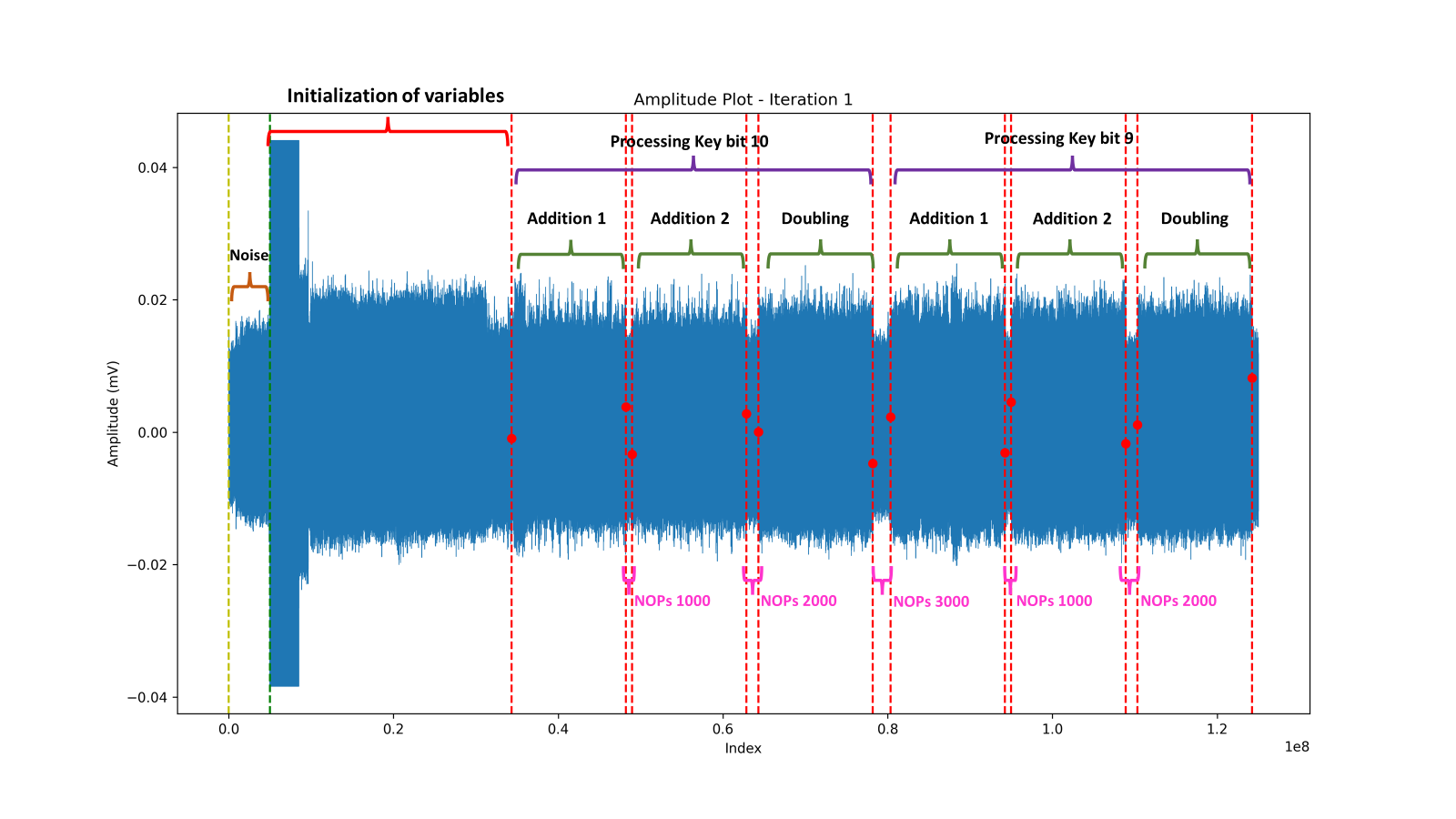}
    \caption{A zoomed-in section of the full \( k\mathbf{P} \) operation showing the noise before start of \( k\mathbf{P} \), the initialization phase and the shape of the point operations while processing the first two key bits.}
    \label{fig:section-of-kP-trace}
\end{figure}

\hyperref[fig:section-of-kP-trace]{Figure~\ref*{fig:section-of-kP-trace}} shows a zoomed-in section of the execution of a part of the whole \( k\mathbf{P} \) operation. It shows the shapes of the different atomic blocks in the \( k\mathbf{P} \) process, including the initial noise before the start of the \( k\mathbf{P} \) operations and the \acp{NOP} that were inserted, which clearly shows how the \acp{NOP} and the atomic blocks for the point operations are structured within the \ac{EM} trace. From \hyperref[fig:section-of-kP-trace]{Figure~\ref*{fig:section-of-kP-trace}}, we observe significant high peaks followed by regular peaks before the main loop of the \( k\mathbf{P} \) operations begin. This section represents the initialisation phase, where all variables are set up before the main \( k\mathbf{P} \) loop starts. In the main loop, 10 scalar bits are processed from right to left, which involves executing 10 point addition operations (with two atomic blocks - Addition 1 and Addition 2) and 10 point doubling operations (with 1 atomic block - Doubling).

Software breakpoints in \ac{CCS} were again used to measure execution time before the start of the main \( k\mathbf{P} \) operation by measuring the number of clock cycles. We also placed breakpoints before the start and after the end of each atomic block for the point operation to capture the execution times of atomic block. With these breakpoints, we were able to place vertical markers (as shown in \hyperref[fig:section-of-kP-trace]{Figure~\ref*{fig:section-of-kP-trace}}) at the start and end of each atomic block to help us easily extract the atomic blocks.

From the full \ac{EM} trace, we extracted the parts referring to each point operation, creating sub-traces for each atomic block, i.e. we obtained 10 “Doubling” sub-traces for 10 point doubling operations, 10 “Addition 1” and 10 “Addition 2” sub-traces for 10 point addition operations. We were sure that the full \( k\mathbf{P} \) execution was recorded correctly by counting these sub-traces and confirming the number of processed key-bits. \hyperref[tab:key_bit_execution_times]{Table~\ref*{tab:key_bit_execution_times}} shows the execution time for each atomic block for each key bit, along with the number of samples that were recorded.

{
\renewcommand{\arraystretch}{1.0} 
\begin{longtable}[H]{|>{\centering\arraybackslash}m{2cm}|l|>{\centering\arraybackslash}m{4.5cm}|>{\centering\arraybackslash}m{4.5cm}|}
\hline
\rowcolor{orange!30} 
\textbf{Key Bit Index} & \textbf{Atomic Block} & \textbf{Number of Clock Cycles} & \textbf{Number of Samples per Atomic Block} \\ \hline
\endfirsthead
\hline
\rowcolor{orange!30} 
\textbf{Index of Key Bit} & \textbf{Atomic Block} & \textbf{Number of Clock Cycles} & \textbf{Number of Samples per Atomic Block} \\ \hline
\endhead

\cellcolor{orange!20}\textbf{1} \multirow{-3}{*}{} & Addition 1 & 278794 & 13939700 \\ \cline{2-4}
\cellcolor{orange!20} & Addition 2 & 278780 & 13939000 \\ \cline{2-4}
\cellcolor{orange!20} & Doubling & 278827 & 13941350 \\ \hline

\cellcolor{blue!20}\textbf{2} \multirow{-3}{*}{} & Addition 1 & 278829 & 13941450 \\ \cline{2-4}
\cellcolor{blue!20} & Addition 2 & 278810 & 13940500 \\ \cline{2-4}
\cellcolor{blue!20} & Doubling & 278827 & 13941350 \\ \hline

\cellcolor{orange!20}\textbf{3} \multirow{-3}{*}{} & Addition 1 & 278829 & 13941450 \\ \cline{2-4}
\cellcolor{orange!20} & Addition 2 & 278825 & 13941250 \\ \cline{2-4}
\cellcolor{orange!20} & Doubling & 278827 & 13941350 \\ \hline

\cellcolor{blue!20}\textbf{4} \multirow{-3}{*}{} & Addition 1 & 278819 & 13940950 \\ \cline{2-4}
\cellcolor{blue!20} & Addition 2 & 278805 & 13940250 \\ \cline{2-4}
\cellcolor{blue!20} & Doubling & 278812 & 13940600 \\ \hline

\cellcolor{orange!20}\textbf{5} \multirow{-3}{*}{} & Addition 1 & 278804 & 13940200 \\ \cline{2-4}
\cellcolor{orange!20} & Addition 2 & 278815 & 13940750 \\ \cline{2-4}
\cellcolor{orange!20} & Doubling & 278827 & 13941350 \\ \hline

\cellcolor{blue!20}\textbf{6} \multirow{-3}{*}{} & Addition 1 & 278819 & 13940950 \\ \cline{2-4}
\cellcolor{blue!20} & Addition 2 & 278835 & 13941750 \\ \cline{2-4}
\cellcolor{blue!20} & Doubling & 278817 & 13940850 \\ \hline

\cellcolor{orange!20}\textbf{7} \multirow{-3}{*}{} & Addition 1 & 278839 & 13941950 \\ \cline{2-4}
\cellcolor{orange!20} & Addition 2 & 278825 & 13941250 \\ \cline{2-4}
\cellcolor{orange!20} & Doubling & 278817 & 13940850 \\ \hline

\cellcolor{blue!20}\textbf{8} \multirow{-3}{*}{} & Addition 1 & 278824 & 13941200 \\ \cline{2-4}
\cellcolor{blue!20} & Addition 2 & 278830 & 13941500 \\ \cline{2-4}
\cellcolor{blue!20} & Doubling & 278852 & 13942600 \\ \hline

\cellcolor{orange!20}\textbf{9} \multirow{-3}{*}{} & Addition 1 & 278854 & 13942700 \\ \cline{2-4}
\cellcolor{orange!20} & Addition 2 & 278840 & 13942000 \\ \hline
\cellcolor{orange!20} & Doubling & 278832 & 13941600 \\ \hline

\cellcolor{blue!20}\textbf{10} \multirow{-3}{*}{} & Addition 1 & 278814 & 13940700 \\ \cline{2-4}
\cellcolor{blue!20} & Addition 2 & 278810 & 13940500 \\ \cline{2-4}
\cellcolor{blue!20} & Doubling & 278822 & 13941100 \\ \hline

\caption{The measured execution times in clock cycles and the number of samples recorded using breakpoints in \ac{CCS} during a \( k\mathbf{P} \) operation with \(k = 1111111111\).}
\label{tab:key_bit_execution_times}
\end{longtable}
}

The execution times presented in \hyperref[tab:key_bit_execution_times]{Table~\ref*{tab:key_bit_execution_times}} are measured in clock cycles. To obtain the actual time in milliseconds (ms), the clock cycle values were converted using the system clock frequency of 100MHz, where each clock cycle corresponds to 10ns. This means, the execution time in milliseconds is obtained by multiplying the number of clock cycles by \(10^{-6}\) (i.e., dividing by 100,000). Additionally, to obtain the actual number of samples, these values were multiplied by 50, as configured on the oscilloscope with a sampling rate of 50 samples per clock cycle.

\begin{figure}[H]
    \centering
    \includegraphics[width=0.75\linewidth]{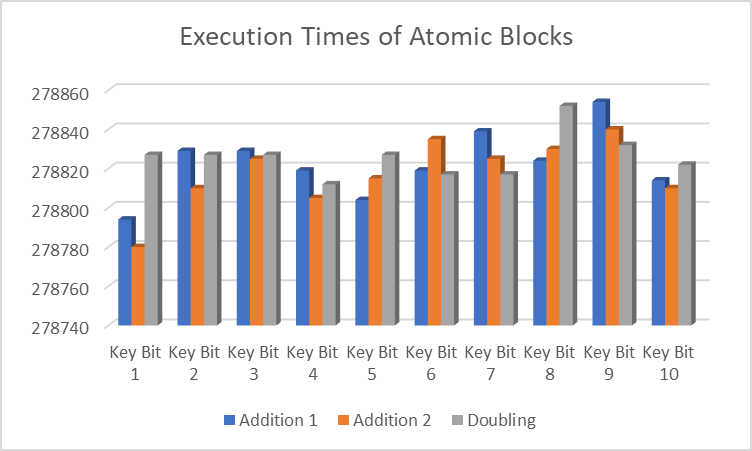}
    \caption{Visual representation of the execution times in clock cycles for all the atomic blocks (Doubling (gray), Addition 1 (blue), Addition 2 (orange)) measured using breakpoints while processing all 10-bit long key.}
    \label{fig:execution-times-in-graph}
\end{figure}

\begin{figure}[H]
    \centering
    \includegraphics[width=1\linewidth]{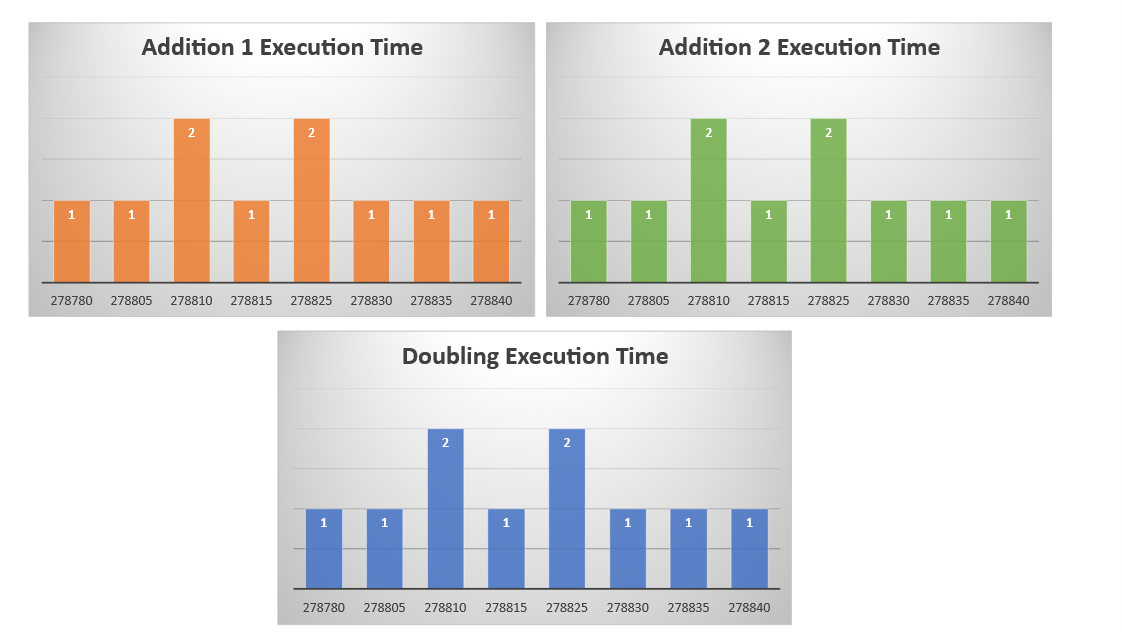}
    \caption{Distribution of the measured execution times for each atomic block in the full \( k\mathbf{P} \) Trace.}
    \label{fig:distribution-of-exec-times}
\end{figure}

\hyperref[fig:execution-times-in-graph]{Figure~\ref*{fig:execution-times-in-graph}} shows a visual representation of the execution times provided in \hyperref[tab:key_bit_execution_times]{Table~\ref*{tab:key_bit_execution_times}}, making it easier to observe timing trends across the atomic blocks. Furthermore, \hyperref[fig:distribution-of-exec-times]{Figure~\ref*{fig:distribution-of-exec-times}} displays the distribution of the occurrences of the measured execution times for each atomic block.

\section{Synchronisation Process of EM Traces}
\label{sec:synchronisation-process}

In this part, we will describe how we synchronised the \ac{EM} sub-traces collected in the previous steps. The goal is to make sure that all the extracted sub-traces are aligned with each other and that each sub-trace aligns correctly at the start of each \( k\mathbf{P} \) operation. Once this step is done, we can directly compare the sub-traces and analyse the patterns of each atomic block more accurately.

For the rest of the work, we will refer to the sub-traces corresponding to each point doubling operation as \texttt{(Doubling)\(_i\)}, where \( i \) is the index of the key bit that was processed. For Point Addition operations, we will refer to the sub-traces as \texttt{(Addition 1)\(_i\)} and \texttt{(Addition 2)\(_i\)}, where Addition 1 and Addition 2 are the atomic blocks for point addition and \( i \) is the index of the key bit that was processed, i.e., \( i \in \{1, 2, \ldots, 10\} \) for the processed 10-bit long binary key \( k = 1111111111 = k_{10}k_9k_8k_7k_6k_5k_4k_3k_2k_1 \).

We used cross-correlation from the Python library to successfully synchronise the 30 sub-traces, which correspond to 10 Doubling sub-traces, 10 Addition 1 sub-traces, and 10 Addition 2 sub-traces. This allowed us to align each sub-trace at the start by maximising the amplitude shape similarity between the sub-traces. We chose \texttt{(Doubling)\(_1\)} as the reference or anchor trace to ensure consistent alignment across all sub-traces. The remaining 29 sub-traces were subsequently aligned to \texttt{(Doubling)\(_1\)}.

\subsection{Loading Sub-Traces and Selecting Window Size}

Before extracting the \( k\mathbf{P} \) operation for each atomic block in the point operation from the previous step, we padded the start and end of each atomic block in the point operation with 50,000 samples (i.e. 1,000 clock cycles) of \acp{NOP} from the already inserted \acp{NOP}. Note that the vertical markers we placed earlier in the \ac{EM} trace did not include these \ac{NOP} samples. Hence, these vertical markers were slightly adjusted to accommodate the additional \ac{NOP} samples added to the original number of samples for the \( k\mathbf{P} \) operation. This padding helped us identify the approximate start and end of the \( k\mathbf{P} \) operation by providing an initial noise buffer. Thus, each extracted sub-trace began with these additional \ac{NOP} samples.

For alignment, we loaded a window of 200,000 samples (i.e. 4,000 clock cycles) from the start of each sub-trace file. This window included the first 50,000 samples of \acp{NOP}, which made sure that we got the full sample points around each \( k\mathbf{P} \) operation for proper synchronisation. \texttt{(Doubling)\(_1\)} served as the fixed anchor trace, and the other 29 sub-traces (i.e., \texttt{(Doubling)\(_i\)}, \texttt{(Addition 1)\(_i\)}, and \texttt{(Addition 2)\(_i\)}) were loaded with the same window size to match this reference.

\subsection{Using Cross-Correlation to Determine The Optimal Shift Position}

To guarantee accurate synchronisation results, we used cross-correlation from Python's \texttt{scipy.signal} library \footnote{\url{https://docs.scipy.org/doc/scipy/reference/generated/scipy.signal.correlate.html}} while focusing on the beginning and around the 50,000\textsuperscript{th} sample from each sub-trace, which marks the end of the \acp{NOP} and the start of the samples corresponding to the execution of the atomic blocks. Including this section provided enough room to shift values left or right without losing any sample points pertaining to the start of the \( k\mathbf{P} \) operation.

Before calculating the cross-correlation, each sub-trace was mean-centred by first computing the mean value of the entire sub-trace and then subtracting the mean value from each sample point. We then computed the cross-correlation of the mean-centred sub-trace with the reference sub-trace, \texttt{(Doubling)\(_1\)}. The peak correlation value was used to determine the optimal shift, which represents the relative shift necessary to align each sub-trace with \texttt{(Doubling)\(_1\)}. Positive shifts moved traces to the right, while negative shifts moved them to the left. \hyperref[fig:cross-corrlation-using-python]{Figure~\ref*{fig:cross-corrlation-using-python}} below shows how this library was used in Python.

\begin{figure}[H]
    \centering
    \includegraphics[width=0.75\linewidth]{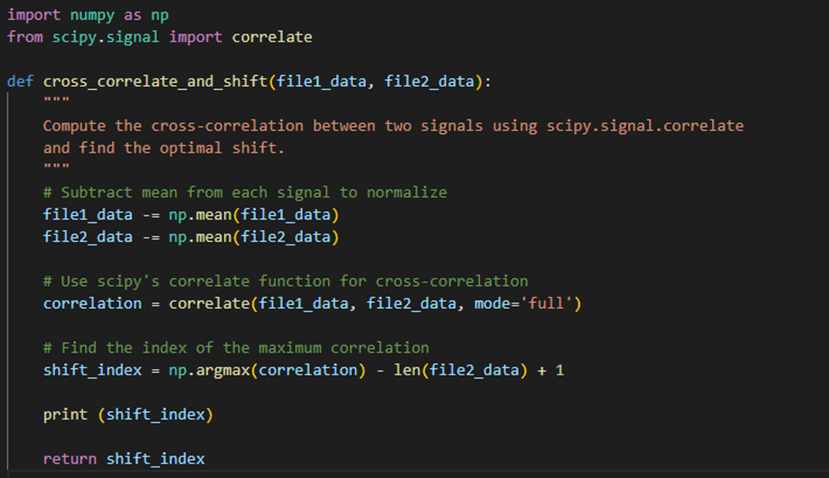}
    \caption{A python function used for computing the mean, cross-correlation and finding the optimal shift}
    \label{fig:cross-corrlation-using-python}
\end{figure}

\begin{figure}[H]
    \centering
    \includegraphics[width=1\linewidth]{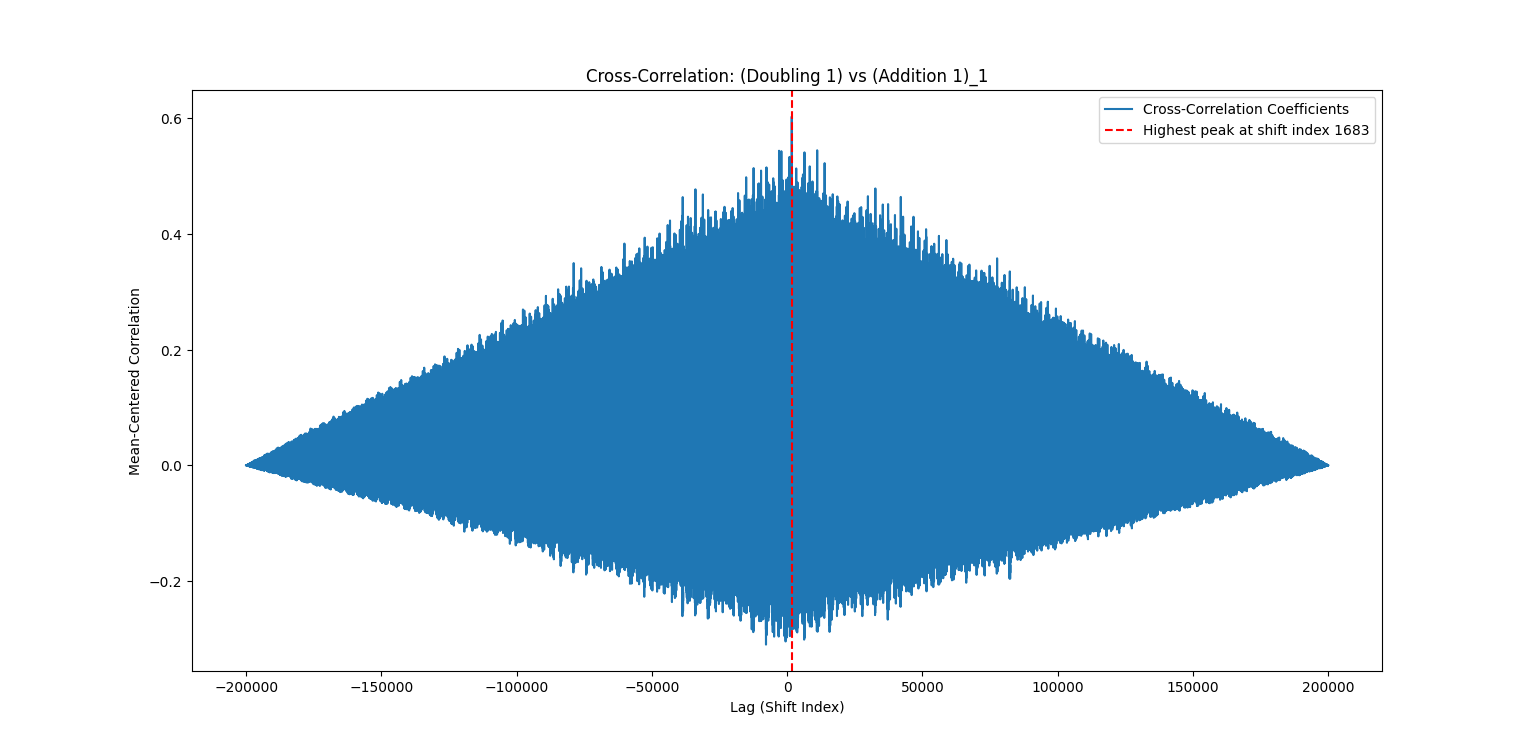}
    \caption{Cross-correlation coefficients between \texttt{(Doubling)\(_1\)} and \texttt{(Addition 1)\(_1\)} showing the highest peak for the optimal shift alignment between the two sub-traces}
    \label{fig:cross-correlation-coefficient}
\end{figure}

\hyperref[fig:cross-correlation-coefficient]{Figure~\ref*{fig:cross-correlation-coefficient}} shows an example of a plot of the results of the cross-correlation coefficients between \texttt{(Doubling)\(_1\)} and \texttt{(Addition 1)\(_1\)} with the red vertical line showing the optimal shift required to align the two sub-traces.

\subsection{Applying The Shift and Synchronisation of Sub-traces}

After determining the best shift position, we adjusted each sub-trace accordingly. For sub-traces that required a left shift, we ignored the first sets of sample points based on the shift value and for sub-traces that required a right shift, we applied the shift directly without padding because the earlier \ac{NOP} samples provided enough buffer to apply the shift without any loss of samples.

\begin{figure}[H]
    \centering

    \begin{subfigure}[b]{0.45\textwidth}
        \caption*{\texttt{(Doubling)\(_1\)} \& \texttt{(Doubling)\(_2\)} - Before Synchronisation.}
        \includegraphics[width=\linewidth]{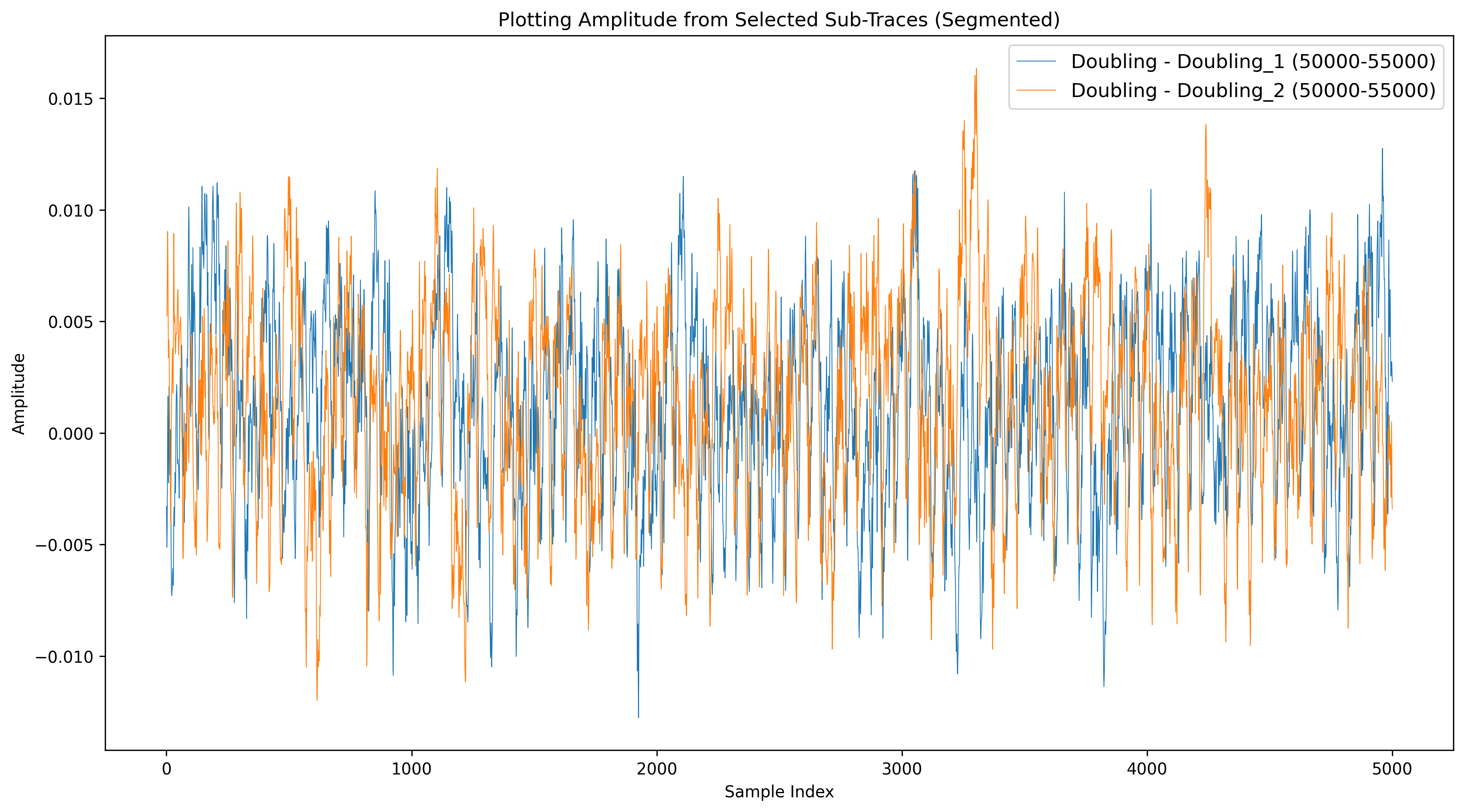} 
    \end{subfigure}
    \hspace{0.05\textwidth}
    \begin{subfigure}[b]{0.45\textwidth}
        \caption*{\texttt{(Doubling)\(_1\)} \& \texttt{(Doubling)\(_2\)} - After Synchronisation.}
        \includegraphics[width=\linewidth]{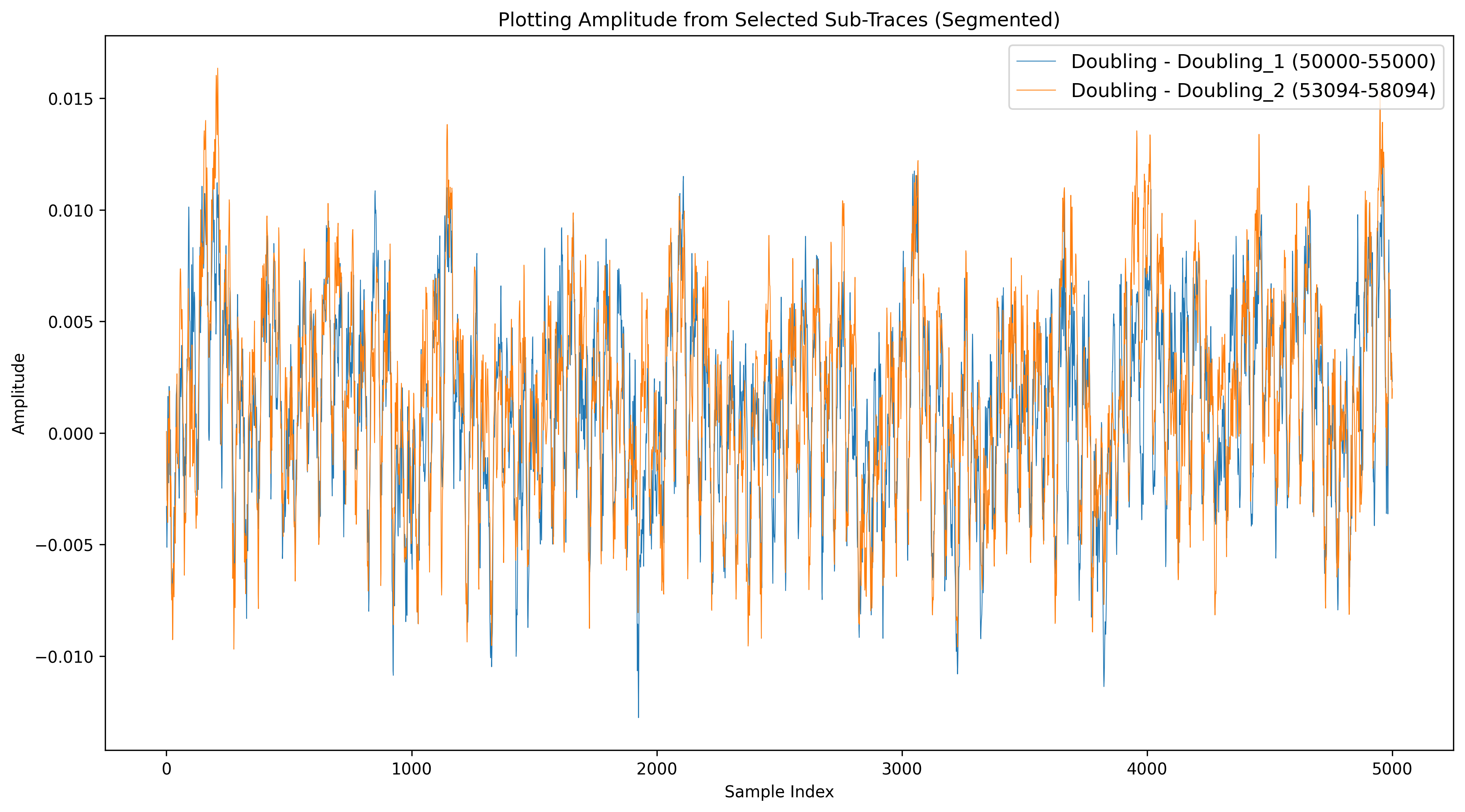}  
    \end{subfigure}

    \caption{Left: 5000 samples of \texttt{(Doubling)\(_1\)} and \texttt{(Doubling)\(_2\)} both starting at index 50000 before the synchronisation. \\
    Right: 5000 samples of \texttt{(Doubling)\(_1\)} starting at index 50000 and \texttt{(Doubling)\(_2\)} starting at index 53094 after synchronisation.}
    \label{fig:synchronisation-effect-doubling}
\end{figure}

\begin{figure}[H]
    \centering

    \begin{subfigure}[b]{0.45\textwidth}
        \caption*{\texttt{(Doubling)\(_1\)} \& \texttt{(Addition 1)\(_2\)} - Before Synchronisation.}
        \includegraphics[width=\linewidth]{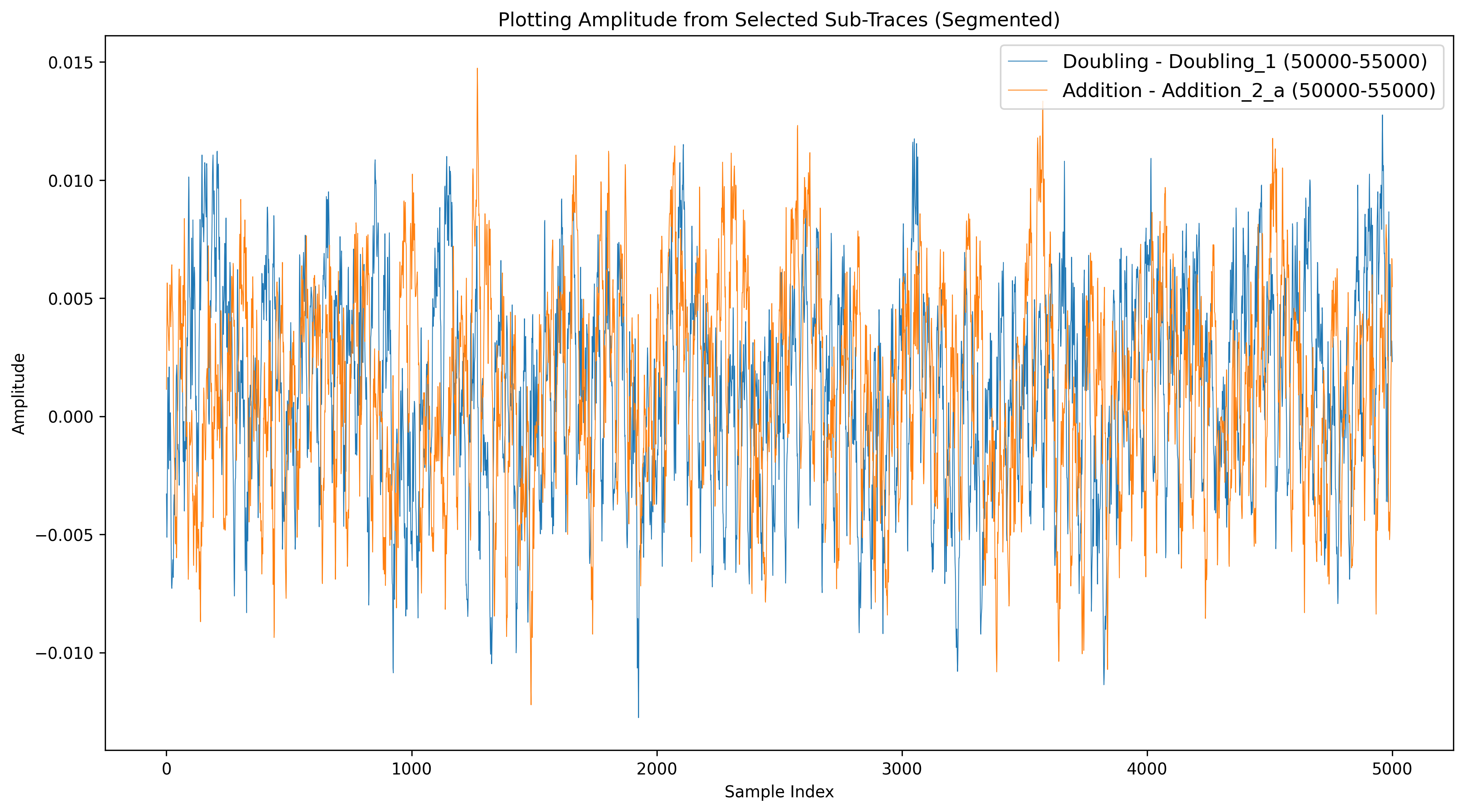} 
    \end{subfigure}
    \hspace{0.05\textwidth}
    \begin{subfigure}[b]{0.45\textwidth}
        \caption*{\texttt{(Doubling)\(_1\)} \& \texttt{(Addition 1)\(_2\)} - After Synchronisation.}
        \includegraphics[width=\linewidth]{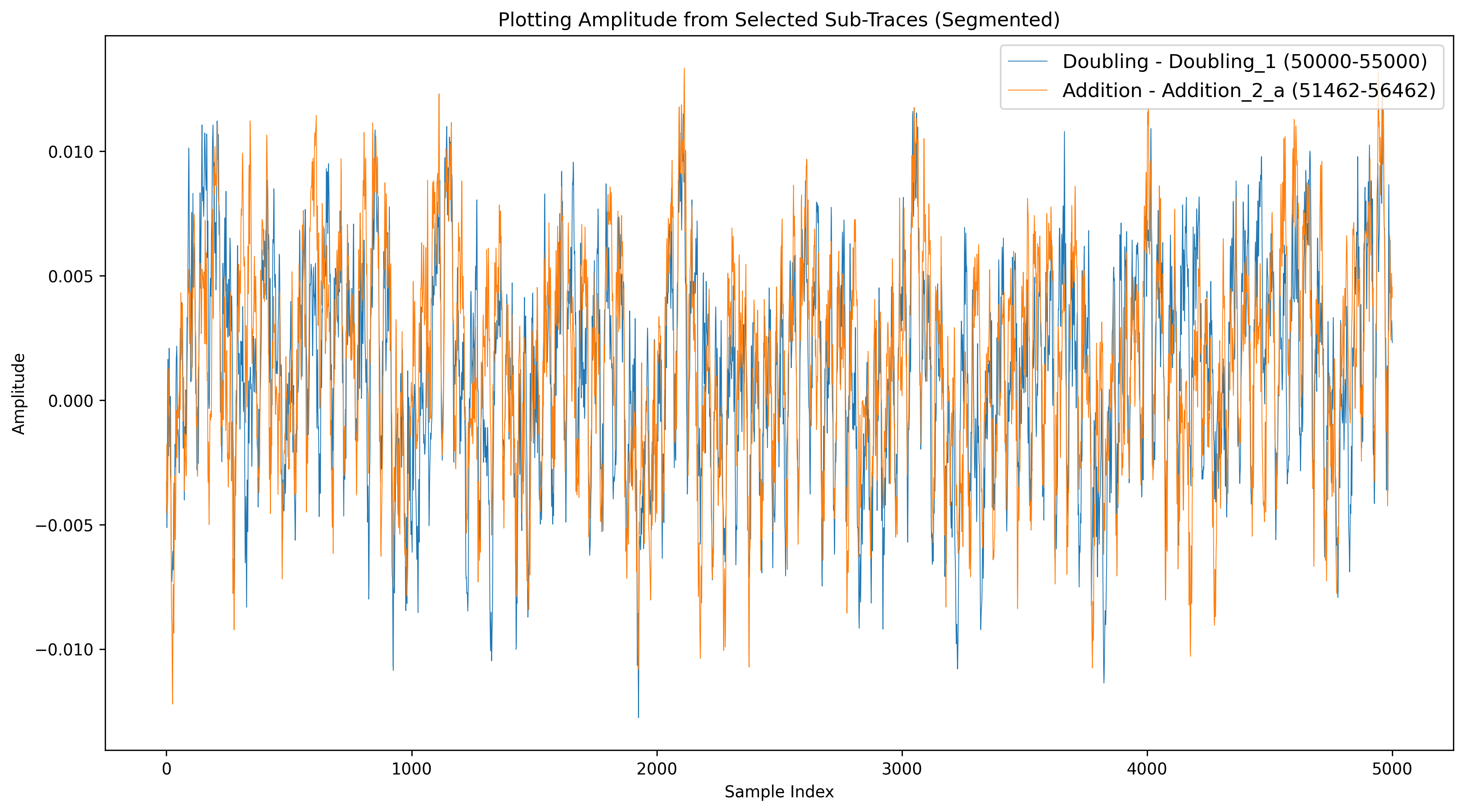}  
    \end{subfigure}

    \caption{Left: 5000 samples of \texttt{(Doubling)\(_1\)} and \texttt{(Addition 1)\(_2\)} both starting at index 50000 before the synchronisation. \\
    Right: 5000 samples of \texttt{(Doubling)\(_1\)} starting at index 50000 and \texttt{(Addition 1)\(_2\)} starting at index 51462 after synchronisation.}
    \label{fig:synchronisation-effect-addition}
\end{figure}

\hyperref[fig:synchronisation-effect-doubling]{Figure~\ref*{fig:synchronisation-effect-doubling}} shows a plot of the alignment process between \texttt{(Doubling)\(_1\)} and \texttt{(Doubling)\(_2\)} before and after the synchronisation. Similarly, \hyperref[fig:synchronisation-effect-addition]{Figure~\ref*{fig:synchronisation-effect-addition}} shows a plot of \texttt{(Doubling)\(_1\)} and \texttt{(Addition 1)\(_2\)} before and after the synchronisation to demonstrate the effect of this process.

\subsection{Plotting and Comparison}

We plotted 12 sub-traces to illustrate the results, with 4 sub-traces from each atomic block (i.e. 4 sub-traces from Doubling, 4 sub-traces from Addition 1, and 4 sub-traces from Addition 2). These sub-traces were randomly chosen to demonstrate the synchronisation process for each atomic block during the \( k\mathbf{P} \) execution.

\begin{figure}[H]
    \centering

    \begin{subfigure}[b]{\textwidth}
        \centering
        \caption*{a}
        \includegraphics[width=\textwidth]{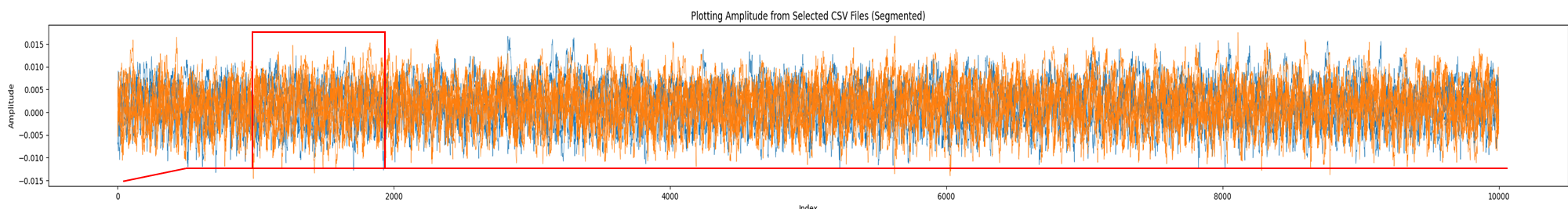}
    \end{subfigure}

    \vspace{1em} 

    \begin{subfigure}[b]{\textwidth}
        \centering
        \caption*{b}
        \includegraphics[width=\textwidth]{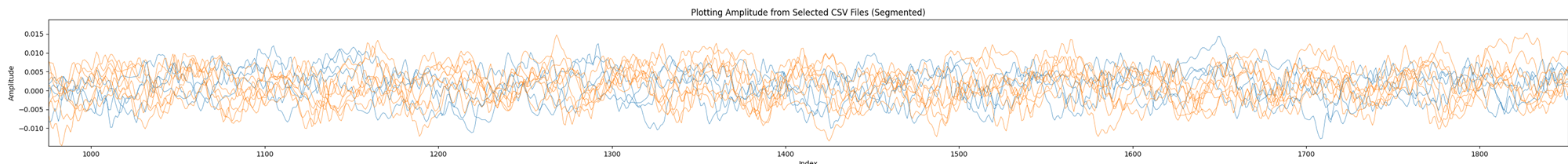}
    \end{subfigure}

    \caption{\texttt{Above:} Plot of a section of 10000 samples (i.e. 200 clock cycles) of unsynchronized sub-traces.\\
    \texttt{Below:} Zoomed-in section of about 900 samples (i.e. 18 clock cycles) of unsynchronised sub-traces.}
    \label{fig:section-of-unsynchronised-traces}
\end{figure}

\hyperref[fig:section-of-unsynchronised-traces]{Figure~\ref*{fig:section-of-unsynchronised-traces}(a)} shows a plot of 10,000 samples (i.e. 200 clock cycles) from the start of the processing of the first key bit for some of the sub-traces mentioned earlier, but with unsynchronised sub-traces. \hyperref[fig:section-of-unsynchronised-traces]{Figure~\ref*{fig:section-of-unsynchronised-traces}(b)} zooms in on a section of \hyperref[fig:section-of-unsynchronised-traces]{Figure~\ref*{fig:section-of-unsynchronised-traces}(a)}, focusing on approximately 900 samples (i.e. 18 clock cycles) to provide a clearer view of the signal behaviour in the unsynchronised sub-traces.

\begin{figure}[H]
    \centering

    \begin{subfigure}[b]{\textwidth}
        \centering
        \caption*{a}
        \includegraphics[width=\textwidth]{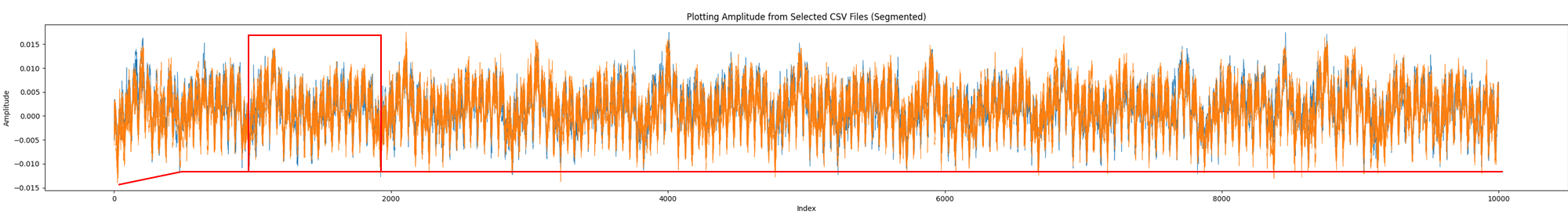}
    \end{subfigure}

    \vspace{1em} 

    \begin{subfigure}[b]{\textwidth}
        \centering
        \caption*{b}
        \includegraphics[width=\textwidth]{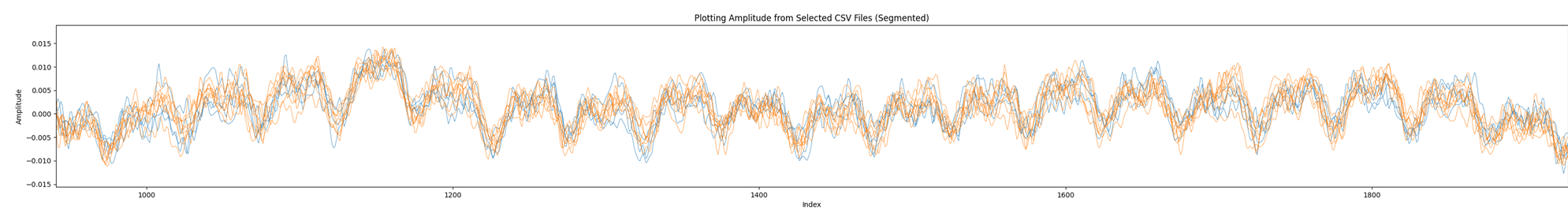}
    \end{subfigure}

    \caption{\texttt{Above:} Plot of a section of 10000 samples (i.e. 200 clock cycles) of synchronized sub-traces.\\
    \texttt{Below:} Zoomed-in section of about 900 samples (i.e. 18 clock cycles) of synchronised sub-traces.}
    \label{fig:section-of-synchronised-traces}
\end{figure}

\hyperref[fig:section-of-synchronised-traces]{Figure~\ref*{fig:section-of-synchronised-traces}a} shows another 10,000 samples, but now with the synchronised traces, in which all the sub-traces are aligned to the reference file, \texttt{(Doubling)\(_1\)}. In this plot, the applied shifts are evident, showing that the selected sub-traces have successfully been synchronised at the start.

The synchronisation process at the start was done to ensure that each sub-trace begins at the same relative position within the execution of the field operations so that we can meaningfully compare the amplitude signals across the atomic blocks.

\subsection{Confirming Synchronisation at The Start of The Sub-Traces}

\begin{figure}[H]
    \centering

    \begin{subfigure}[b]{\textwidth}
        \centering
        \caption*{a}
        \includegraphics[width=0.75\textwidth]{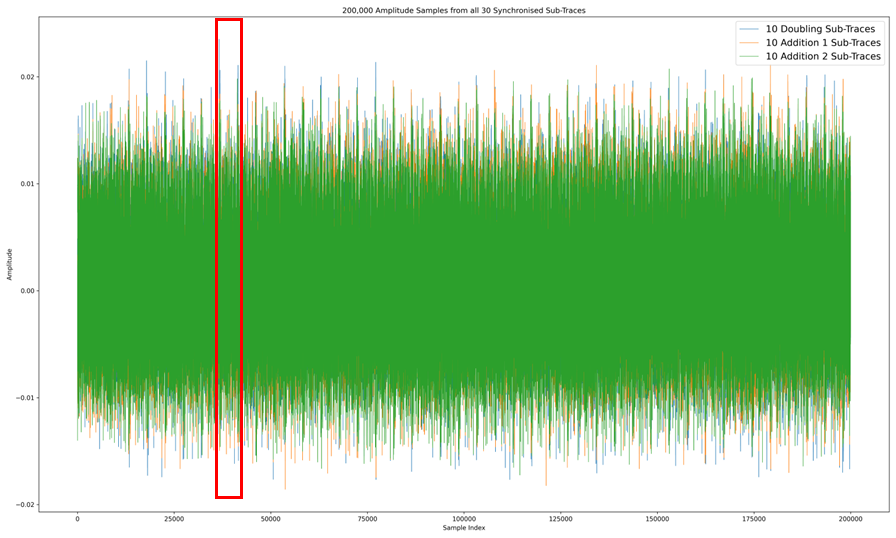}
    \end{subfigure}

    \vspace{1em} 

    \begin{subfigure}[b]{\textwidth}
        \centering
        \caption*{b}
        \includegraphics[width=0.75\textwidth]{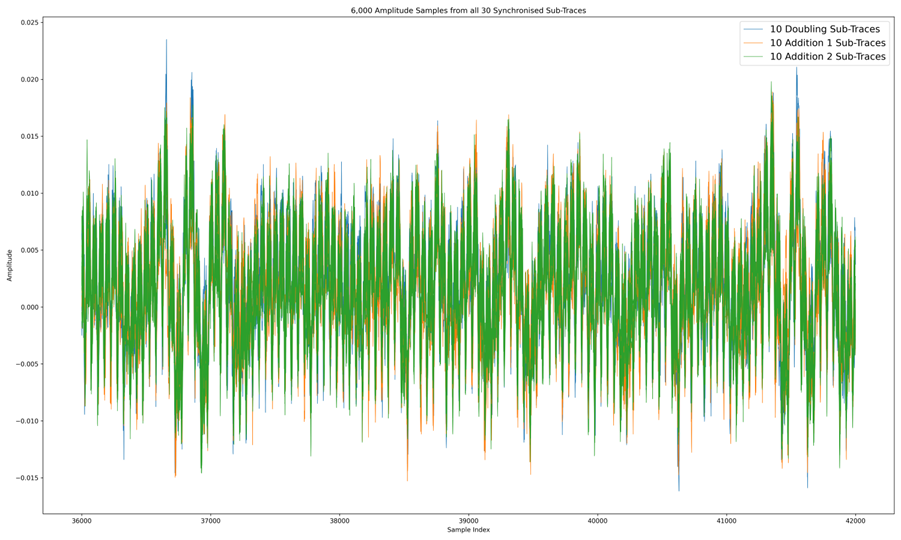}
    \end{subfigure}

    \caption{\texttt{Above:} First 200,000 samples at the start of all 30 atomic block patterns. i.e. all Doubling (blue), Addition 1 (orange), Addition 2 (green) sub-traces.\\
    \texttt{Below:} Zoomed-in section of about 6,000 samples from all 30 atomic block sub-traces.}
    \label{fig:section-of-samples-from-start-of-kp}
\end{figure}

Before proceeding with any form of analysis, it was important for us to confirm that the sub-traces were properly synchronised at the start of each atomic block. \hyperref[fig:section-of-samples-from-start-of-kp]{Figure~\ref*{fig:section-of-samples-from-start-of-kp}a} shows the first 200,000 samples (i.e. 4,000 clock cycles) from the start of each sub-trace corresponding to all the 10 key bits processed, which captures sub-traces from each atomic block.

\hyperref[fig:section-of-samples-from-start-of-kp]{Figure~\ref*{fig:section-of-samples-from-start-of-kp}b} also shows a zoomed-in region of approximately 6,000 samples (i.e. 120 clock cycles). Here, we can observe that the amplitude values across the different sub-traces are mostly well aligned, especially around the high amplitude peaks. The blue amplitude signals represent all 10 sub-traces corresponding to the Doubling atomic block, the orange represents all 10 sub-traces from the Addition 1 atomic block, while the green represents all 10 sub-traces from the Addition 2 atomic block. This shows that our initial synchronisation at the start of the sub-traces was successful.

\section{Identifying Desynchronised Regions in Previously Synchronised Sub-Traces}

Each clock cycle within the sub-trace of any atomic block corresponds to a specific operation or process. As we showed in \hyperref[sec:eval-flecc-cr-ops]{Section~\ref*{sec:eval-flecc-cr-ops}}, there was a variation of 5-clock cycle processes or delays in the execution times of the different atomic blocks, even though they all execute the same sequence of field operations. Hence, it is likely that these variations could cause desynchronisation within the sub-traces. To ensure that the results from our analysis are correct, it was necessary to ensure that the previously synchronised sub-traces discussed in \hyperref[sec:synchronisation-process]{Section~\ref*{sec:synchronisation-process}} remain synchronised not only at their start but also throughout the entire execution time or at least during the execution time of the specific region selected for the analysis. The goal of this investigation was to find out if the atomic blocks show any distinguishable patterns or differences that an attacker could exploit.

To begin our search for any such desynchronised regions, we computed and plotted the mean of all 10 sub-traces corresponding to each atomic block, that is, the sub-traces from Doubling, Addition 1 and Addition 2 atomic blocks while processing all 10 key-bit values collected during the experiments. The mean was calculated across the amplitude values at each individual sample index. Specifically, for each sample point, the corresponding amplitude values from all sub-traces were averaged. Figure 7.14 shows the mean plot of all the sub-traces. In this plot, the blue amplitude signals represents the mean for the Doubling sub-traces, the orange for the Addition 1 sub-traces and the green represents the mean for all Addition 2 sub-traces.

\begin{figure}[H]
    \centering
    \includegraphics[width=0.75\linewidth]{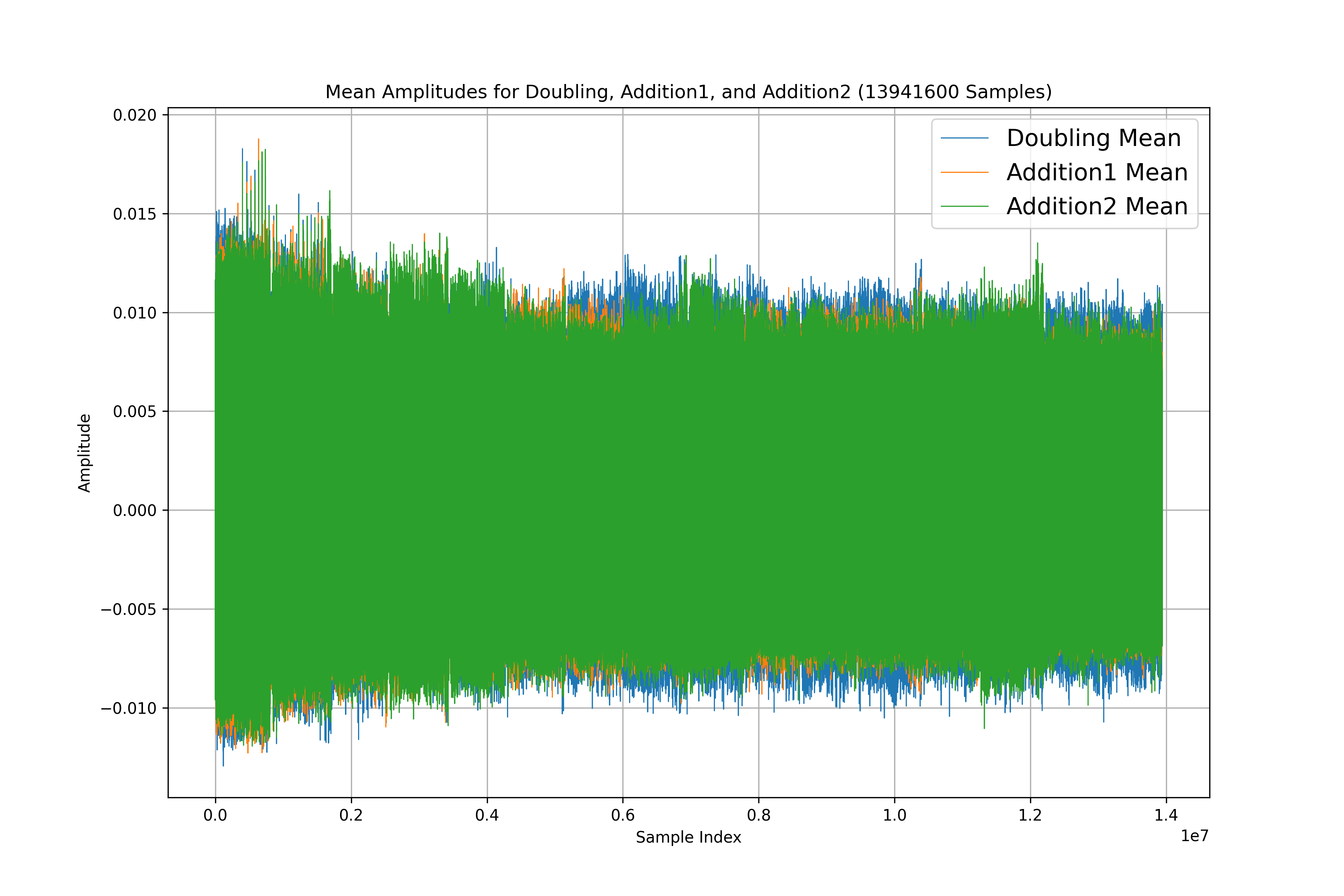}
    \caption{The overlaid mean plots of 13,941,600 samples (i.e. 278,832 clock cycles) from each Doubling, Addition 1 and Addition 2 sub-traces at each index before identifying any desynchronised regions.}
    \label{fig:mean-plot-13million}
\end{figure}

\begin{figure}[H]
    \centering

    \begin{subfigure}[b]{\textwidth}
        \centering
        \caption*{a}
        \includegraphics[width=0.65\textwidth]{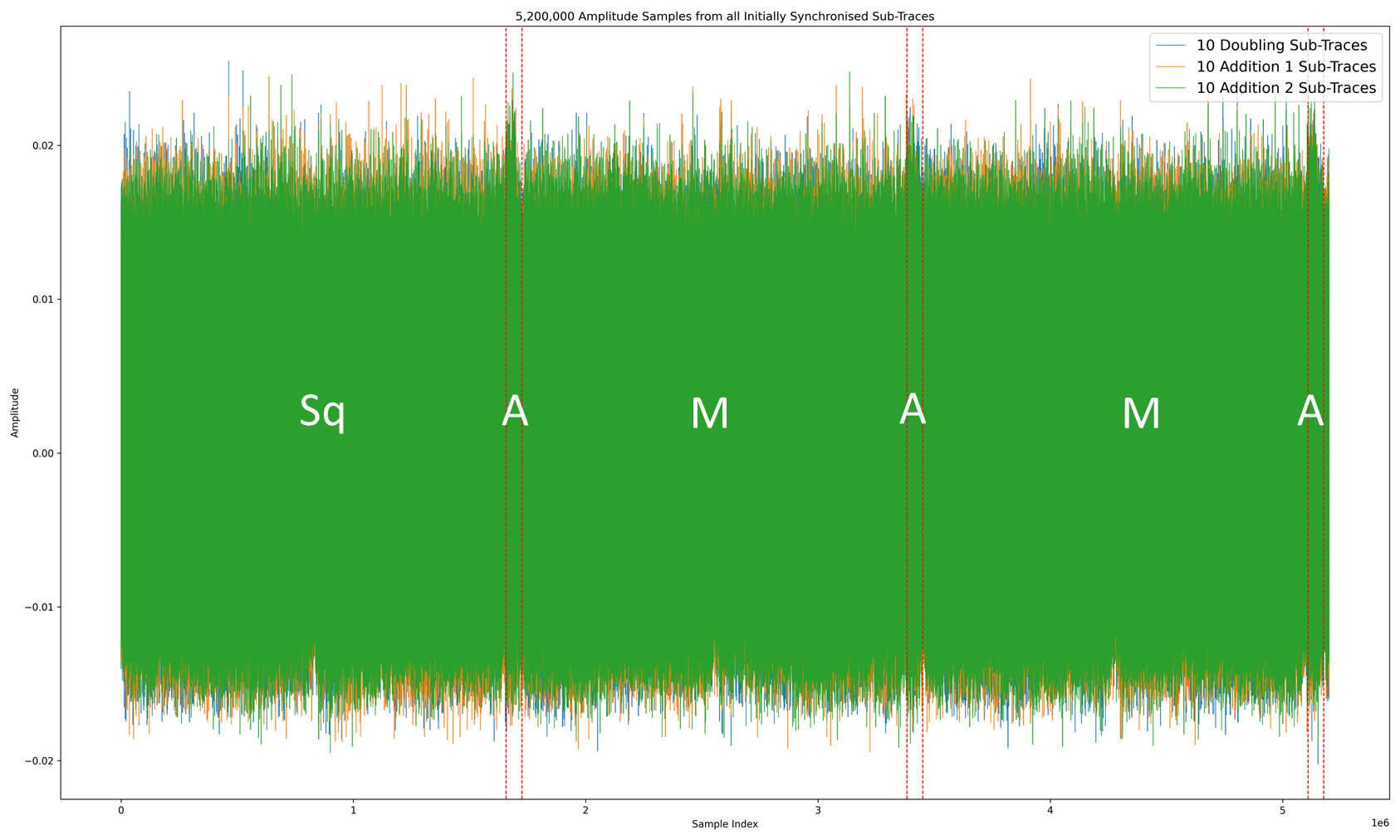}
    \end{subfigure}

    \vspace{1em} 

    \begin{subfigure}[b]{\textwidth}
        \centering
        \caption*{b}
        \includegraphics[width=0.70\textwidth]{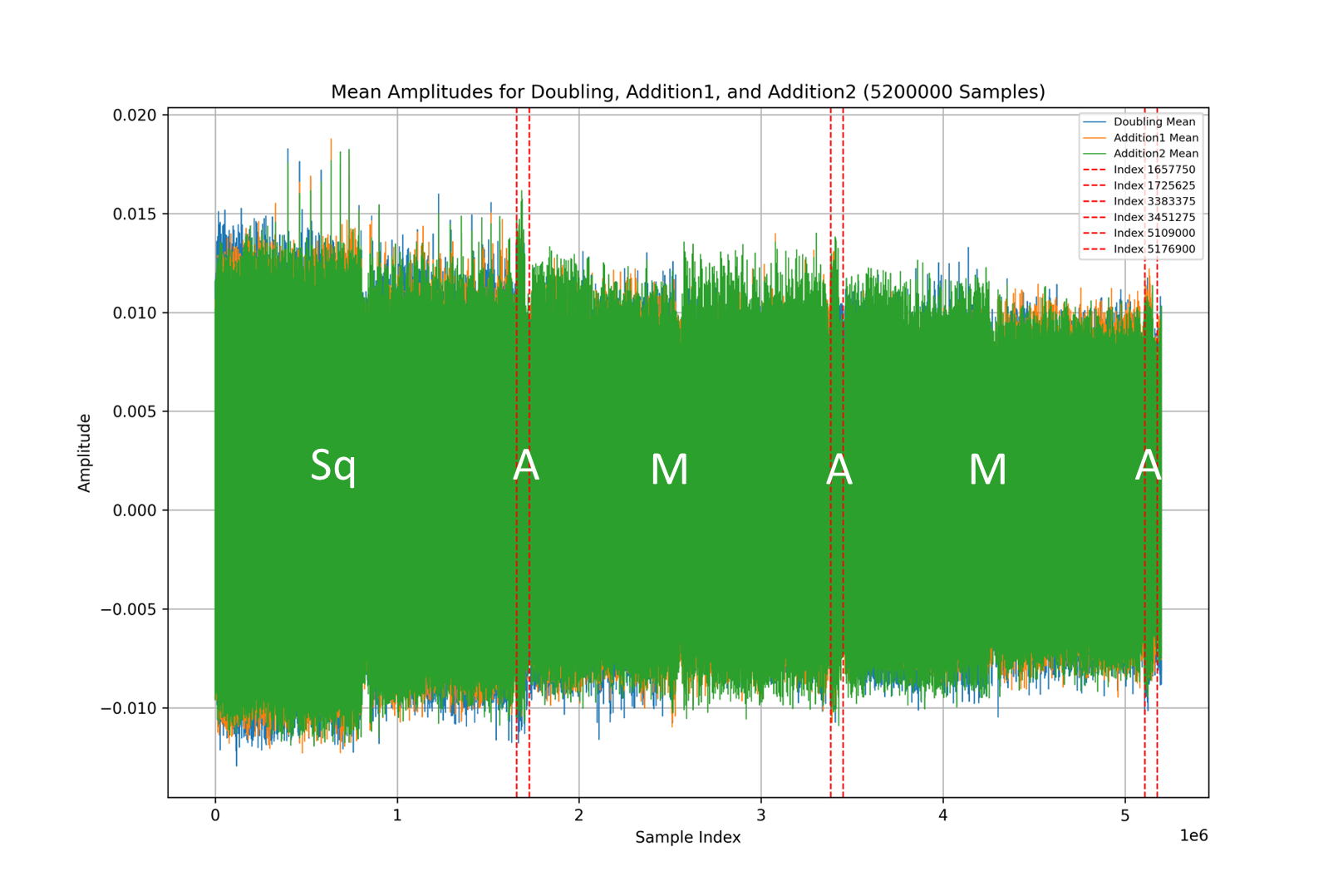}
    \end{subfigure}

    \caption{\texttt{Above:} Overlaid plot from our focus region of the first 5,200,000 samples from all 10 Doubling, 10 Addition 1 and 10 Addition 2 sub-traces.\\
    \texttt{Below:} Mean plot of our focus region of the first 5,200,000 samples from all 10 Doubling, 10 Addition 1 and 10 Addition 2 sub-traces at each index before identifying any desynchronised regions.}
    \label{fig:mean-plot-5million}
\end{figure}

For the rest of our analysis and experiments, we focused on the first 5,200,000 samples (i.e. 104,000 clock cycles) of the mean traces, which is approximately 37.29\% of the full length of the atomic blocks. The sequence of the field operations present in this region is: \acf{Sq}, \acf{A}, \acf{M}, \acf{A}, \acf{M}, \acl{A}. The rest of the atomic patterns consists of \ac{Sq}, \ac{A}, \ac{M} and \ac{S} field operations, i.e. the same type of field operations present in our focus region. Thus, we assume that the selected region is long enough to observe the distinguishability of the atomic blocks if it exists. \hyperref[fig:mean-plot-5million]{Figure~\ref*{fig:mean-plot-5million}a} shows the raw plot of 5,200,000 samples (i.e. 104,000 clock cycles) from all 10 Doubling sub-traces, 10 Addition 1 sub-traces and 10 Addition 2 sub-traces overlaid on top of each other with the marked field operations while \hyperref[fig:mean-plot-5million]{Figure~\ref*{fig:mean-plot-5million}b} shows the mean plot of this same region with the same set of sub-traces and the various field operation. The amplitude of the sub-traces is relatively constant over time, but that of the mean trace decreased over time which could mean a potential desynchronisation. Hence, it is necessary to identify the start of any potential desynchronisation.

To identify regions within our sub-traces that were desynchronised, we compared sub-trace triplets derived from processing each individual key bit. Each triplet typically consisted of one doubling and two addition sub-traces, for example, \texttt{(Doubling)\(_1\)}, \texttt{(Addition 1)\(_1\)} and \texttt{(Addition 2)\(_1\)} or \texttt{(Doubling)\(_2\)}, \texttt{(Addition 1)\(_2\)} and \texttt{(Addition 2)\(_2\)}.

\begin{figure}[H]
    \centering

    \begin{subfigure}[b]{0.50\textwidth}
        \caption*{a}
        \includegraphics[width=\textwidth]{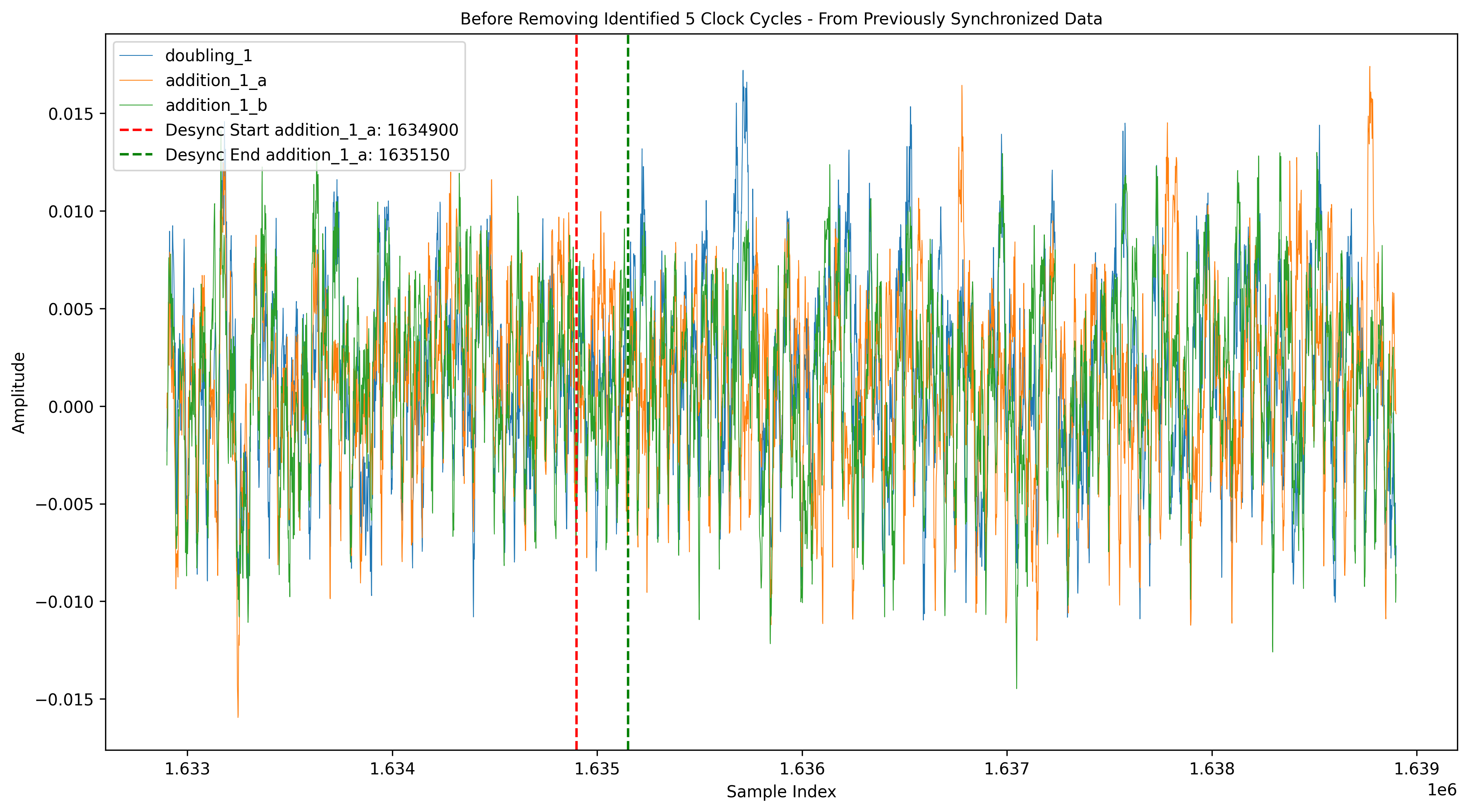} 
    \end{subfigure}
    
    \vspace{1em} 
    
    \begin{subfigure}[b]{0.50\textwidth}
        \caption*{b}
        \includegraphics[width=\textwidth]{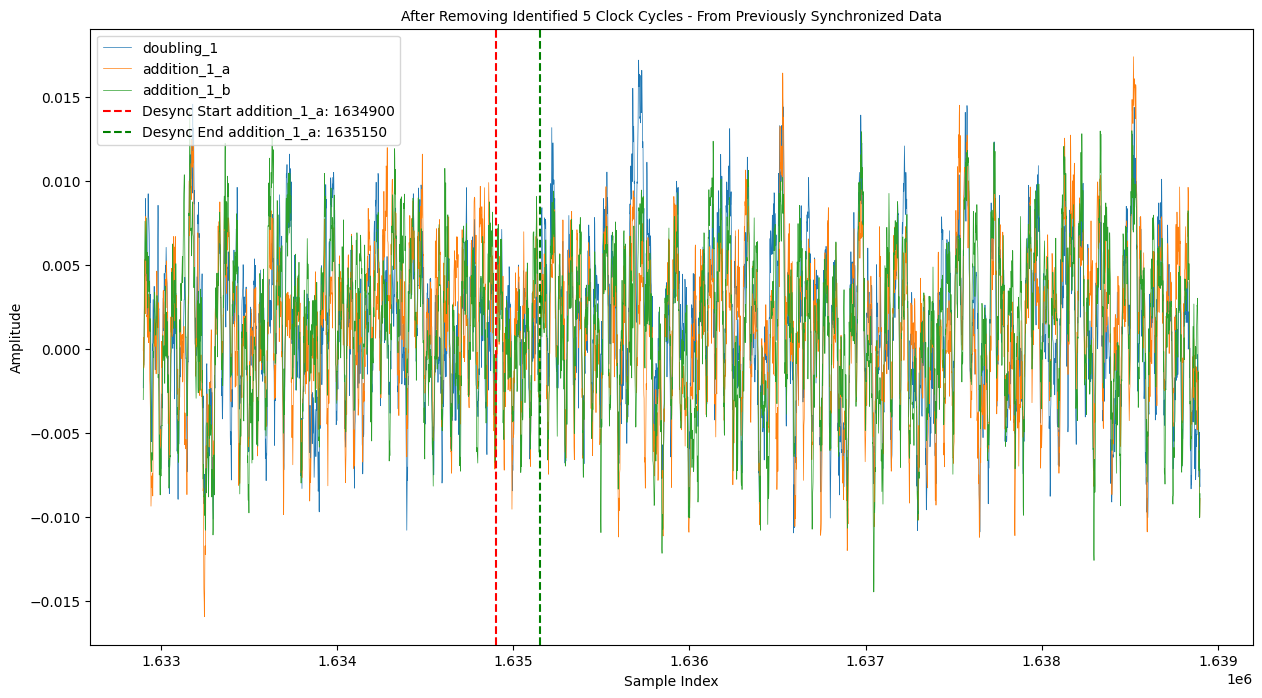}  
    \end{subfigure}

    \caption{\texttt{Above:} Showing 6,000 (120 clock cycles) samples around the start of desynchronisation between \texttt{(Doubling)\(_1\)} (blue), \texttt{(Addition 1)\(_1\)} (orange) and \texttt{(Addition 2)\(_1\)} (green) at about 1,634,900 samples (32,698 clock cycles) from \texttt{(Addition 1)\(_1\)} in the first main loop of the first key bit. \\
    \texttt{Below:} Showing realignments between \texttt{(Doubling)\(_1\)} (blue), \texttt{(Addition 1)\(_1\)} (orange) and \texttt{(Addition 2)\(_1\)} (green) after removing 250 samples (5 clock cycles) from \texttt{(Addition 1)\(_1\)} (orange) in the first main loop of the first key bit.}
    \label{fig:desync-OP1}
\end{figure}

For instance, while comparing the sub-traces corresponding to the processing of the first key bit, we overlaid and plotted \texttt{(Doubling)\(_1\)}, \texttt{(Addition 1)\(_1\)} and \texttt{(Addition 2)\(_1\)} on top of each other. At around roughly sample 1,634,900 (i.e. 32,698 clock cycles), we identified some desynchronisation between \texttt{(Doubling)\(_1\)} and \texttt{(Addition 1)\(_1\)} sub-traces. We also determined that removing 5-clock cycles, equivalent to 250 samples from \texttt{(Addition 1)\(_1\)}  starting from sample point 1,634,900 (indicated by a red vertical dash line in \hyperref[fig:desync-OP1]{Figure~\ref*{fig:desync-OP1}}) to sample point 1,635,150 (indicated by a green vertical dash line in \hyperref[fig:desync-OP1]{Figure~\ref*{fig:desync-OP1}}) from the sub-trace fixed and realigned the sub-traces. We also identified other regions within this sub-trace triplet where there was desynchronisation and ensured that we removed all such 5-clock cycle processes. \hyperref[fig:desync-OP1]{Figure~\ref*{fig:desync-OP1} (a)} shows a plot of 6,000 samples (i.e. 120 clock cycles) around the sub-traces from \texttt{(Doubling)\(_1\)}, \texttt{(Addition 1)\(_1\)} and \texttt{(Addition 2)\(_1\)} before removing the desynchronised samples (i.e. 5-clock cycle processes) starting from sample 1,634,900 and \hyperref[fig:desync-OP1]{Figure~\ref*{fig:desync-OP1} (b)} shows the same region in the sub-traces after removing the desynchronised samples.

\begin{figure}[H]
    \centering

    \begin{subfigure}[b]{0.50\textwidth}
        \caption*{a}
        \includegraphics[width=\linewidth]{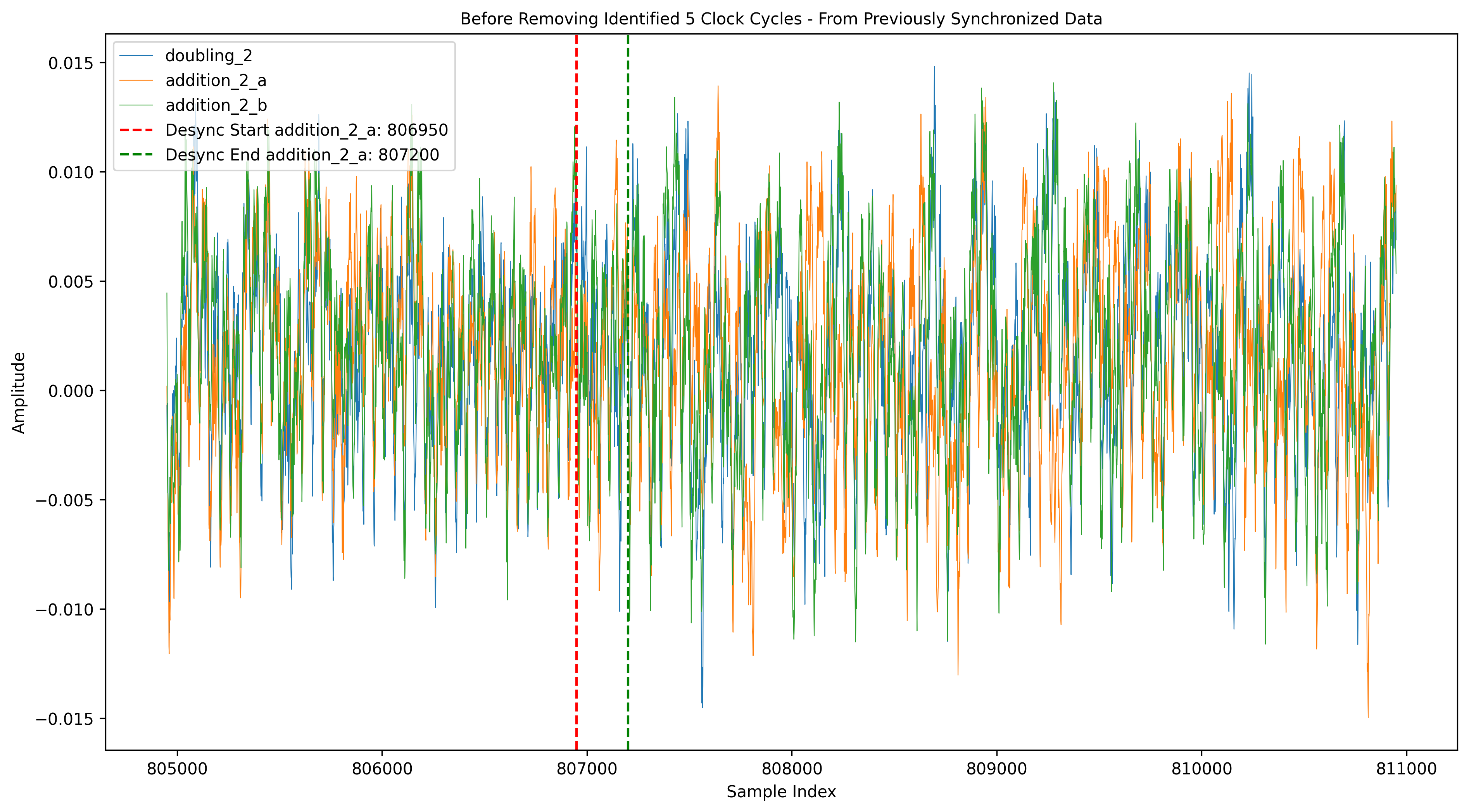} 
    \end{subfigure}

    \vspace{1em} 
    
    \begin{subfigure}[b]{0.50\textwidth}
        \caption*{b}
        \includegraphics[width=\linewidth]{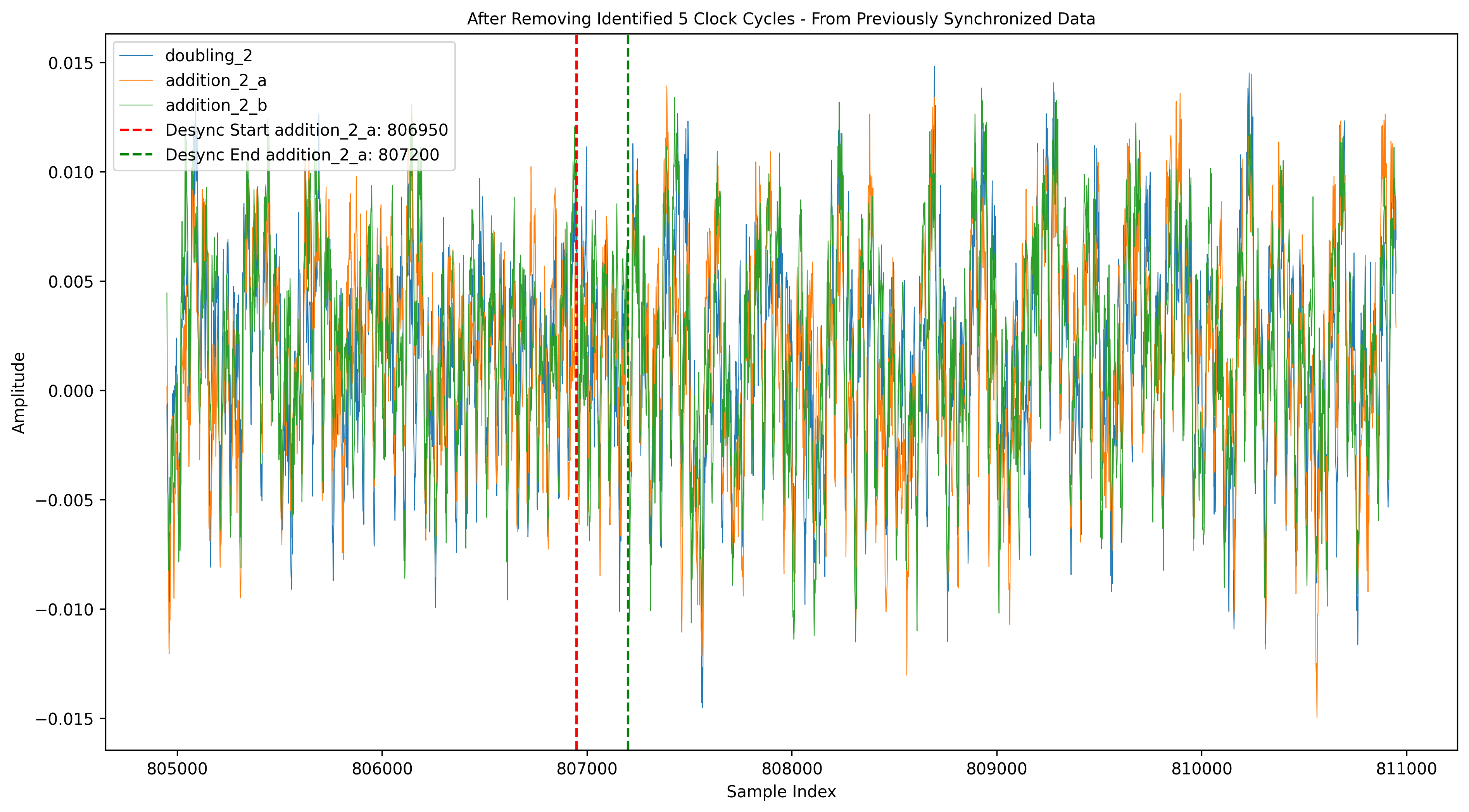}  
    \end{subfigure}

    \caption{\texttt{Above:} Showing 6,000 (120 clock cycles) samples around the start of desynchronisation between \texttt{(Doubling)\(_2\)} (blue), \texttt{(Addition 1)\(_2\)} (orange) and \texttt{(Addition 2)\(_2\)} (green) at about 806,950 samples (16,139 clock cycles) from \texttt{(Addition 1)\(_2\)} in the first main loop of the second key bit. \\
    \texttt{Below:} Showing realignments between \texttt{(Doubling)\(_2\)} (blue), \texttt{(Addition 1)\(_2\)} (orange) and \texttt{(Addition 2)\(_2\)} (green) after removing 250 samples (5 clock cycles) from \texttt{(Addition 1)\(_2\)} (orange) in the first main loop of the second key bit.}
    \label{fig:desync-OP2}
\end{figure}

A similar desynchronisation was observed while comparing the sub-traces from the second key bit triplet. When \texttt{(Doubling)\(_2\)}, \texttt{(Addition 1)\(_2\)} and \texttt{(Addition 2)\(_2\)} sub-traces were overlaid and plotted on top of each other, we identified some desynchronisation between \texttt{(Doubling)\(_2\)} and \texttt{(Addition 1)\(_2\)} at around sample 806,950 (i.e. 16,139 clock cycles). We again removed 5-clock cycles, which is equivalent to 250 samples from the \texttt{(Addition 1)\(_2\)} sub-trace, starting from sample 806,950 (indicated by the red vertical line in \hyperref[fig:desync-OP2]{Figure~\ref*{fig:desync-OP2}}) to sample 807,200 (indicated by the green vertical line in \hyperref[fig:desync-OP2]{Figure~\ref*{fig:desync-OP2}}), which realigned the sub-traces at the index. Here again, we identified multiple regions within this triplet with desynchronised regions and removed the 5-clock cycle processes that caused the desynchronisation. \hyperref[fig:desync-OP2]{Figure~\ref*{fig:desync-OP2} (a)} also shows a plot of 6,000 samples (i.e. 120 clock cycles) around the region in \texttt{(Doubling)\(_2\)}, \texttt{(Addition 1)\(_2\)} and \texttt{(Addition 2)\(_2\)} sub-traces where we identified the desynchronisation and before removing the 5-clock cycle processes while \hyperref[fig:desync-OP2]{Figure~\ref*{fig:desync-OP2} (b)} shows the same region after removing the 5-clock cycle processes.

In our investigations, we also observed the existence of a “safe zone” or a buffer region of approximately 7 to 10 clock cycles within which the precise start and end of the desynchronisation region could not be determined. This means that, within this buffer region, we could remove the 5-clock cycle processes without introducing further desynchronisation in the sub-traces. While this may seem helpful or negligible, it could also potentially cause incorrect results in our analysis, for example, the identification of false \ac{SCA} leakage.

\begin{figure}[H]
    \centering

    \begin{subfigure}[b]{1.0\textwidth}
        \centering
        \caption*{a}
        \includegraphics[width=0.7\textwidth, height=6cm]{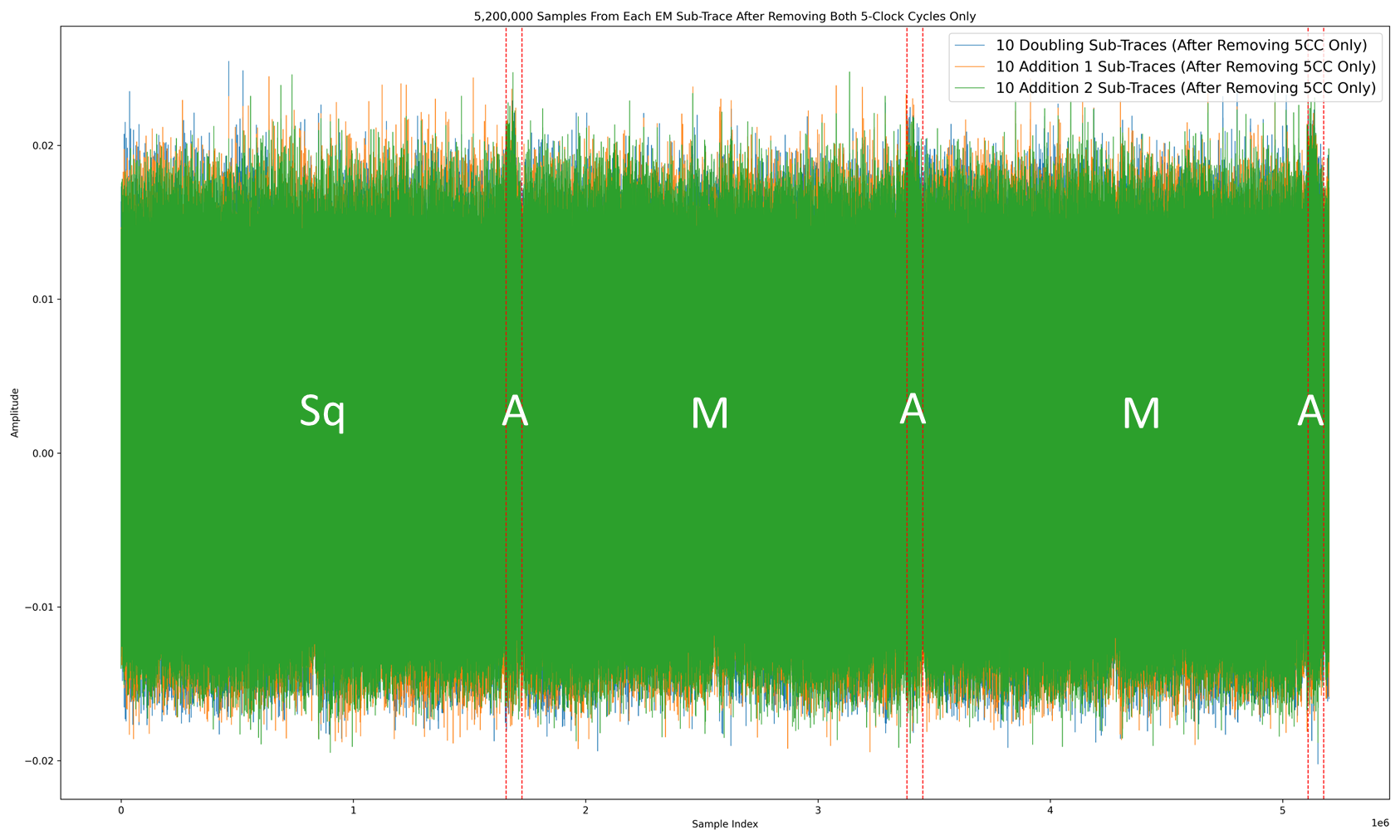}
    \end{subfigure}

    \vspace{0.5em}

    \begin{subfigure}[b]{1.0\textwidth}
        \centering
        \caption*{b}
        \includegraphics[width=0.7\textwidth, height=6cm]{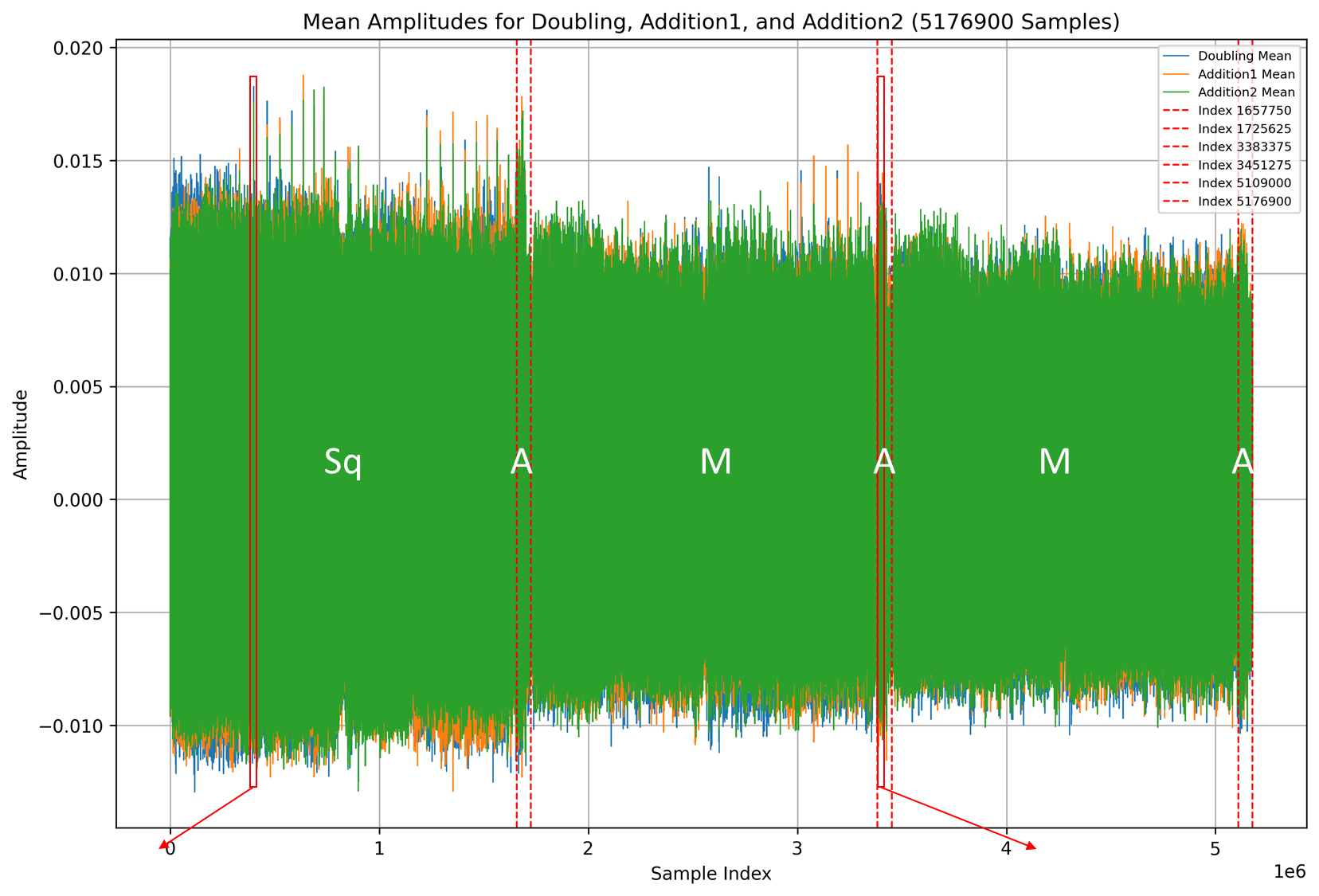}
    \end{subfigure}

    \vspace{0.5em}

    \begin{subfigure}[b]{0.45\textwidth}
        \centering
        \includegraphics[width=1.0\linewidth, height=4cm]{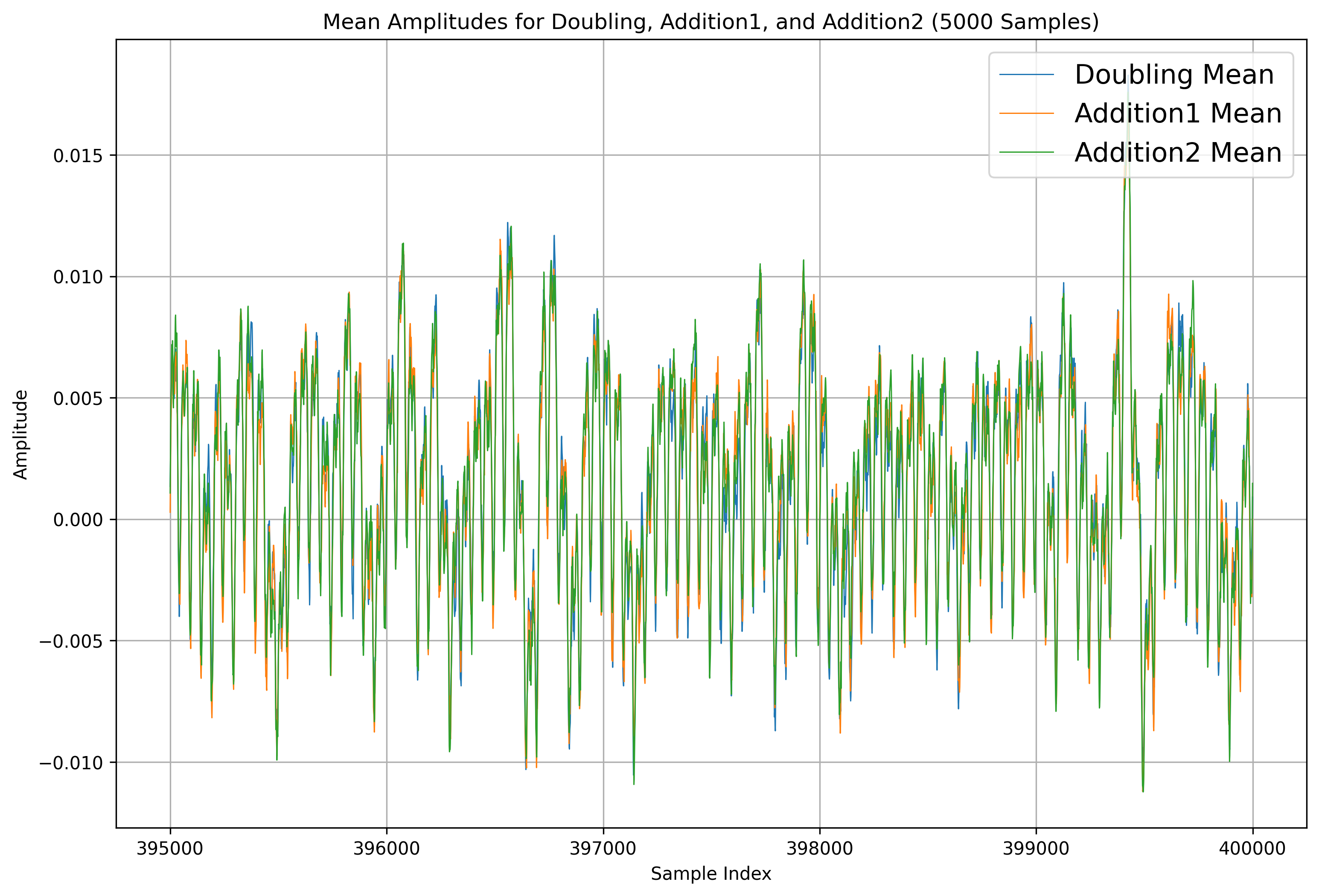}
    \end{subfigure}
    \hspace{0.05\textwidth}
    \begin{subfigure}[b]{0.45\textwidth}
        \centering
        \includegraphics[width=1.0\linewidth, height=4cm]{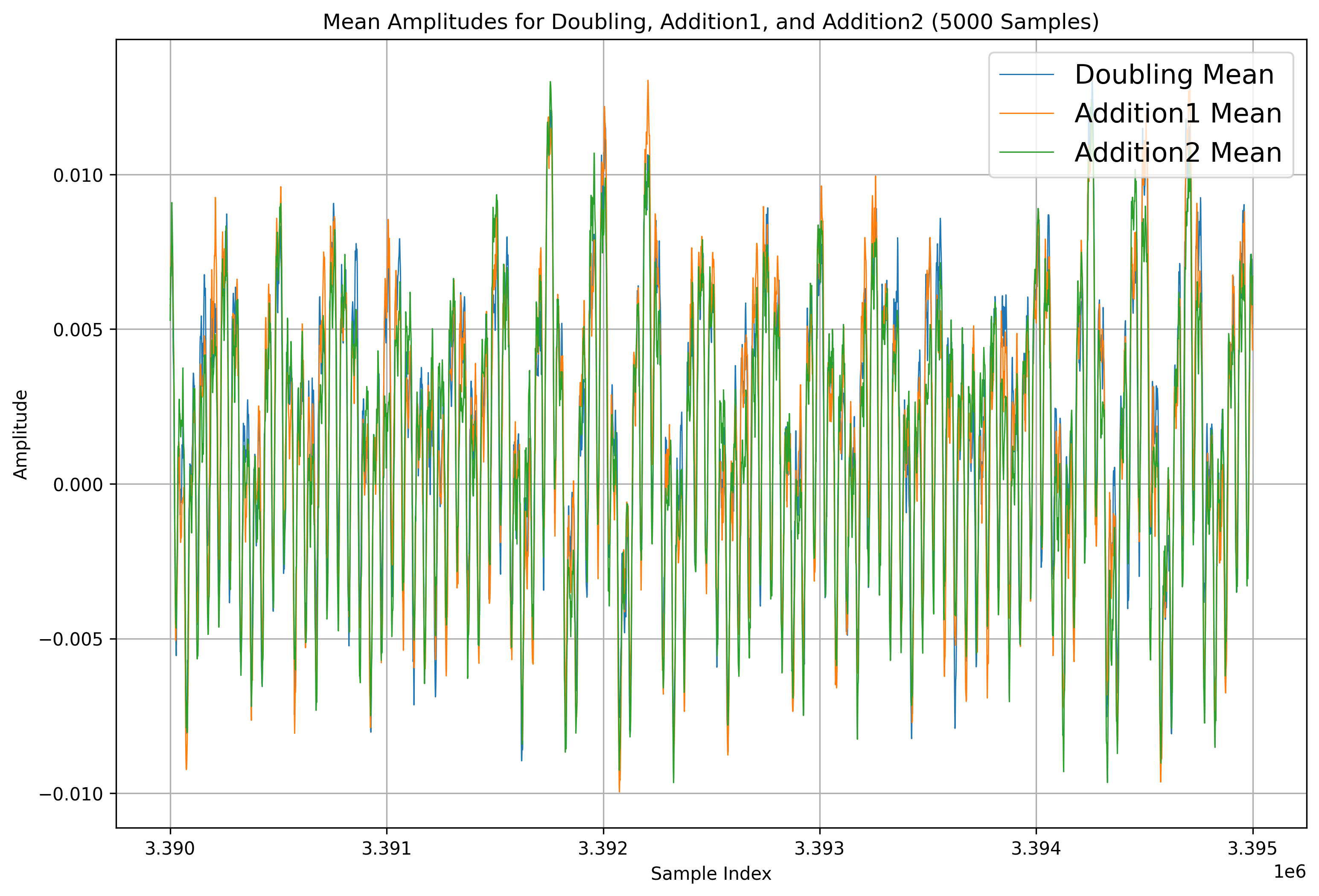}
    \end{subfigure}

    \caption{\textbf{(a)}. Showing the overlaid plot of the first 5,200,000 samples from all 10 Doubling, 10 Addition 1 and 10 Addition 2 sub-traces after removing only the 5-clock cycle processes.\\
    \textbf{(b)} The mean plot of our focus region of 5,200,000 samples from all 10 Doubling, 10 Addition 1 and 10 Addition 2 sub-traces at each index after identifying and removing only the 5-clock cycle processes with zoomed-in sections.}
    \label{fig:mean-plot-5M-5CC}
\end{figure}

For the correctness of our analysis that will be performed later in this work, we identified and removed the 5-clock cycle processes (i.e. 250 samples) that caused the desynchronisation in the sub-traces in every sub-trace triplet corresponding to the processing of all 10 key bits. We ensured that each sub-trace triplet was properly synchronised within our focus area of 5,200,000 samples. In \hyperref[fig:mean-plot-5M-5CC]{Figure~\ref*{fig:mean-plot-5M-5CC} (a)}, we show the raw plot of 5,200,000 samples from the 10 Doubling sub-traces, 10 Addition 1 sub-traces and 10 Addition 2 sub-traces overlaid on top of each other with the marked field operations after removing the 5-clock cycle processes while \hyperref[fig:mean-plot-5M-5CC]{Figure~\ref*{fig:mean-plot-5M-5CC} (b)} shows a mean plot of the same set of sub-traces from the same region with zoomed-in sections within the plot to show how well the sub-traces are synchronised.

We cannot definitively determine the exact cause of these repeated 5-clock cycle delays at this time. Further research is required to determine the precise factors responsible for this behaviour. However, we can establish that this consistent 5-clock-cycle delay corresponds to the duration measured using breakpoints, as shown in \hyperref[tab:execution_time_analysis]{Table~\ref*{tab:execution_time_analysis}}. We also assume that these delays or changes in timing might be caused by the fact that operations that access different registers or memory addresses may use different amounts of power, which could affect the time it takes to address memory. In any case, more research is needed to confirm this assumption.

We identified and removed about 52 of these 5-clock cycle processes across all the sub-trace triplets within our focus region. More details about this can be found in \hyperref[tab:desync_regions]{Table~\ref*{tab:desync_regions}} in \hyperref[appendix:D]{Appendix~\ref*{appendix:D}}.

After identifying and removing all the 5-clock-cycle delays from each sub-trace triplet, we observed additional regions of desynchronisation within some of the triplets. We found that removing 1-clock cycle (i.e. 50 samples), from specific areas in the sub-traces, could further improve the synchronisation. This process was done similarly to removing the 5-clock cycle processes by identifying and removing the 1-clock cycle process from all the sub-trace triplets corresponding to the processing of all the 10 key bits. In this process, we identified and removed 5 of such 1-clock cycle processes across the sub-trace triplets. Although these smaller delays were not as frequent or as consistent as the 5-clock-cycle ones, they still contributed to minor desynchronisation that could affect our analysis if not removed. In fact, after removing these 1-clock cycle processes, we further identified other regions that still required removing 5-clock cycle processes that may not have been identified if these 1-clock cycle processes were not removed. 

\begin{figure}[H]
    \centering

    \begin{subfigure}[b]{0.50\textwidth}
        \caption*{a}
        \includegraphics[width=\linewidth]{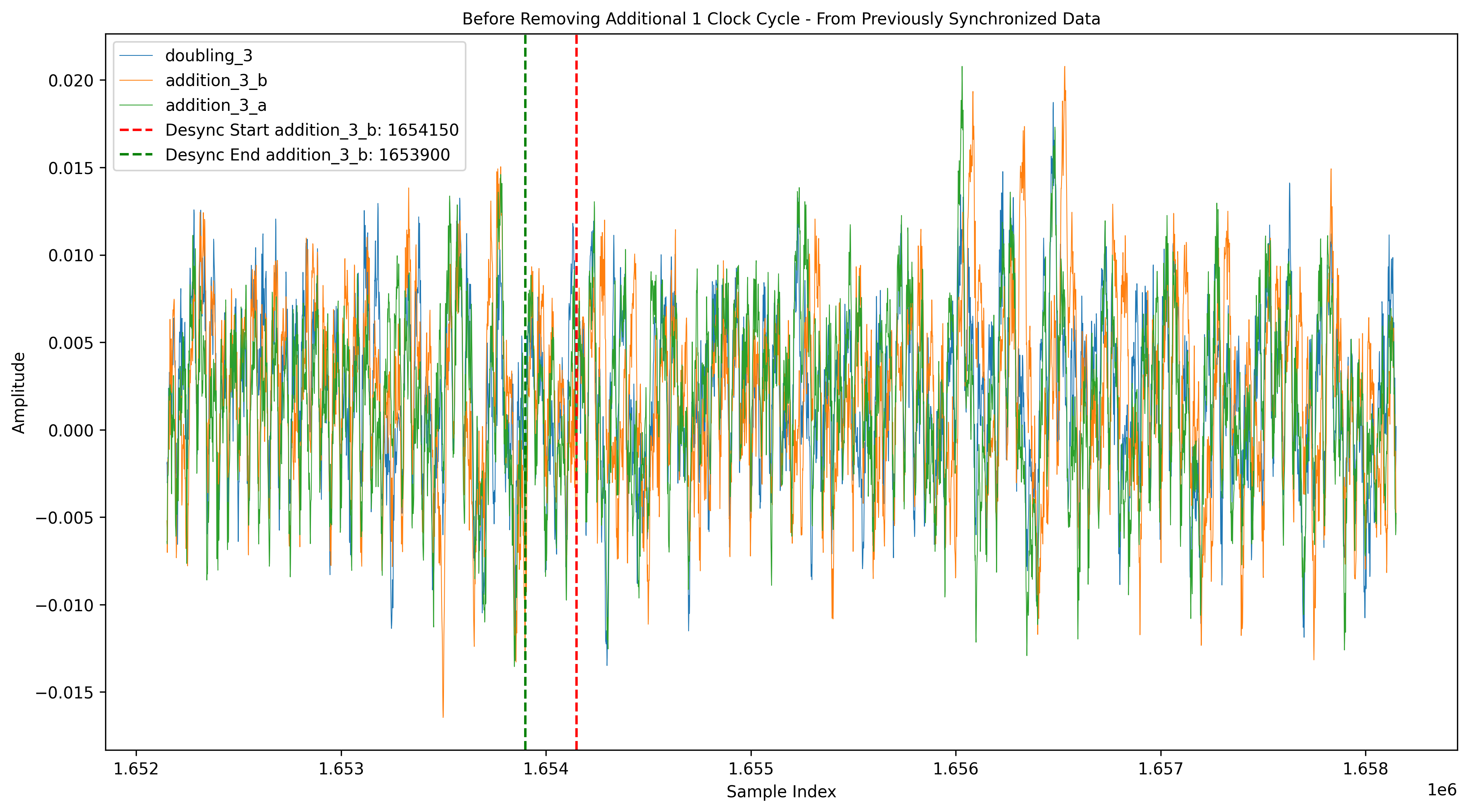} 
    \end{subfigure}

    \vspace{1em} 
    
    \begin{subfigure}[b]{0.50\textwidth}
        \caption*{b}
        \includegraphics[width=\linewidth]{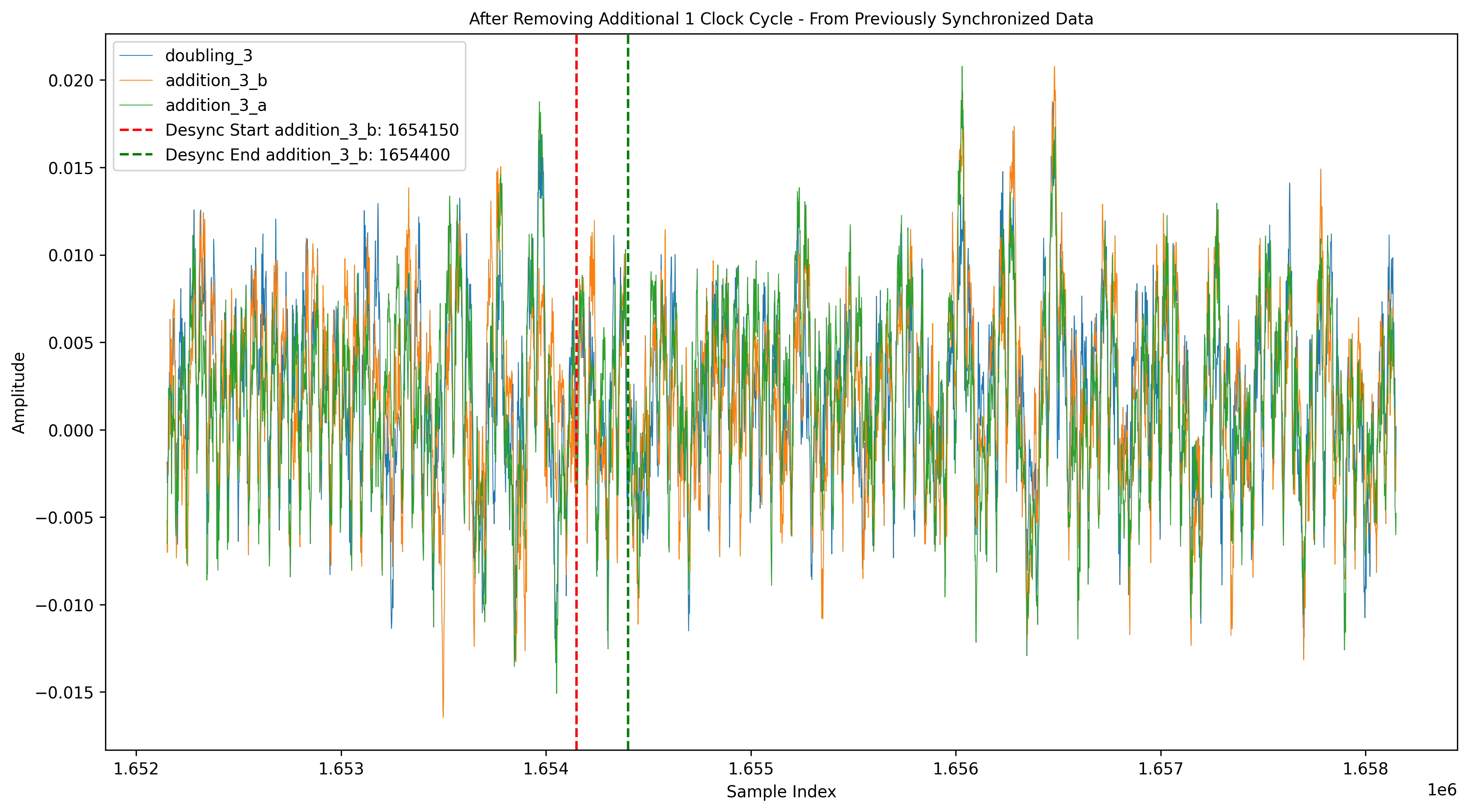}  
    \end{subfigure}

    \caption{\texttt{(a)}. Showing 6000 samples at the start of the desynchronisation between \texttt{(Doubling)\(_3\)} (blue), \texttt{(Addition 2)\(_3\)} (orange) and \texttt{(Addition 1)\(_3\)} (green) at about 1,654,150 samples (33,083 clock cycles) from \texttt{(Addition 2)\(_3\)} in the third main loop of the third key bit. \\
    \texttt{(b)}. showing realignments between \texttt{(Doubling)\(_3\)} (blue), \texttt{(Addition 2)\(_3\)} (orange) and \texttt{(Addition 1)\(_3\)} (green) after removing 50 samples from \texttt{(Addition 2)\(_3\)} (orange).}
    \label{fig:desync-OP3-1CC}
\end{figure}

In \hyperref[fig:desync-OP3-1CC]{Figure~\ref*{fig:desync-OP3-1CC}} we show an example from plotting \texttt{(Doubling)\(_3\)}, \texttt{(Addition 1)\(_3\)} and \texttt{(Addition 2)\(_3\)} sub-traces from within the synchronised sub-traces (where the 5-clock cycles have already been removed) of the third key bit where we identified such an occurrence. From \hyperref[fig:desync-OP3-1CC]{Figure~\ref*{fig:desync-OP3-1CC} (a)}, starting from sample 1,654,150 (i.e. 33,083 clock cycles), we can see that removing 50 samples from (Addition 2)3 (orange) further improved the synchronisation of the sub-trace triplet as shown in \hyperref[fig:desync-OP3-1CC]{Figure~\ref*{fig:desync-OP3-1CC} (b)}.

\begin{figure}[H]
    \centering

    \begin{subfigure}[b]{1.0\textwidth}
        \centering
        \caption*{a}
        \includegraphics[width=0.7\textwidth, height=6cm]{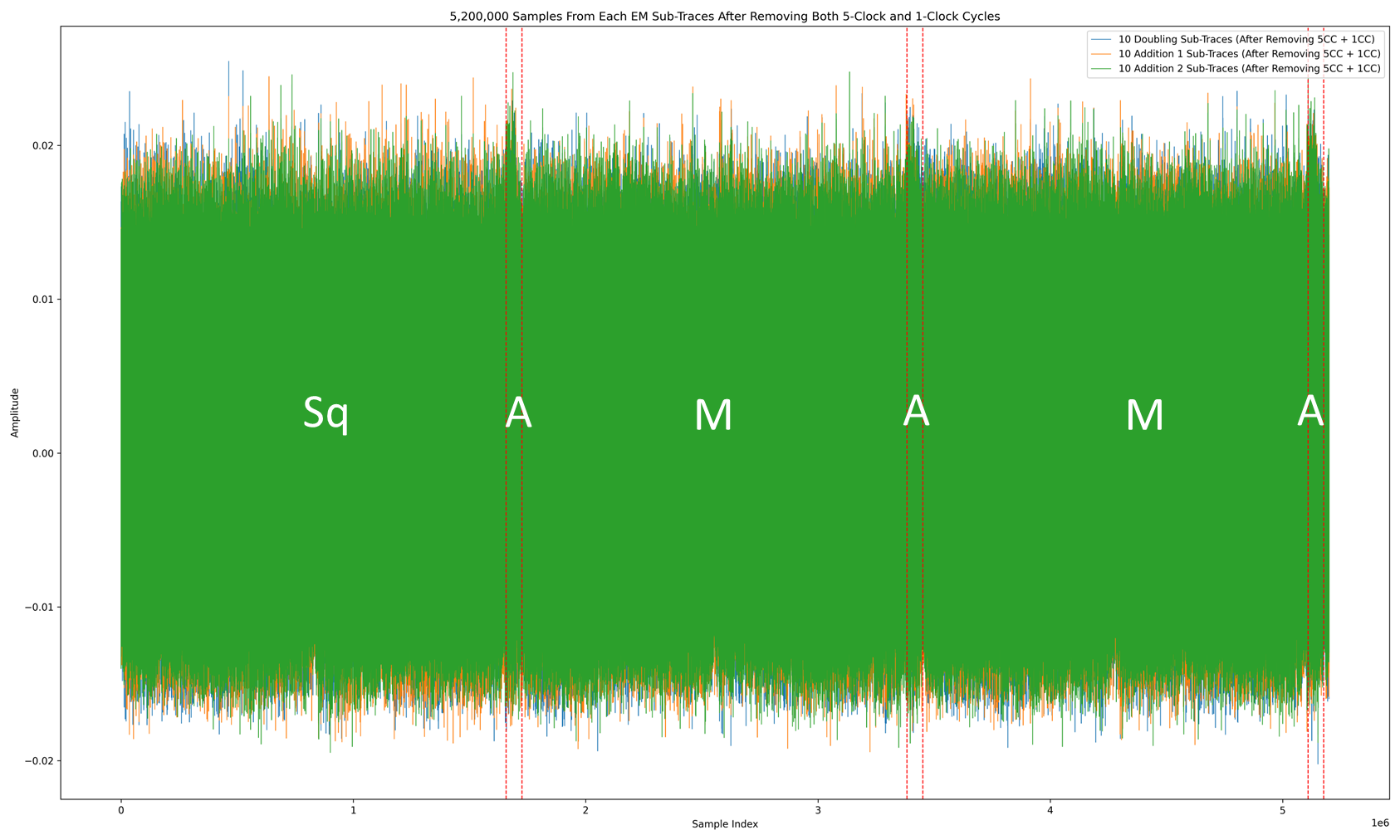}
    \end{subfigure}

    \vspace{0.5em}

    \begin{subfigure}[b]{1.0\textwidth}
        \centering
        \caption*{b}
        \includegraphics[width=0.7\textwidth, height=6cm]{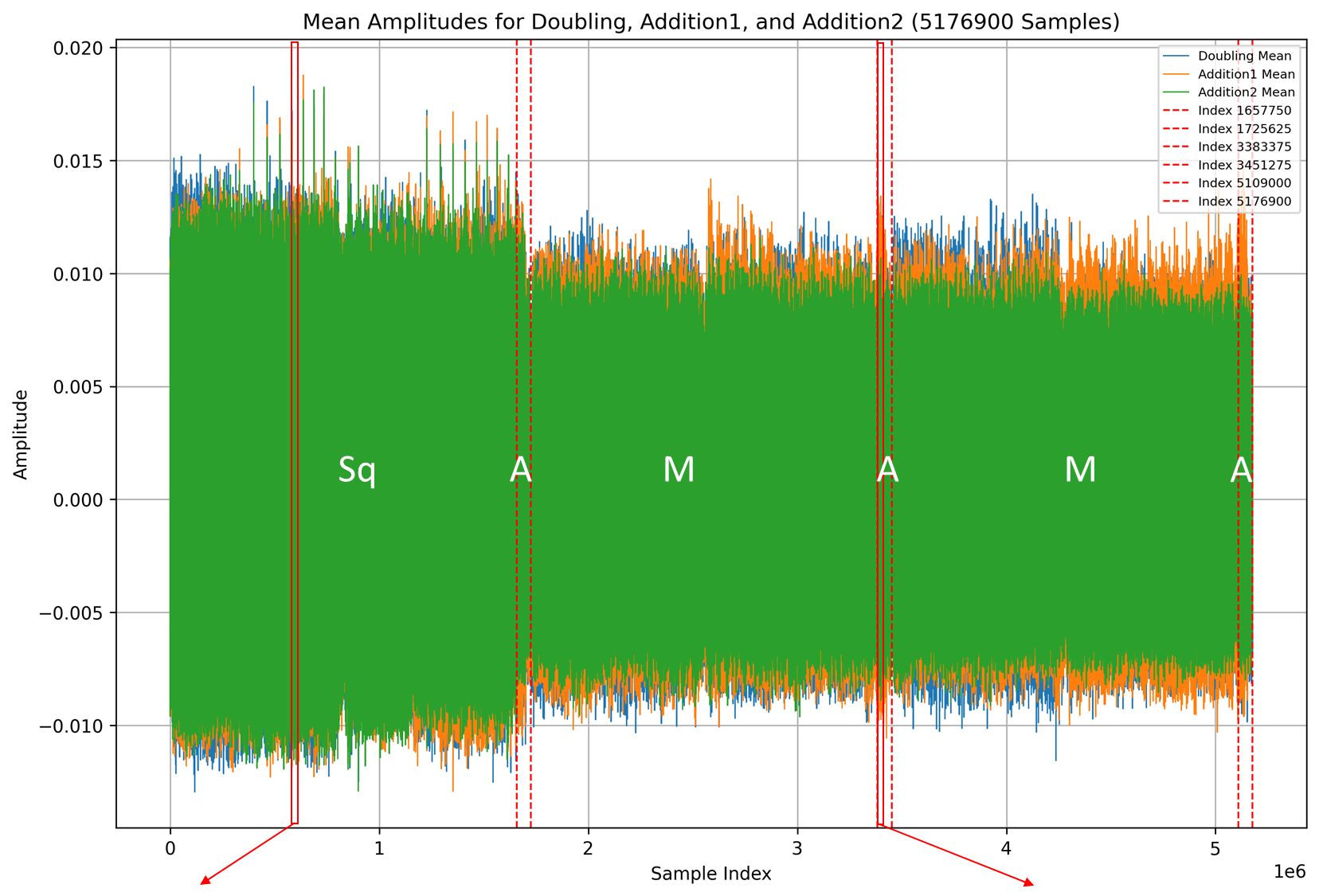}
    \end{subfigure}

    \vspace{0.5em}

    \begin{subfigure}[b]{0.45\textwidth}
        \centering
        \includegraphics[width=1.0\linewidth, height=4cm]{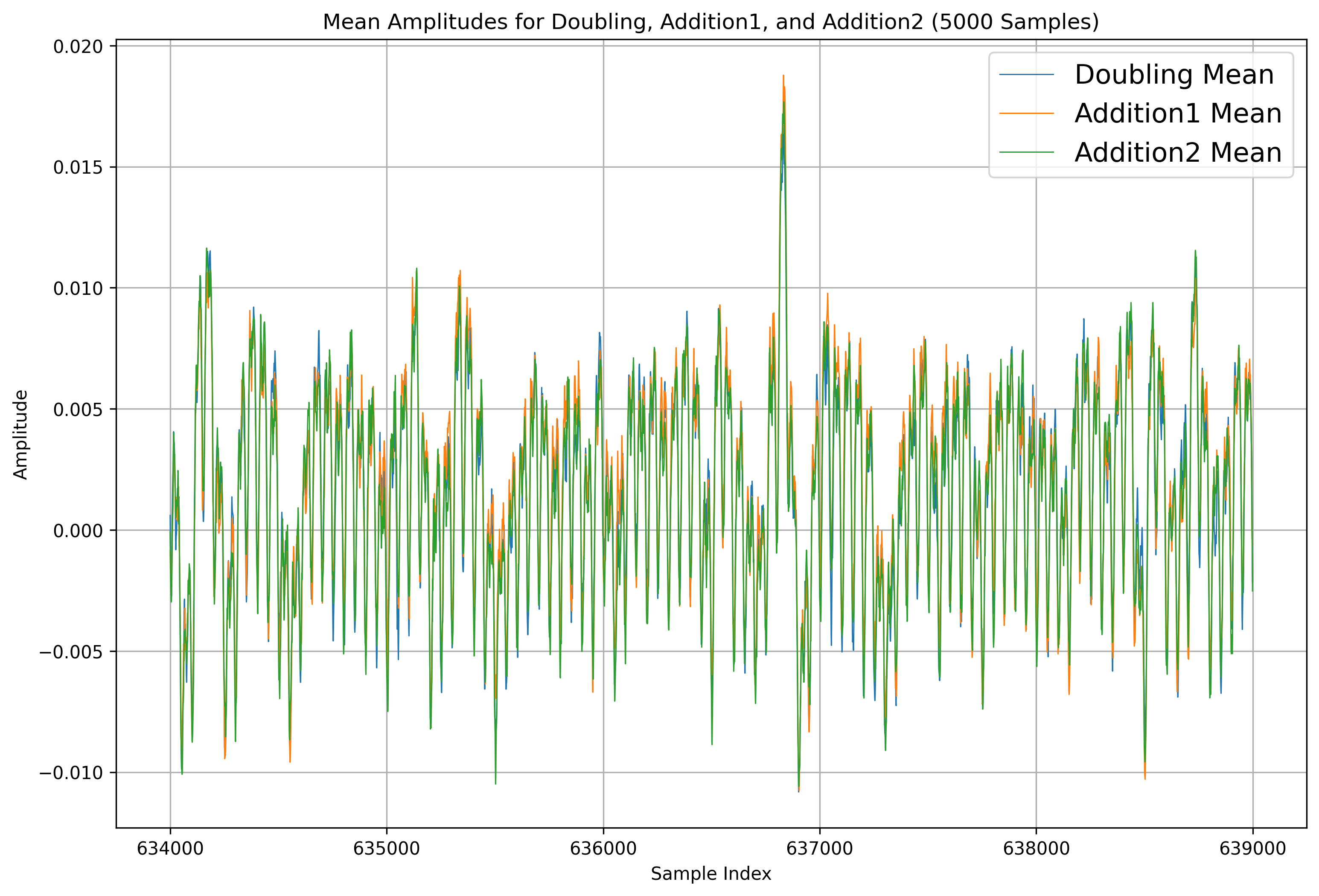}
    \end{subfigure}
    \hspace{0.05\textwidth}
    \begin{subfigure}[b]{0.45\textwidth}
        \centering
        \includegraphics[width=1.0\linewidth, height=4cm]{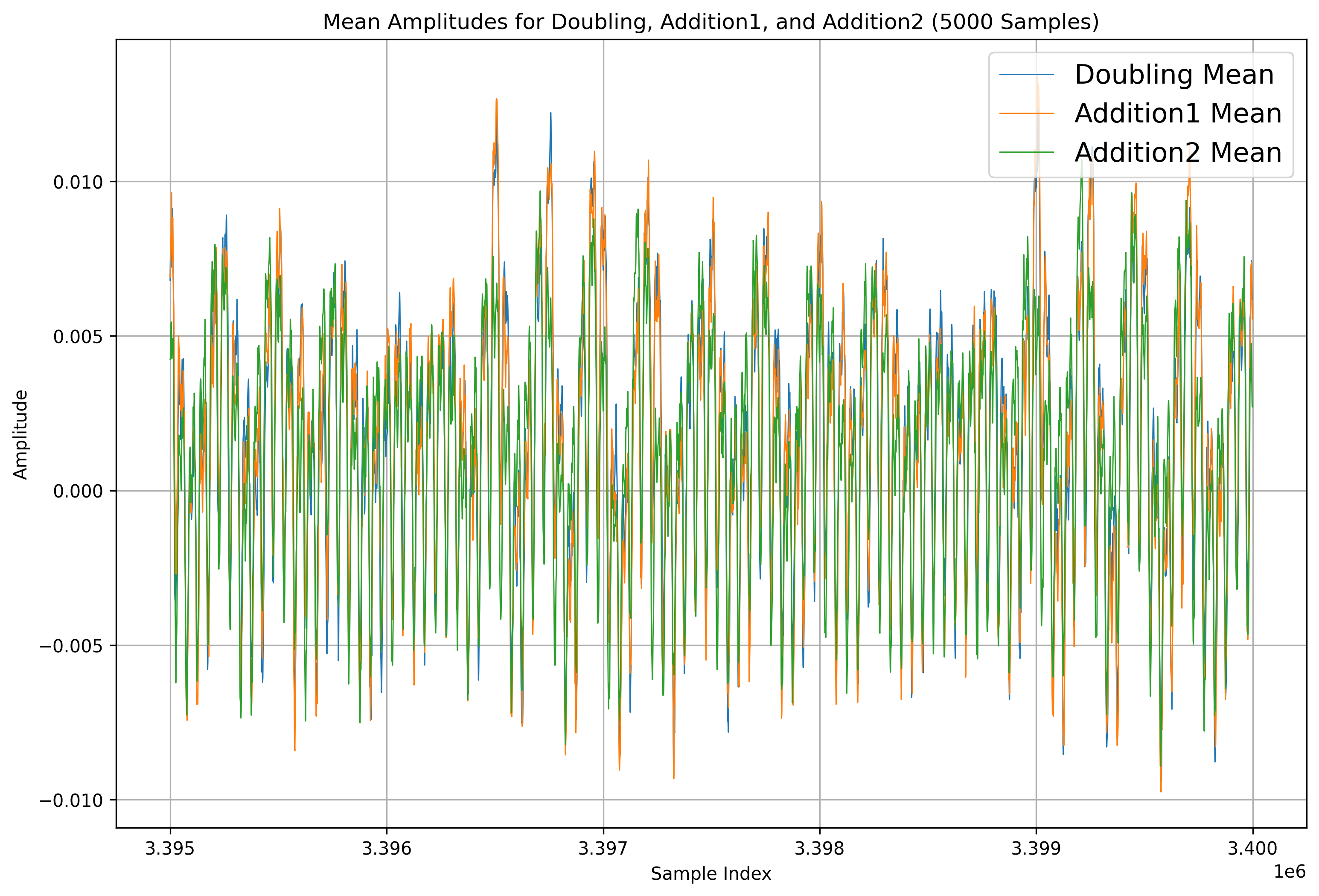}
    \end{subfigure}

    \caption{\textbf{(a)}. Showing the overlaid plot of the first 5,200,000 samples from all 10 Doubling, 10 Addition 1 and 10 Addition 2 sub-traces after removing both 5-clock and 1-clock cycle processes.\\
    \textbf{(b)} The mean plot of our focus region of 5,200,000 samples from all 10 Doubling, 10 Addition 1 and 10 Addition 2 sub-traces at each index after identifying and removing both 5-clock and 1-clock cycle processes with zoomed-in sections.}
    \label{fig:mean-plot-5M-1CC}
\end{figure}

Further research and measurements may be required to understand the processes that contribute to the additionally inserted 5-clock cycle or the 1-clock cycle processes. Without such measurements, these additional processes appear to be random which makes it difficult to identify and automate the search for all such additional processes. \hyperref[fig:mean-plot-5M-1CC]{Figure~\ref*{fig:mean-plot-5M-1CC} (a)} again shows a plot of 5,200,000 samples from all the 10 Doubling sub-traces, 10 Addition 1 sub-traces and 10 Addition 2 sub-traces overlaid on top of each other with the marked field operations after removing both 5-clock and 1-clock cycle processes while \hyperref[fig:mean-plot-5M-1CC]{Figure~\ref*{fig:mean-plot-5M-1CC} (b)} shows the mean plot of the same set of sub-traces from the same this region with zoomed-in sections after removing both the 5-clock cycle processes and the 1-clock cycle process.

\hyperref[tab:desync_regions]{Table~\ref*{tab:desync_regions}} in \hyperref[appendix:D]{Appendix~\ref*{appendix:D}} shows the sample index numbers from which all 5-clock cycle processes including the 1 clock-cycle processes were removed for each sub-trace triplet corresponding to each key bit.

\section{Automated Simple Electromagnetic Analysis Attack}

After the synchronisation, we implemented the simple \ac{SCA} method described by Kabin in \cite{kabin_horizontal_2023} and extended it with thresholds to investigate the distinguishability in the amplitude ranges for each atomic block. We applied this method pairwise (i.e. between sub-traces from Doubling vs Addition 1, Doubling vs Addition 2, and Addition 1 vs Addition 2) to two different sets of sub-traces, focusing on samples from our focus region only.

\begin{itemize}
    \item \textbf{Sub-trace set 1:} 30 sub-traces where we identified and removed both 5-clock and 1-clock cycle processes within our focus region.
    
    \item \textbf{Sub-trace set 2:} Another set of 30 sub-traces where we identified and removed only 5-clock cycles processes within our focus region.
\end{itemize}

At each index across the combined sub-traces for each atomic block, from the beginning to the end of each sub-trace, we calculated the maximum and minimum amplitude values for each operation, resulting in three separate sub-traces - one for each atomic block and further denoted as “max-min sub-traces”. Specifically, we selected all Doubling sub-traces (10 sub-traces) and computed the maximum and minimum values at each index across all the sub-traces focusing on the first 5,200,000 samples only. We did the same for Addition 1 and Addition 2 sub-traces as well. In each of these max-min sub-traces, the maximum amplitude at every index was stored in one column and the minimum amplitude in the other. This was done for all 30 sub-traces corresponding to all 10 key bits.

To detect anomalies, we applied a fixed threshold of 0.003 V as a margin for identifying non-overlapping amplitude ranges, suggesting potential distinguishability. This means that an anomaly is flagged when the difference between the maximum and minimum amplitude voltage of the sample index exceeds 0.003 V. This threshold value seemed like a fairly significant indicator of distinguishability to identify meaningful differences between the amplitude ranges in the atomic blocks. Using higher threshold could result in failing to identify any distinguishability in the atomic blocks especially when differences in the amplitude are small but still meaningful. On the other hand, using a much smaller threshold could lead to identifying smaller but irrelevant differences between the amplitude ranges.

Our detection was automated through a Python script that flagged an anomaly between any two datasets (max-min sub-trace from any two atomic blocks) when:

\begin{itemize}
    \item The maximum amplitude voltage at an index of one dataset was lower than the minimum amplitude voltage of the other at the same index by more than the threshold, i.e.,
    \[
    \text{max}_{\text{dataset\_1}}(i) < \text{min}_{\text{dataset\_2}}(i) - 0.003
    \] \\

    OR
    
    \item The reverse was true, with the maximum amplitude voltage of the second dataset at an index lower than the minimum voltage of the first at the same index by more than the threshold, i.e.,
    \[
    \text{max}_{\text{dataset\_2}}(i) < \text{min}_{\text{dataset\_1}}(i) - 0.003
    \]
\end{itemize}

Should either of these conditions hold at a given index, this non-overlapping range indicated an anomaly, suggesting possible side-channel distinguishability. We also handled comparisons among three datasets using a similar approach whereby the threshold condition was applied over all possible combinations among the three sets (max-min sub-traces from all three atomic blocks).

Additionally, we implemented a similar function to handle comparisons among three datasets, where the threshold condition applied across all combinations within the three sets:
\begin{itemize}
    \item The maximum amplitude voltage at an index \( i \) of one dataset was lower than the minimum amplitude voltage of the other two datasets at the same index by more than the threshold, i.e.,
    \[
    \text{max}_{\text{dataset\_1}}(i) < \text{min}_{\text{dataset\_2}}(i) - 0.003 \quad \textbf{OR}
    \]
    \[
    \text{max}_{\text{dataset\_1}}(i) < \text{min}_{\text{dataset\_3}}(i) - 0.003.
    \]

    \item The maximum amplitude voltage at an index \( i \) of the second dataset was lower than the minimum amplitude voltage of the first and third datasets at the same index by more than the threshold, i.e.,
    \[
    \text{max}_{\text{dataset\_2}}(i) < \text{min}_{\text{dataset\_1}}(i) - 0.003 \quad \textbf{OR}
    \]
    \[
    \text{max}_{\text{dataset\_2}}(i) < \text{min}_{\text{dataset\_3}}(i) - 0.003.
    \]

    \item The maximum amplitude voltage at an index \( i \) of the third dataset was lower than the minimum amplitude voltage of the first and second datasets at the same index by more than the threshold, i.e.,
    \[
    \text{max}_{\text{dataset\_3}}(i) < \text{min}_{\text{dataset\_1}}(i) - 0.003 \quad \textbf{OR}
    \]
    \[
    \text{max}_{\text{dataset\_3}}(i) < \text{min}_{\text{dataset\_2}}(i) - 0.003.
    \]
\end{itemize}

\subsection{Findings From Sub-trace Set 1}

\begin{figure}[H]
    \centering
    \includegraphics[width=0.7\linewidth]{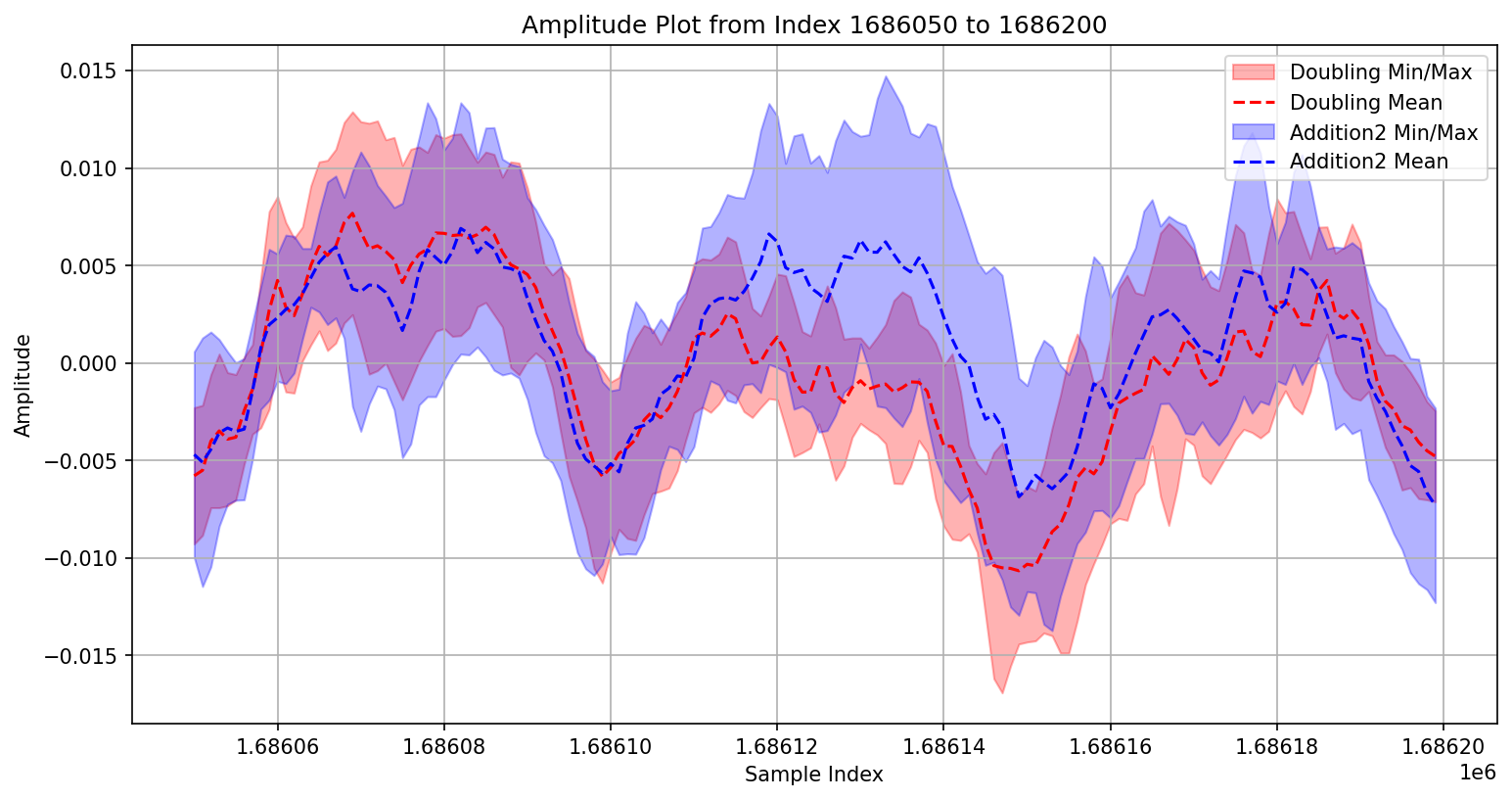}
    \caption{Comparison of the maximum and minimum amplitude ranges for Doubling (orange) and Addition 2 (blue) max-min sub-traces from sub-trace set 1 (with both 5 and 1 clock cycles removed) showing indistinguishability in the atomic blocks.}
    \label{fig:no-distinguishability-D-A2}
\end{figure}

\begin{figure}[H]
    \centering
    \includegraphics[width=0.7\linewidth]{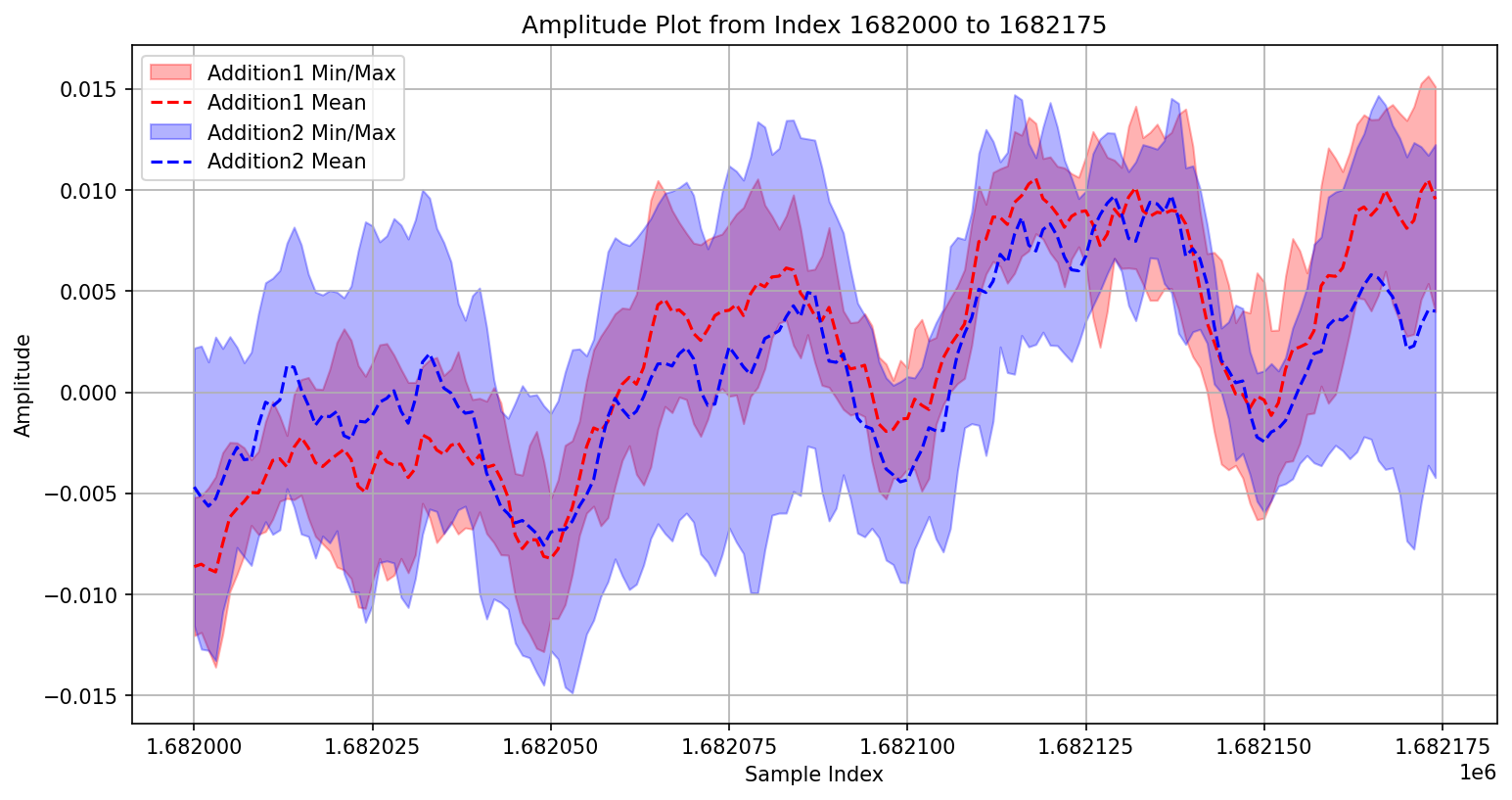}
    \caption{Comparison of the maximum and minimum amplitude range for Addition 1 (orange) and Addition 2 (blue) max-min sub-traces from sub-trace set 1 (with both 5 and 1 clock cycles removed) showing indistinguishability in the atomic blocks.}
    \label{fig:no-distinguishability-A1-A2}
\end{figure}

For “sub-trace set 1” (where both 5-clock cycles and 1-clock cycles were removed), our analysis we did not identify any gaps in the amplitude ranges while analysing the max-min sub-traces from any of the two pairs of atomic blocks neither did we identify any gaps while analysing max-min sub-traces from all three atomic blocks simultaneously using a threshold of 0.003 V. This indicates that, when both the additionally inserted 5-clock and 1-clock cycles were removed, the amplitude ranges for all atomic blocks (Doubling, Addition 1, and Addition 2) overlapped properly which makes it difficult to distinguish between them from the attacker’s perspective. To further assess the synchronisation, we selected and plotted the regions from comparing Doubling (orange) and Addition 2 (blue) max-min sub-traces at sample index 1,686,050 to 1,686,200 as shown in \hyperref[fig:no-distinguishability-D-A2]{Figure~\ref*{fig:no-distinguishability-D-A2}} and also from Addition 1 (orange) and Addition 2 (blue) max-min sub-traces at sample index 1,682,000 to 1,682,175 as shown in \hyperref[fig:no-distinguishability-A1-A2]{Figure~\ref*{fig:no-distinguishability-A1-A2}}. We analysed these regions to check how well the max-min sub-traces overlapped after we removed both 5-clock and 1-clock cycle processes. From the plots, it is clear that the sub-traces mostly overlap in these regions which confirms that there is no distinguishability in these regions. Similar results were also observed across the entire max-min sub-trace that was analysed.

\subsection{Findings From Sub-trace Set 2}

\begin{figure}[H]
    \centering
    \includegraphics[width=0.7\linewidth]{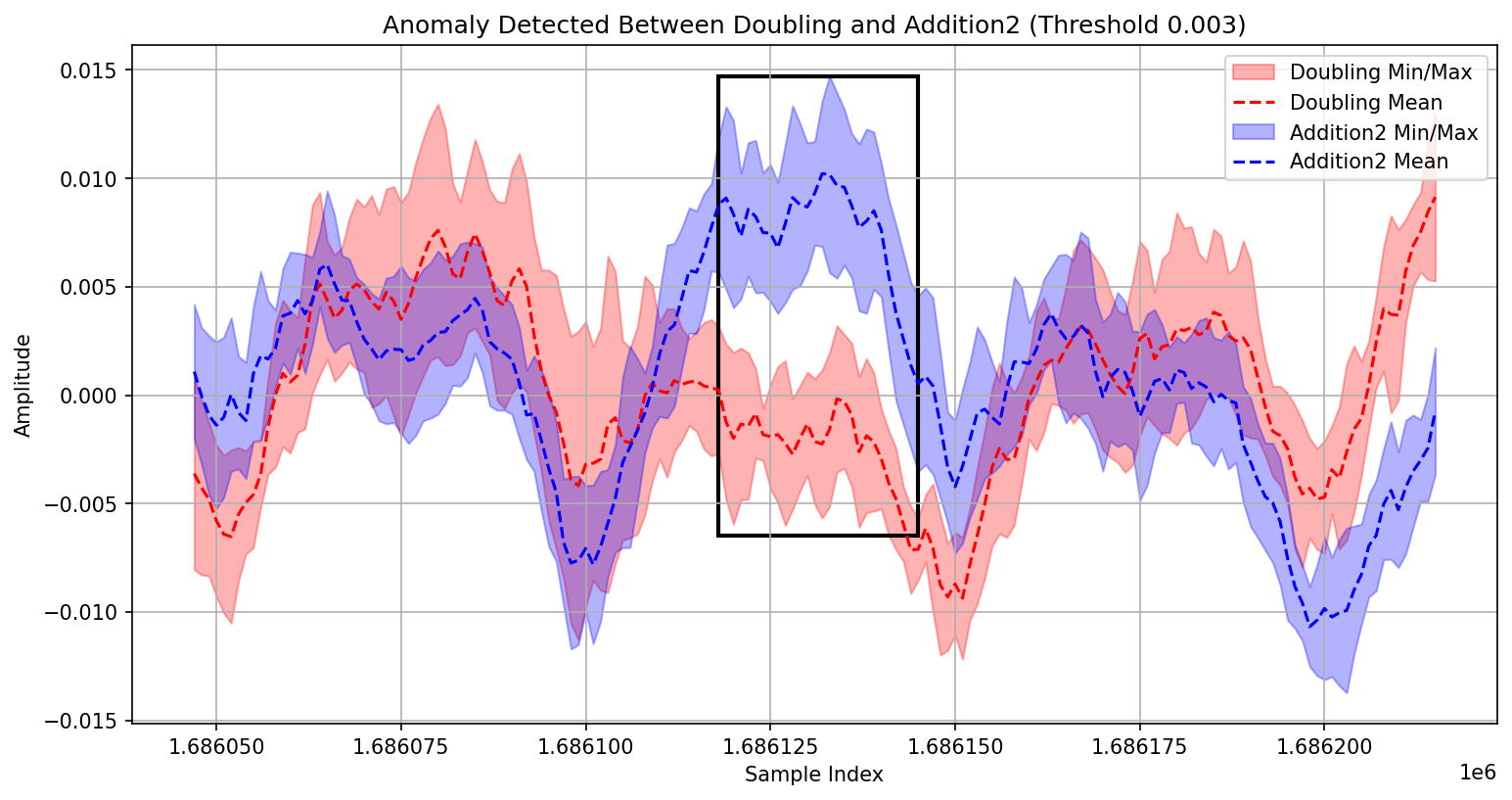}
    \caption{An example of distinguishability analysis between the maximum and minimum amplitude voltage in max-min sub-traces from Doubling (orange) and Addition 2 (blue) atomic blocks with $v > 0.003$ at around sample index 1,682,122 to 1,682,141.}
    \label{fig:distinguishability-D-A2}
\end{figure}

Focusing on “sub-trace set 2” (where we removed only 5-clock cycle processes), we only analysed this sub-trace set to demonstrate how the desynchronisation of the sub-traces could result in incorrect analysis result. After analysing the results from comparing sub-traces of two atomic blocks, we found several instances where the amplitude ranges of the different atomic blocks did not overlap, which indicates distinguishable characteristics. In \hyperref[fig:distinguishability-D-A2]{Figure~\ref*{fig:distinguishability-D-A2}}, while comparing the maximum and minimum amplitude voltages between the max-min sub-traces of the Doubling and Addition 2 atomic blocks, a clear gap is observed around sample index 1,686,122 to 1,686,141. The pink region in \hyperref[fig:distinguishability-D-A2]{Figure~\ref*{fig:distinguishability-D-A2}} represents the range between the maximum and minimum amplitude voltages of the sub-traces from the Doubling atomic block, while the blue region represents the range for the Addition 2 atomic block. The purple region shows the region where there are no gaps between the maximum or minimum amplitude voltages of either atomic blocks or how they overlap.

\begin{figure}[H]
    \centering

    \begin{subfigure}[b]{0.45\textwidth}
        \caption*{\textbf{(a):} Showing Distinguishability.}
        \includegraphics[width=\linewidth]{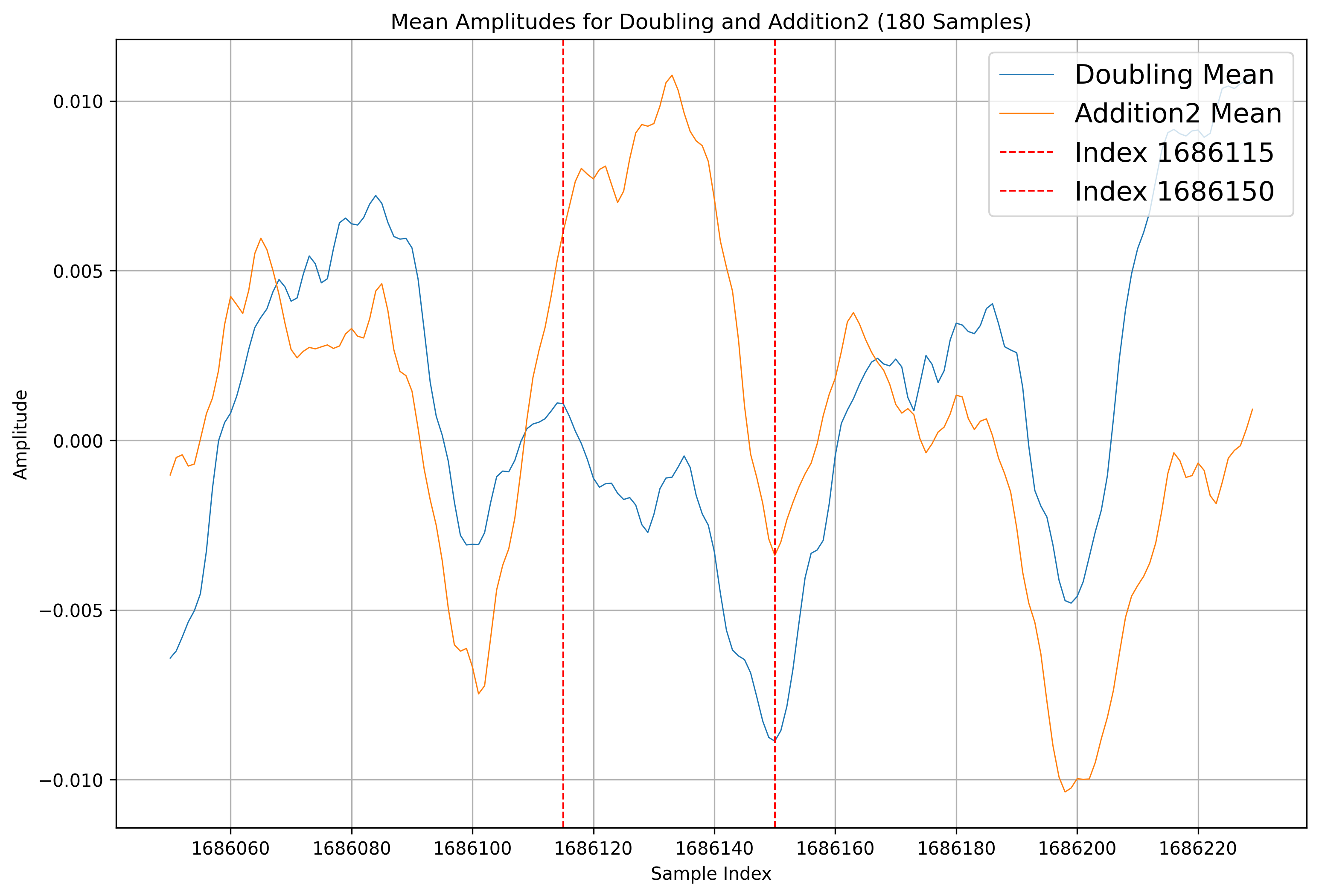} 
    \end{subfigure}
    \hspace{0.05\textwidth}
    \begin{subfigure}[b]{0.45\textwidth}
        \caption*{\textbf{(b):} Showing No Distinguishability.}
        \includegraphics[width=\linewidth]{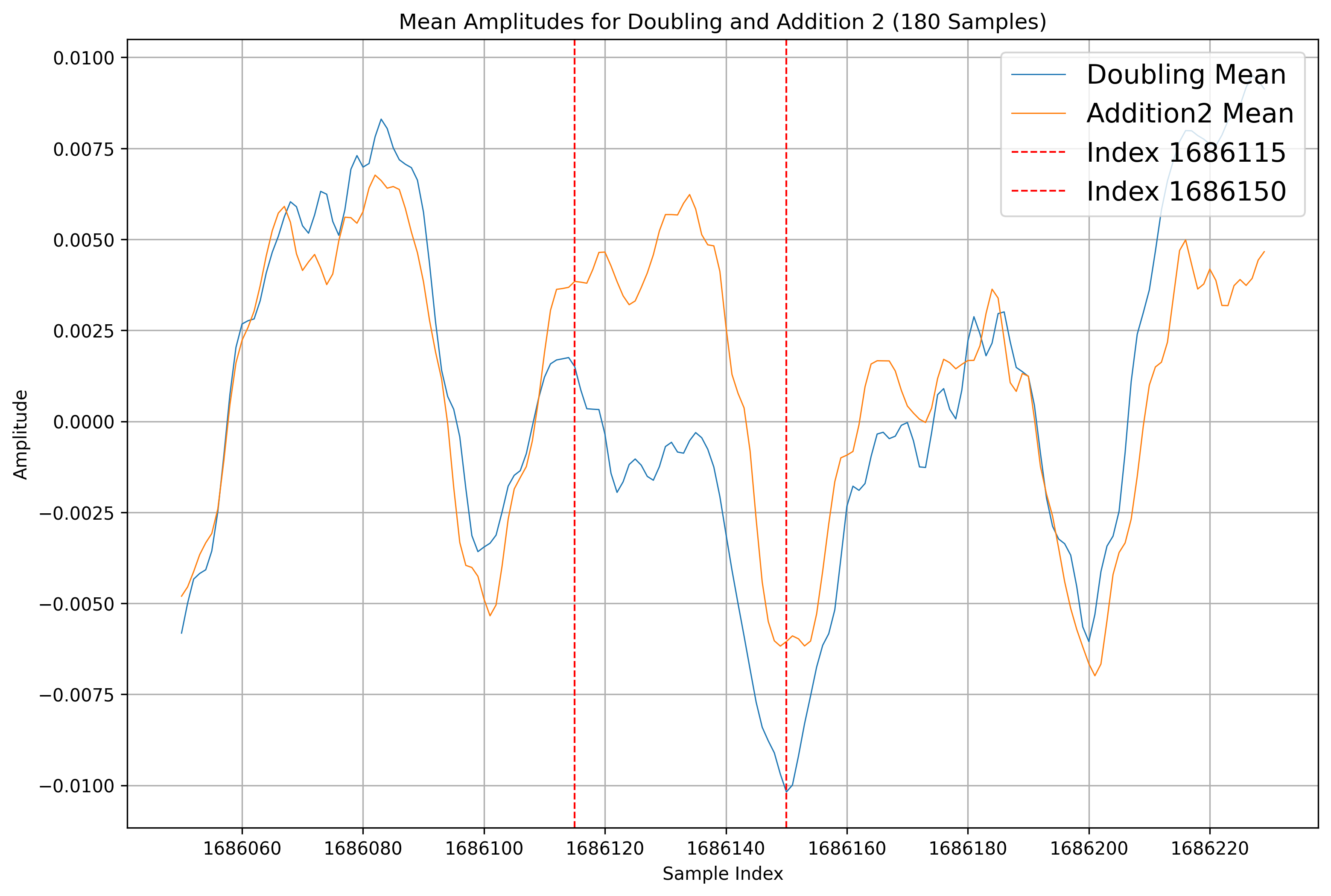}  
    \end{subfigure}

    \caption{\textbf{Left:} Mean amplitude plot of the Doubling and Addition 2 sub-traces from \hyperref[fig:distinguishability-D-A2]{Figure~\ref*{fig:distinguishability-D-A2}} (with only 5-clock cycle processes removed), showing a visibly large gap between the amplitude signals at index 1,686,122 to 1,686,141 due to the desynchronisation of sub-traces and also showing the reason for the distinguishability. \\
    \textbf{Right:} Mean amplitude plot of the same sub-traces used in \hyperref[fig:no-distinguishability-A1-A2]{Figure~\ref*{fig:no-distinguishability-A1-A2}} (with both 5-clock and 1-clock cycle processes removed), showing how close the amplitude signals overlap in the same region and showing the reason for no distinguishability due to improved synchronisation.}
    \label{fig:mean-plot-D-A2}
\end{figure}

To further show the distinguishability, \hyperref[fig:mean-plot-D-A2]{Figure~\ref*{fig:mean-plot-D-A2} (a)} shows the mean plot of the same sub-traces from \hyperref[fig:distinguishability-D-A2]{Figure~\ref*{fig:distinguishability-D-A2}} (i.e. max-min of Doubling and Addition 2 in sub-trace set 2) while focusing on the same region where we observed the distinguishaibility i.e. from sample index 1,686,122 to 1,686,141. The mean plot shows the large gap between the amplitude signals from the Doubling and Addition 2 atomic blocks in this region. On the other hand, \hyperref[fig:mean-plot-D-A2]{Figure~\ref*{fig:mean-plot-D-A2} (b)} shows the mean plot from the same sub-traces used in \hyperref[fig:no-distinguishability-A1-A2]{Figure~\ref*{fig:no-distinguishability-A1-A2}} i.e. where both 5-clock and 1-clock cycle processes were removed. In this plot, the mean amplitude signals from the 2 atomic blocks closely overlap in the same region, which explains why there was no distinguishability observed. The two plots demonstrate the effect of synchronisation in the amplitude signals in the two sub-trace sets when both 5-clock and 1-clock cycle processes were removed from the sub-traces, which further confirms that the desynchronisation of the sub-traces, due to the additionally inserted clock processes was a key factor in the distinguishability we observed in the analysis.

\begin{figure}[H]
    \centering
    \includegraphics[width=0.7\linewidth]{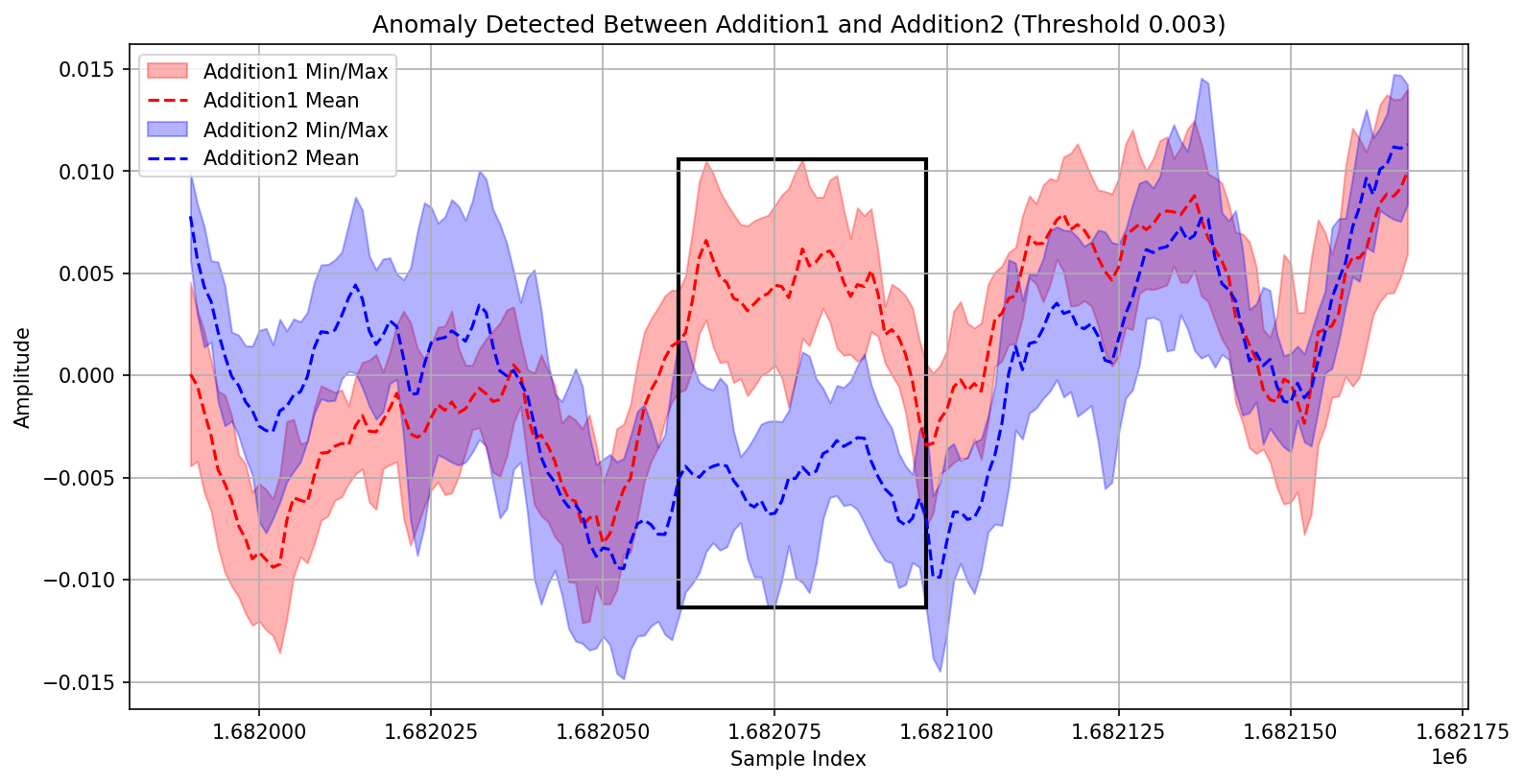}
    \caption{An example of distinguishability analysis between the maximum and minimum amplitude ranges in max-min sub-traces from Addition 1 (orange) and Addition 2 (blue) atomic blocks with $v > 0.003$ at around sample index 1,682,065 to 1,682,093.}
    \label{fig:distinguishability-A1-A2}
\end{figure}

Similarly, in \hyperref[fig:distinguishability-A1-A2]{Figure~\ref*{fig:distinguishability-A1-A2}}, while comparing the sub-traces from the Addition 1 and Addition 2 atomic blocks, at around sample index 1,682,065 to 1,682,093, there were also instances where the atomic blocks showed patterns that can be distinguished, with a clear gap between the maximum and minimum amplitudes.

\begin{figure}[H]
    \centering

    \begin{subfigure}[b]{0.45\textwidth}
        \caption*{\textbf{(a):} Showing Distinguishability.}
        \includegraphics[width=\linewidth]{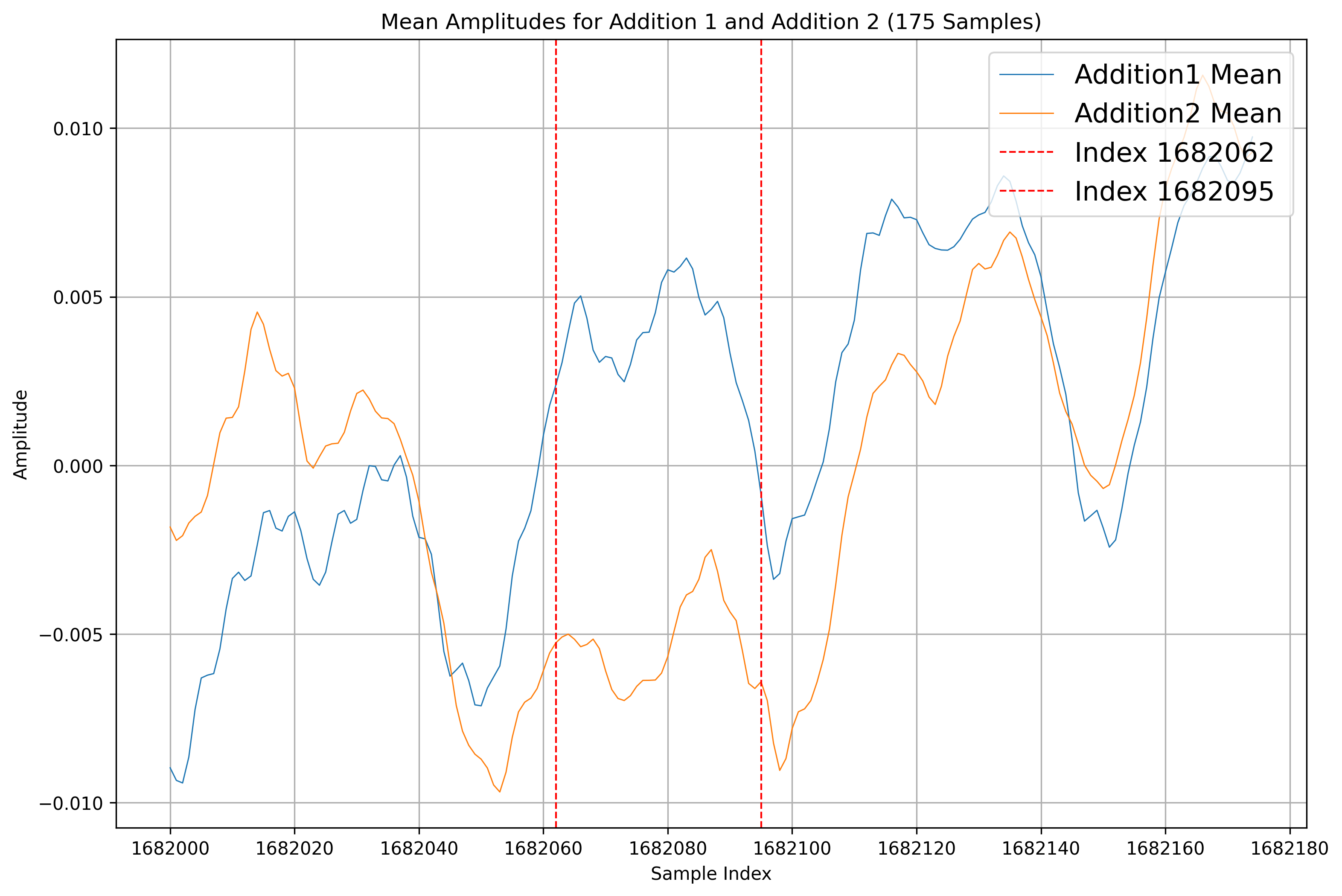} 
    \end{subfigure}
    \hspace{0.05\textwidth}
    \begin{subfigure}[b]{0.45\textwidth}
        \caption*{\textbf{(b):} Showing No Distinguishability.}
        \includegraphics[width=\linewidth]{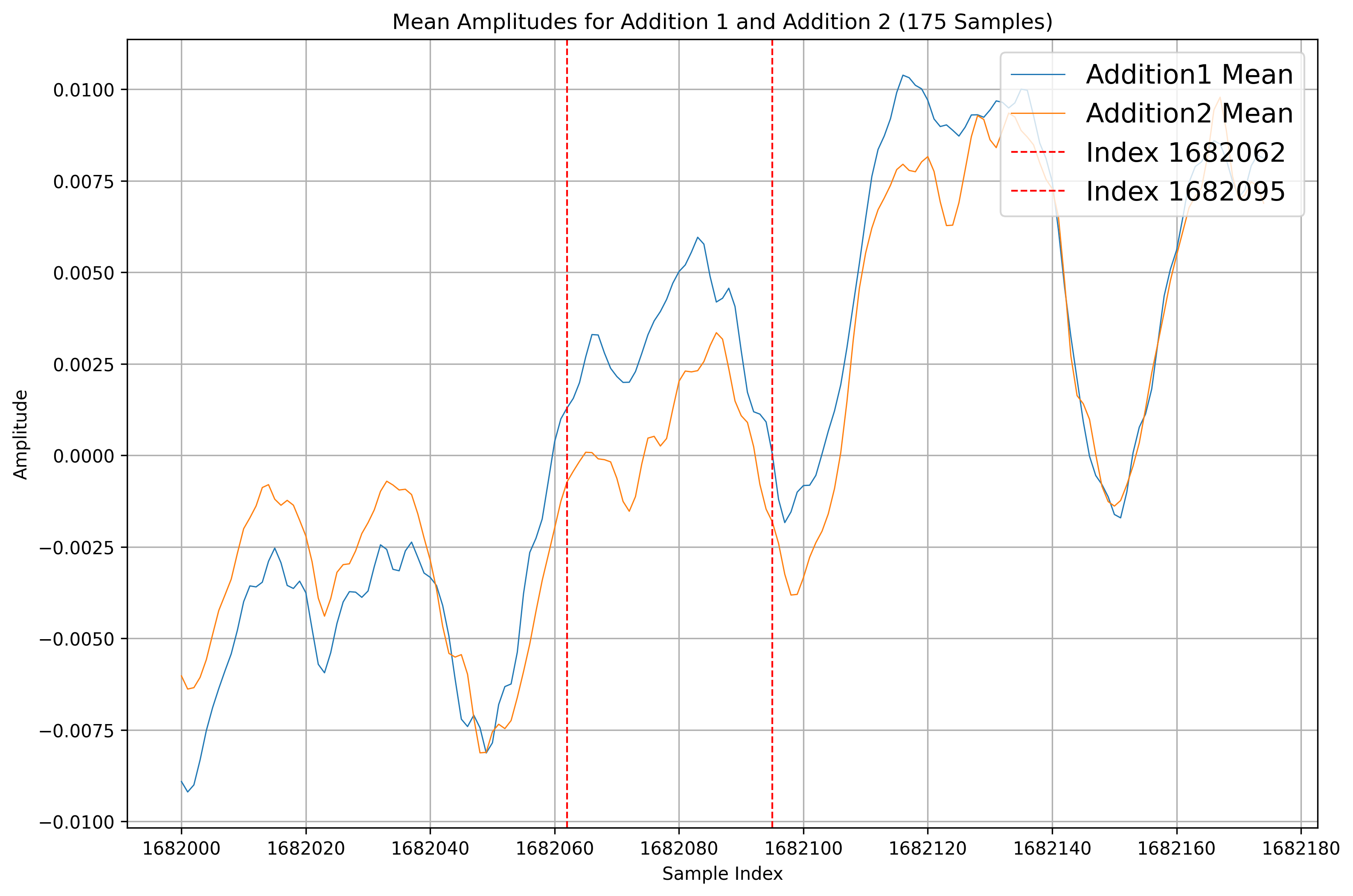}  
    \end{subfigure}

    \caption{\textbf{Left:} Mean amplitude plot of the Addition 1 and Addition 2 sub-traces from \hyperref[fig:distinguishability-A1-A2]{Figure~\ref*{fig:distinguishability-A1-A2}} (with only 5-clock cycle processes removed), showing a visibly large gap in the amplitude signals at sample index 1,682,065 to 1,682,093 due to the desynchronisation of the sub-traces in this region and showing the reason for the distinguishability. \\
    \textbf{Right:} Mean amplitude plot of the same sub-traces used in \hyperref[fig:no-distinguishability-A1-A2]{Figure~\ref*{fig:no-distinguishability-A1-A2}} (with both 5-clock and 1-clock cycle processes removed), showing how close the amplitude signals overlap in the same region and showing the reason for no distinguishability due to improved synchronisation.}
    
    \label{fig:mean-plot-A1-A2}
\end{figure}

We show again in \hyperref[fig:mean-plot-A1-A2]{Figure~\ref*{fig:mean-plot-A1-A2} (a)}, the mean plot of the same sub-traces from \hyperref[fig:distinguishability-A1-A2]{Figure~\ref*{fig:distinguishability-A1-A2}} while focusing on the same region where the distinguishability was observed. The mean plot here also shows the large gap in the amplitude signals between the Addition 1 and Addition 2 atomic blocks at index 1,682,065 to 1,682,093 and explains the reason for the distinguishability observed in this region. Similarly, \hyperref[fig:mean-plot-A1-A2]{Figure~\ref*{fig:mean-plot-A1-A2} (b)} shows the mean plot from the same sub-traces used in \hyperref[fig:no-distinguishability-A1-A2]{Figure~\ref*{fig:no-distinguishability-A1-A2}} where both the 5-clock and 1-clock cycle processes were removed. In this case, the mean amplitude signals from Addition 1 and Addition 2 closely overlap within the same region, which explains why there was no distinguishability observed in this region. This again confirms that removing both additionally inserted 5-clock and 1-clock cycle processes helped synchronise the sub-traces more effectively.

We did not identify any obvious gaps in the amplitude ranges after simultaneously comparing the results from the max-min sub-traces of all three atomic blocks. They all seemed to be exactly overlapping. We even set the threshold to \(0.000\) to ensure that we could detect the smallest minor gaps in the amplitude ranges. Even with this change, no gaps were seen between the different operations.

Based on these results, it seems possible to tell the difference between any two given atomic blocks but unlikely when comparing all three different atomic blocks together using their \ac{EM} traces. In \hyperref[tab:num-desync-samples]{Table~\ref*{tab:num-desync-samples}}, we show the number of distinguishable samples we identified while using the threshold 0.003 V and comparing the max-min sub-traces from the different atomic blocks.

\begin{table}[H]
\renewcommand{\arraystretch}{1.2} 
\centering
\begin{tabular}{|p{6.5cm}|p{7cm}|}
\hline
\rowcolor{orange!20} 
    \textbf{Pairwise Atomic Blocks} & \textbf{Number of Distinguishable Samples With Threshold \(> 0.003\, V\)} \\ \hline
    Doubling vs   Addition 1 & 0 \\ \hline
    Doubling vs   Addition 2 & 46 \\ \hline
    Addition 1   vs Addition 2 & 31 \\ \hline
    Doubling vs   Addition 1 vs Addition 2 & 0 \\ \hline
\end{tabular}
\caption{Number of distinguishable samples identified while using threshold 0.003 V and comparing samples from the different atomic blocks.}
\label{tab:num-desync-samples}
\end{table}

While we were unable to identify the exact processes that caused the 1-clock cycle delays, removing them clearly had an impact on the overlap of the amplitude ranges. Also, the fact that these gaps only appeared because this sub-trace set included the additional 1-clock cycle processes suggests that these processes were likely contributing to the distinguishability observed in this analysis. However, further investigation is needed to determine the precise cause of these delays.
\section{Comparison of Experimental Results With Related Works.}

In this part of the work, we will compare the findings of this thesis work to that of similar research works. As we mentioned earlier, there have been no prior experimental investigations into the distinguishaibility of the atomic blocks in Giraud and Verneuil's atomic pattern. As such, the findings of this study can only be compared with other works that have investigated similar atomic patterns. Among these are the work of Li et al. \cite{li_distinguishability_2024} \cite{li_practical_2024}, who investigated Longa's \cite{longa_accelerating_2008} atomic pattern, and Sigourou et al. \cite{sigourou_successful_2023}, who focused on investigating Rondepierre's \cite{rondepierre_revisiting_2014} atomic pattern. \hyperref[tab:summary-comparison-results]{Table~\ref*{tab:summary-comparison-results}} summarises the findings of this work alongside the results of Sigourou et al. and Li et al.

{\footnotesize
\renewcommand{\arraystretch}{1.2}
\begin{longtable}{|>{\raggedright\arraybackslash}m{3.5cm}|>{\raggedright\arraybackslash}m{3.5cm}|>{\raggedright\arraybackslash}m{3.5cm}|>{\raggedright\arraybackslash}m{3.5cm}|}

\hline
\rowcolor{orange!20}
\textbf{Parameter} & \textbf{This work} & \textbf{Sigourou et al. \cite{sigourou_successful_2023}} & \textbf{Li et al. \cite{li_distinguishability_2024}} \\
\hline
\endfirsthead

\hline
\rowcolor{orange!20}
\textbf{Parameter} & \textbf{This work} & \textbf{Sigourou et al. \cite{sigourou_successful_2023}} & \textbf{Li et al. \cite{li_distinguishability_2024}} \\
\hline
\endhead

The implemented \( k\mathbf{P} \) algorithm & Right-to-left double-and-add & Left-to-right double-and-add & Left-to-right double-and-add \\
\hline
Implemented Atomic Pattern & Giraud and Verneuil \cite{giraud_atomicity_2010} & Rondepierre \cite{rondepierre_revisiting_2014} & Longa \cite{longa_accelerating_2008} \\
\hline
Cryptographic Library & FLECC\_IN\_C & FLECC\_IN\_C & FLECC\_IN\_C \\
\hline
Board for Experiment & LAUNCHXL-F28379D & LAUNCHXL-F280025C & LAUNCHXL-F28379D \\
\hline
Scalar bit length & 10 & 22 & 22 \\
\hline
Sampling Rate of Oscilloscope & 5 GS/s & 500 MS/s & 1 GS/s \\
\hline
Number of samples per clock cycle & 50 & 5 & 10 \\
\hline
Number of samples per point operation & \ac{PD}: \textasciitilde13,941,300 & \textasciitilde2 million & \ac{PD}: \textasciitilde3,200,000 \\
& \ac{PA}: \textasciitilde27,881,950 & & \ac{PA}: \textasciitilde4,800,000 \\
\hline
No. of samples for full \( k\mathbf{P} \) & 500 million & 100 million & 200 million \\
\hline
\ac{SCA} Type & Horizontal & Horizontal & Horizontal \\
\hline
\ac{SCA} Method & Automated Simple \ac{SCA} & Simple \ac{EM} Analysis & Automated Simple \ac{SCA} \\
\hline
Major Findings & - Some distinguishable patterns between atomic blocks after removing additionally inserted 5-clock cycle processes. & - \ac{SCA} leakage in 1\textsuperscript{st} and 4\textsuperscript{th} OPs & - Negative simple \ac{SCA} results \\
& - When additionally inserted 1-clock cycle processes were also removed, no distinguishable gaps were observed. & - Desynchronisation found & - Desynchronisation found \\
& - No distinguishability from analysis of all 3 atomic blocks simultaneously. & - FLECC is not constant-time & - FLECC is constant-time \\
& - FLECC is constant-time & & \\
\hline

\caption{Summary of the comparison of results between our work, the work by Sigourou et al. \cite{sigourou_successful_2023} and the work by Li et al. \cite{li_distinguishability_2024}} 
\label{tab:summary-comparison-results}
\end{longtable}
}

All three works used the FLECC cryptographic library to implement the atomic pattern \( k\mathbf{P} \) algorithm. While Sigourou et al. \cite{sigourou_successful_2023} and Li et al. \cite{li_distinguishability_2024}, \cite{li_practical_2024} implemented the Double-and-add Left-to-Right \( k\mathbf{P} \) algorithm in \hyperref[sec:left-to-right-SMA]{Section~\ref*{sec:left-to-right-SMA}}, this work implemented the Double-and-add Right-to-Left \( k\mathbf{P} \) algorithm described in \hyperref[sec:right-to-left-SMA]{Section~\ref*{sec:right-to-left-SMA}}.

All works inserted \acp{NOP} in the implementation. These helped create distinguishable gaps in the trace, aiding in identifying and separating the different atomic blocks.

Evaluating the execution time of the constant-time operations in the FLECC library was an important part of all three works. Sigourou et al. \cite{sigourou_successful_2023} reported that the FLECC library’s operations were not truly constant-time, citing inconsistencies in execution times. However, in this thesis, the functions were evaluated to confirm their constant-time behaviour. We used breakpoints to repeatedly measure the execution times of the first field multiplication, the first field addition and the field subtraction in the Doubling atomic block and we obtained identical results consistently. This confirmed that the functions in the FLECC library indeed execute in constant time, a finding consistent with the results from Li et al. \cite{li_distinguishability_2024}, \cite{li_practical_2024}.

Similar to the findings by Sigourou et al. and Li et al., this work identified some desynchronisation from already synchronised sub-traces. Specifically, a desynchronisation of approximately 5-clock cycles was observed in both this work and that of Li et al. In our analysis, we identified and removed all the 5-clock cycle processes in our focus region. Additionally, we identified another 1-clock cycle process which, when removed, further improved the synchronisation.

All three works also performed a simple \ac{EM} \ac{SCA} attack. Sigourou et al. \cite{sigourou_successful_2023} were able to find parts of the \( k\mathbf{P} \) trace that allowed them to clearly distinguish between \ac{PA} and \ac{PD} operations. Li et al. \cite{li_distinguishability_2024}, \cite{li_practical_2024}, on the other hand, could not show a high degree of distinguishability because of the small number of samples per clock cycle they obtained in their measurements. This limitation meant that they could neither confirm nor rule out the vulnerability of their atomic pattern implementation to horizontal address-bit \ac{SCA}. However, their implementation was still resistant to their applied simple \ac{SCA} techniques. In this work, we did not identify any distinguishable patterns in the atomic blocks when we removed both additionally inserted 5-clock and 1-clock cycle processes from the sub-traces. This indicates that the Giraud-Verneuil atomic pattern is indistinguishable and resistant to our applied horizontal \ac{SCA} techniques in this study.
        \chapter{Conclusion and Future Works}
\label{chap:conclusion}

In this thesis work, we investigated the distinguishability of the atomic blocks in Giraud and Verneuil’s atomic pattern \cite{giraud_atomicity_2010} when implemented using the right-to-left double-and-add \( k\mathbf{P} \) algorithm. Using horizontal \ac{EM} \ac{SCA}, we investigated whether the execution of the atomic blocks that have the same sequence of field operations could leak any side-channel information. Giraud and Verneuil’s atomic pattern was implemented, adapted, and experimentally evaluated on a microcontroller board, after which the measured \ac{EM} traces were analysed.

The main goal of this work was to find out if the \ac{EM} emanations from the execution of the \( k\mathbf{P} \) operations showed any distinguishable characteristics in the atomic blocks that could be used to find the secret scalar used in scalar multiplication. As a first step, we reviewed all the literature that mentioned or cited Giraud and Verneuil's atomic pattern. Out of the 58 papers found, none had experimentally evaluated Giraud and Verneuil's proposed countermeasure. To address this gap, we implemented the Giraud-Verneuil atomic pattern using the right-to-left scalar multiplication algorithm with EC P-256, a curve recommended by \ac{NIST}. The implementation was executed on the F28379D LaunchPad \ac{MCU} using the FLECC\_IN\_C cryptographic library.  The \( k\mathbf{P} \) was executed using a 10-bit scalar key (i.e. \(k=1111111111\)). This made it easier to measure and evaluate the results. We measured once the \ac{EM} emanation of the execution of a full \( k\mathbf{P} \) operation that resulted in a trace with 500 million samples at a 5 GS/s sampling rate. To evaluate the resistance of the implemented algorithm against simple \ac{SCA}, we conducted a horizontal \ac{EM} \ac{SCA} attack.

Our investigations were based on an automated simple \ac{SCA} technique described in \cite{kabin_horizontal_2023}, where the goal was to identify regions within the combined sub-traces corresponding to each atomic block one could identify distinguishable gaps between the maximum and minimum amplitude voltages of the different atomic blocks. We analysed two distinct sets of sub-traces:

\begin{itemize}
    \item Sub-trace set 1: Consisted of 30 sub-traces (from the processing of all 10 key bits) from the atomic blocks, where we identified and removed both 5-clock and 1-clock cycle processes from the sub-traces within our focus region.
    
    \item Sub-trace set 2: Also consisted of 30 sub-traces (from the processing of all 10 key bits) from the atomic blocks, where we identified and removed only the 5-clock cycles from the sub-traces within our focus region.
\end{itemize}

For each sub-trace set, we performed our analysis on pairs of atomic blocks (Doubling vs Addition 1, Doubling vs Addition 2, and Addition 1 vs Addition 2) as well as between all three atomic blocks simultaneously. Our investigations revealed several important findings. 

For sub-trace set 1, where we removed both 5-clock and 1-clock cycle processes, there were no distinguishable gaps in their amplitude signals when analysing either pairs of atomic blocks or all three atomic blocks simultaneously under our given conditions. This indicated that after removing both additionally inserted 5-clock and 1-clock cycle processes from the sub-traces, there was an improved synchronisation in the amplitude ranges for the Doubling, Addition 1, and Addition 2 atomic blocks, which made them indistinguishable.

For sub-trace set 2, where we removed only 5-clock cycle processes, we only analysed this set of sub-traces to demonstrate the effect of the desynchronisation on the analysis results. In our investigations, we found multiple instances of distinguishable gaps in the amplitude signals when comparing any two atomic blocks, for example, between the maximum and minimum sub-traces of Doubling and Addition 2 as shown in \hyperref[fig:distinguishability-D-A2]{Figure~\ref*{fig:distinguishability-D-A2}} and \hyperref[fig:distinguishability-A1-A2]{Figure~\ref*{fig:distinguishability-A1-A2}}. This meant that the additionally inserted clock cycle processes significantly impacted the distinguishability of the atomic blocks. If these processes are not correctly identified and removed, the results of the analysis may be incorrect. Therefore, further research is needed to determine which specific processes are responsible for introducing these additionally inserted clock cycle delays.

Our findings from this work indicate that although the Giraud-Verneuil atomic pattern has been successfully implemented and includes the correct implementation of dummy operations, there is still a concern regarding the subtle distinguishability in the atomic blocks, especially when there is the presence of the additional clock processes in their \ac{EM} trace. Future research work could focus on previous assumptions, such as those from Sigourou et al. \cite{sigourou_successful_2023}, which propose that this distinguishability in atomic patterns could be a result of how different registers are addressed in memory and the hamming distances between them. This investigation could help clarify the exact processes that caused the distinguishability in the atomic pattern.


	\cleardoublepage{}
	\addtocontents{toc}{\protect\vspace*{\baselineskip}}
	\cleardoublepage
	\phantomsection
	
	\printbibliography

    \let\origdoublepage\cleardoublepage
    \renewcommand{\cleardoublepage}{\clearpage}

    \appendix
    \crefname{chapter}{Appendix}{Appendices}  
    \Crefname{chapter}{Appendix}{Appendices}  
    
    \cleardoublepage{}
    \phantomsection

    \begin{shiftedchapter}
    \refstepcounter{chapter}                                
    \chapter*{Appendix \thechapter}                         
    \label{appendix:A}
    \addcontentsline{toc}{chapter}{Appendix \thechapter}    
    \markboth{Appendix \thechapter}{Appendix \thechapter}   
    \label{appendixA:LiteratureReview}

\section{Detailed Analysis of State-of-The-Art Literature Review}
{\scriptsize
\begin{longtable}{|>{\centering\arraybackslash}p{1.5cm}|>{\raggedright\arraybackslash}m{2cm}|>{\raggedright\arraybackslash}m{2cm}|>{\centering\arraybackslash}m{1.5cm}|>{\centering\arraybackslash}m{2cm}|>{\raggedright\arraybackslash}m{2cm}|>{\centering\arraybackslash}m{2cm}|>{\centering\arraybackslash}m{2cm}|}
\caption{Analysis of Literature Review}\\
\hline
\textbf{} & \textbf{} & \textbf{} & \multicolumn{2}{|>{\raggedright\arraybackslash}m{4cm}|}{\textbf{Key-Dependent Addressing as a vulnerability}} & \textbf{} & \textbf{} & \textbf{} \\ \hline
\textbf{Paper} & \textbf{Main Idea and results} & \textbf{Context of Reference to Giraud} & {\textbf{Mentioned}} & \textbf{The indistinguishability of registers is mentioned} & \textbf{Relation to Giraud} & \textbf{Is a detailed review of paper required?} & \textbf{Relevance to Thesis} \\ \hline
\endfirsthead
\caption[]{Analysis of Literature Review (continued)} \\
\hline
\textbf{} & \textbf{} & \textbf{} & \multicolumn{2}{|>{\raggedright\arraybackslash}m{4cm}|}{\textbf{Key-Dependent Addressing as a vulnerability}} & \textbf{} & \textbf{} & \textbf{} \\ \hline
\textbf{Paper} & \textbf{Main Idea and results} & \textbf{Context of Reference to Giraud} & {\textbf{Mentioned}} & \textbf{The indistinguishability of registers is mentioned} & \textbf{Relation to Giraud} & \textbf{Is a detailed review of paper required?} & \textbf{Relevance to Thesis} \\ \hline
\endhead

\rowcolor{midblue_table}
\cite{abdulrahman_new_2013} & Introduces a novel method for elliptic curve scalar multiplication, processing three bits of the scalar consistently with five arithmetic operations (3 DBLs and 2 ADDs) for all eight bit combinations without employing dummy operations or pre-computed point lookup tables. & Giraud is referenced in the context of smart cards, focusing on the experimental cost ratio between modular addition and modulo multiplication, which is considered to optimize elliptic curve operations. & {No} & No & - & No & - \\ \hline

\rowcolor{midblue_table}
\cite{baldwin_co-ecc_2012} & Analyzes co-Z-based versions of the Montgomery ladder and Joye’s double-add algorithm, noting their inherent resistance to SPA attacks and comparing their performance across software, hardware, and co-design implementations. & References Giraud concerning the algorithmic cost of field operations, emphasizing that field addition costs, though often deemed negligible, can significantly impact hardware devices, particularly when combined with other small operations. & {No} & No & - & No & - \\ \hline

\rowcolor{midblue_table}
\cite{baldwin_hardware_2013} & The paper explores flexible hardware architectures for cryptographic systems, analyzing dedicated accelerators' performance in speed, area, power, and energy. It specifically discusses implementations for EC point scalar multiplication and hash functions on FPGAs, along with a hardware-software co-design architecture supporting various cryptographic algorithms. & Giraud is referenced in the context of discussing the significance of field addition and subtraction costs in hardware devices, particularly in comparison to multiplications or inversions. The paper highlights the importance of considering these costs, especially when combined with other operations, despite their general neglect in one-to-one conversions. & {No} & No & - & No & - \\ \hline

\rowcolor{midblue_table}
\cite{duquesne_choosing_2018} & Explicitly describing how to generate optimal parameters for pairing-based cryptography. It covers aspects such as pairing algorithms, tower field construction, Fp arithmetic, and system coordinates. & Giraud is referenced in the context of discussing available choices for extension field and elliptic curve arithmetic. Author discusses the impact of Fp addition costs on these choices & {No} & No & - & No & - \\ \hline

\rowcolor{midblue_table}
\cite{guerrini_randomized_2017} & The paper introduces a uniformly randomized scalar multiplication algorithm using mixed-radix representation and exact n-covers, which enhances security against side-channel attacks while maintaining efficiency. This algorithm is proposed as an efficient alternative to Coron’s scalar randomization method, especially advantageous for curves defined modulo primes close to powers of two. & Giraud’s reference contributes to the discussion on potent attacks against ECC implementations, mainly focusing on atomic schemes and unified formulae, which are commonly regarded as effective countermeasures against SPA attacks. & {No} & No & - & No & - \\ \hline

\rowcolor{midblue_table}
\cite{hasan_abdulrahman_efficient_2013} & The paper presents an optimized, side-channel-attack-resistant implementation of the NIST Curve P-256, offering 128-bit security, and examines trade-offs between operation time and storage for different methods. & Giraud is referenced in the context of providing the experimental ratio of the cost of modular addition to modular multiplication (A/M) on smart cards, which is used to analyze the cost of EC operations. Giraud is again referenced in the context of presenting different atomic block structures for EC point arithmetic operations. & {No} & No & - & No & - \\ \hline

\rowcolor{midblue_table}
\cite{nitaj_memory-constrained_2011} & Presents new formulas for efficient point addition and doubling using the Montgomery ladder, reducing memory usage in low-resource environments by utilizing a common projective Z-coordinate representation while performing all computations. & Referenced in the context of evaluating the complexity of differential addition-and-doubling formulas for elliptic curves. Authors discuss how the complexity ratio of field-arithmetic operations can vary based on implementation and usage context, and it considers both squaring and multiplication operations. & {No} & No & - & No & - \\ \hline

\rowcolor{midblue_table}
\cite{kim_speeding_2017} & The paper introduces Montgomery scalar multiplication algorithms for short Weierstrass curves, aiming to improve efficiency and security in elliptic curve cryptography. These algorithms require only 12 field multiplications plus 12 to 20 field additions per scalar bit using 8 to 10 field registers, outperforming the binary NAF method on average. & Giraud is referenced in the context of proposing methods to improve atomicity in ECSM algorithms, focusing on doubling and mixed addition operations, with results indicating the significance of addition operations in ECSM algorithms on smart cards with cryptocoprocessors. & {No} & No & - & No & - \\ \hline

\rowcolor{midblue_table}
\cite{al-somani_method_2014} & The paper introduces a method that enhances the efficiency of elliptic curve scalar multiplication by replacing pre-computation overhead with parallelizable post-computations, offering a significant improvement in performance for elliptic curve cryptography. & - & {No} & No & - & No & - \\ \hline

\rowcolor{midblue_table}
\cite{zulberti_script-based_2022} & Introduces a framework automates verification in hardware and software co-design, focusing on cycle-true simulations, netlist analysis, and design space exploration, thereby improving team collaboration and speeding up evaluation for Elliptic-Curve-Cryptography accelerators & - & {No} & No & - & No & - \\ \hline

\rowcolor{orange_table}
\cite{sako_side-channel_2016} & A study of SCA resistance on ECDSA implementation in the Android cryptographic library shows that the recovery of secret keys is efficiently possible on smartphones using electromagnetic SCA and lattice reduction techniques. & As an example of an SCA atomicity countermeasure & {No} & No & - & No & - \\ \hline

\rowcolor{orange_table}
\cite{danger_synthesis_2013} & Provides a comprehensive overview of various side-channel attacks targeting elliptic curve cryptography in embedded systems, particularly smart-cards. It evaluates the effectiveness of different countermeasures against these attacks, considering their cost in terms of time and space & Referenced in the context of SCA atomicity countermeasures. Giraud and Verneuil enhanced this countermeasure, making it more effective against SCA & {No} & No & Not directly relate to Giraud’s specific work & No & - \\ \hline

\rowcolor{orange_table}
\cite{danger_dynamic_2013} & The paper explores vulnerabilities in ECC due to Zero Power Analysis (ZPA) attacks and proposes updating elliptic curve formulas dynamically to mitigate these risks. ZPA exploits special points that yield zero values during ECC point operations. & Giraud is referenced in the context of performance analysis, particularly related to the cost ratios of different field operations (multiplication, addition, and subtraction) in elliptic curve computations. & {No} & No & - & No & - \\ \hline

\rowcolor{orange_table}
\cite{ryan_improving_2016} & The authors improve an attack introduced by Bauer et al. The improved method allows for comparing multiple multiplications in elliptic curve implementations, leading to a higher success rate. Experimental results from real targets validate the soundness of their approach & Referenced as a side-channel atomicity countermeasure on which the Big Mac attack attack target & {No} & No & - & No & - \\ \hline

\rowcolor{orange_table}
\cite{el_mrabet_contributions_2017} & The paper delves into pairing-based cryptography, discussing algorithmic complexity, efficient evaluation, and security aspects of pairings. & Giraud is referenced in the context of discussing the cost of addition operations in extension field arithmetic, particularly highlighting their significance in real-world implementations, especially at the 128-bit security level. & {No} & No & - & No & - \\ \hline

\rowcolor{orange_table}
\cite{meier_side-channel_2014} & Investigates new attack paths for side-channel analysis on elliptic curves embedded scalar multiplications, which are protected by regular algorithms and scalar blinding. It presents vertical and horizontal attacks that can recover the entire secret scalar, exploiting the sparsity of standardized curves’ moduli for efficiency & Giraud is referenced in the context of “Atomic block” countermeasures for scalar multiplication on embedded devices. & {No} & No & - & No & - \\ \hline

\rowcolor{orange_table}
\cite{lemke-rust_single-trace_2017} & Authors show that scalar multiplication algorithms with precomputations are vulnerable to attacks correlating measurements with precomputed values and evidence from simulations and 8-bit AVR experiments. They also demonstrate that attacks can succeed without knowing the precomputed values by using clustering techniques. & Referenced in the context of balancing point addition and doubling formulas to implement atomicity in scalar multiplications. It also focuses on extension field arithmetic and the relative cost of Fp operations. & {No} & No & - & No & - \\ \hline

\rowcolor{orange_table}
\cite{jin_enhancing_2023} & The paper introduces a method to enhance Deep-Learning based Side-Channel Analysis (DL-SCA) by simultaneously training on multiple bytes, significantly improving robust profiling and increasing the success rate of recovering a secret AES key by 250\% with a limited number of traces. & When discussing countermeasures employed in modern embedded cryptographic devices for asymmetric cryptographic primitives like RSA and ECC, highlighting techniques such as Montgomery ladder, double-and-add always, and randomization to resist side-channel attacks. & {No} & No & - & No & - \\ \hline

\rowcolor{orange_table}
\cite{lee_single-trace_2022} & Introduces a new non-profiling attack method for asymmetric cryptosystems, using a single trace and short attack time to recover a full private key, overcoming past limitations. Leveraging one-shot learning with a convolutional Siamese network, it achieves up to 100\% accuracy in leaking private keys from protected public-key cryptosystems using just one trace in a non-profiled setting. & Giraud is cited in relation to "atomicity-based implementations" for safeguarding public-key cryptosystems against non-profiling attacks in embedded devices. & {No} & No & - & No & - \\ \hline

\rowcolor{orange_table}
\cite{loai_review_2017} & The paper reviews recent side-channel attacks on ECC, RSA, and AES cryptosystems and discusses effective countermeasures to protect these systems from cyber threats. & Giraud is referenced in the context of side-channel atomicity as a countermeasure against SPA (Side-Channel Attacks). & {No} & No & The paper reviews side channel attacks and countermeasures on ECC, specifically discussing atomicity improvements for elliptic curve scalar multiplication and referencing Giraud's related work. & No & - \\ \hline

\rowcolor{orange_table}
\cite{murdica_physical_2014} & The paper investigates physical attacks on ECC, discusses various attacks and countermeasures, and evaluates their effectiveness, particularly in the context of smart cards. It highlights ECC's advantages over RSA in terms of speed and memory usage while addressing the vulnerabilities and proposing solutions for physically accessible devices. & Giraud is referenced in the context of improving the Side-Channel Atomicity countermeasure for elliptic curve operations, specifically by optimizing the atomic patterns for point doubling and addition in modified coordinates. & {No} & No & - & No & - \\ \hline

\rowcolor{orange_table}
\cite{murdica_securite_2014} & The paper investigates physical attacks on ECC, discusses various attacks and countermeasures, and evaluates their effectiveness, particularly in the context of smart cards. It highlights ECC's advantages over RSA in terms of speed and memory usage while addressing the vulnerabilities and proposing solutions for physically accessible devices. & Giraud is referenced in the context of improving the Side-Channel Atomicity countermeasure for elliptic curve operations, specifically by optimizing the atomic patterns for point doubling and addition in modified coordinates. & {No} & No & - & No & - \\ \hline

\rowcolor{orange_table}
\cite{russon_exploiting_2020} & The paper highlights the vulnerability of ECC implementations to C safe-error fault attacks, which can reveal secret data by exposing dummy operations and through fault injection. This risk is demonstrated on libraries like OpenSSL and its forks. They emphasize the need to protect cryptographic libraries, particularly in embedded devices or environments prone to physical access by malicious actors. & Giraud is referenced in the context of countermeasures that introduce dummy operations to mask the difference between a point doubling and a point addition in ECC, which are vulnerable to C safe-error attacks. & {No} & No & - & No & - \\ \hline

\rowcolor{orange_table}
\cite{tawalbeh_towards_2016} & Authors review recent side-channel attacks on Elliptic Curve Cryptography (ECC) and discuss effective countermeasures to improve communication system security. They emphasize the need for combined countermeasures to address various attacks and advise users to follow best security practices when sharing sensitive information online. & Giraud is referenced in the context of side-channel atomicity as a countermeasure against SPA (Side-Channel Attacks). & {No} & No & The paper reviews side channel attacks and countermeasures on ECC, specifically discussing atomicity improvements for elliptic curve scalar multiplication and referencing Giraud's related work. & No & - \\ \hline

\rowcolor{gray_table}
\cite{abarzua_survey_2019} & Examines the balance between security and performance in countermeasures against passive side-channel attacks on Elliptic Curve Cryptography (ECC), particularly focusing on scalar multiplication algorithms without precomputation & Referenced in relation to the efficiency benefits of the Atomic Blocks methodology. Specifically, that squaring is less expensive than multiplication, which is relevant to Giraud’s work & {No} & No & - & No & - \\ \hline

\rowcolor{gray_table}
\cite{abarzua_survey_2021} & Examines the balance between security and performance in countermeasures against passive side-channel attacks on Elliptic Curve Cryptography (ECC), particularly focusing on scalar multiplication algorithms without precomputation & Referenced in relation to the efficiency benefits of the Atomic Blocks methodology. Specifically, that squaring is less expensive than multiplication, which is relevant to Giraud’s work & {No} & No & - & No & - \\ \hline

\rowcolor{gray_table}
\cite{chabrier_arithmetic_2013} & The paper focuses on developing robust arithmetic operators for elliptic curve cryptosystems (ECC) that protect the secret key in scalar multiplication through redundant number representations and behavior-based algorithms, enhancing resistance to side-channel attacks. & Giraud is referenced in the context of atomic pattern improvement methods in ECC. & {No} & No & - & No & - \\ \hline

\rowcolor{gray_table}
\cite{houssain_elliptic_2012} & The paper highlights the importance of ECC for resource-constrained devices like WSN, proposing novel architectures and algorithms to enhance security against SCA attacks, especially Power Analysis Attacks and DPA. It conducts a comparative evaluation of eight FPGA-synthesized architectures, showcasing their advantages in terms of cost complexity. & Giraud is referenced in the context of proposing a new atomic pattern to protect ECSM implementations against PAA, focusing on optimizing squarings and minimizing field additions and negations. & {No} & No & - & No & - \\ \hline

\rowcolor{gray_table}
\cite{lu_general_2013} & The paper introduces a general framework called "side-channel atomicity" to protect scalar multiplication algorithms from Simple Power Attack (SPA), offering security and flexibility across various coding methods. It is demonstrated to outperform existing methods in terms of efficiency and interoperability, with implementation results showing its superiority on AMD Athlon X2 245-based hardware. & Giraud is referenced in the context of proposing an improved side-channel atomicity technique to support the NAF coding method, addressing limitations in previous atomicity-based algorithms related to single scalar multiplication and handling access to the additive inverse of precomputed points and corresponding point addition. & {No} & No & - & No & - \\ \hline

\rowcolor{gray_table}
\cite{nascimento_comparison_2014} & The paper evaluates algorithmic countermeasures against side-channel attacks on elliptic curve cryptosystems over prime fields, focusing on the trade-offs between security and computational cost for variable-base scalar multiplication algorithms without precomputation. & Giraud is referenced in the context of discussing countermeasures against SSCA, specifically mentioning the need for multiplication and squaring operations to be indistinguishable, as utilized in atomic blocks and unified formulas. & {No} & No & - & No & - \\ \hline

\rowcolor{gray_table}
\cite{rivain_fast_2011} & Discussion on the regular implementation of cryptographic algorithms with different trade-offs, alongside introducing a new binary algorithm with comparable performance. & Cited when discussing the computational cost of field operations in scalar multiplication algorithms & {No} & No & - & No & - \\ \hline

\rowcolor{gray_table}
\cite{al-somani_system_2014} & The paper proposes a countermeasure method for securing scalar multiplication in ECCs against DPA and safe-error attacks, focusing on enhancing security for low-cost applications. & - & {No} & No & - & No & - \\ \hline

\rowcolor{yellow_table}
\cite{abarzua_method_2015} & The paper presents new sets of atomic blocks to enhance protection against SSCA and C-safe fault attacks, structured with specific field operations, and demonstrates unified Jacobian doubling with mixed Jacobian-affine addition, avoiding "dummy" operations for potential performance gains. & Giraud is referenced for presenting new atomic blocks for Jacobian Addition and Modified Jacobian Doubling, enhancing the efficiency of scalar multiplication in right-to-left algorithms. & {No} & No & Building upon Giraud’s work, the paper proposes more compact and efficient atomic blocks that protect against C-safe fault attacks, enhancing the safety of elliptic curve scalar multiplication. & No & - \\ \hline

\rowcolor{yellow_table}
\cite{abarzua_complete_2012} & The authors propose new atomic blocks to enhance safety against side-channel attacks and C-safe fault attacks during scalar multiplication for elliptic curves over prime fields, offering both compactness and enhanced protection for improved performance. & Giraud is referenced when discussing atomic block structures for scalar multiplication algorithms, mainly focusing on right-to-left scalar multiplication and avoiding the Doubling attack by Fouque and Valette. & {No} & No & - & No & - \\ \hline

\rowcolor{yellow_table}
\cite{bauer_horizontal_2015} & The authors reveal a new side-channel attack that compromises elliptic curve implementations using point/scalar randomization with atomicity, challenging previous security assumptions and underscoring the need to address overlooked vulnerabilities. & The paper details an attack on elliptic curve atomic implementations with input randomization, combining horizontal modus operandi and collision correlation analysis, specifically targeting state-of-the-art implementations like those by Giraud-Verneuil. & {No} & No & It presents an attack that could potentially invalidate the countermeasures Giraud proposed in his paper on improving atomicity in elliptic curve scalar multiplication. & No & - \\ \hline

\rowcolor{yellow_table}
\cite{das_improved_2016} & The paper discusses the Big Mac attack, which exploits operand sharing in RSA and ECC implementations despite conventional countermeasures. It proposes a method to prevent high-cost cache attacks (HCCA) on NIST curves in ECC algorithms by rearranging operands in field multiplications while maintaining atomicity. & Giraud is referenced in the context of proposing improved low-cost approaches to implementing side-channel atomicity for scalar multiplication algorithms. & {No} & No & - & No & - \\ \hline

\rowcolor{yellow_table}
\cite{ryan_safe-errors_2016} & The paper addresses the atomicity protection countermeasure for ECC against SPA, highlighting a new vulnerability to C Safe-Error attacks in ECDSA implementations. It proposes an effective solution involving dummy operations to detect faults, demonstrating that recovering just two bits of the scalar can compromise the secret key. & Giraud is referenced in the context of discussing a pattern for elliptic curve point addition and doubling operations, specifically highlighting the use of modified Jacobian coordinates and the application of Right-to-Left Double-and-Add algorithms. & {No} & No & - & No & - \\ \hline

\rowcolor{yellow_table}
\cite{francillon_revisiting_2014} & Introduces new formulas based on the atomicity principle for enhanced side-channel resistance, optimizing the cost of both doubling and addition for efficient double scalar multiplications using the Straus-Shamir trick. & Referenced in the context of improving the atomicity principle as a countermeasure against SSCA. & {No} & No & - & No & - \\ \hline

\rowcolor{lightblue_table}
\cite{kabin_horizontal_2023} & This study explores vulnerabilities in kP implementations based on the Montgomery ladder to horizontal side-channel attacks, successfully revealing secret values using statistical methods and exploiting key-dependent addressing. Countermeasures proposed include implementing field multipliers with hiding abilities and regular scheduling for block addressing to enhance resistance against side-channel attacks. & Giraud is referenced in the context of presenting an implementation of the atomicity principle for ECSM algorithms, particularly focusing on embedded devices, with specific details about the atomic patterns proposed. & Yes & No & - & No & - \\ \hline

\rowcolor{lightblue_table}
\cite{kabin_ec_2021} & kP algorithm based on atomic patterns proposed by Rondepierre was implemented for an ASIC in the 250 nm IHP technology Hardware. A simple SCA attack was successful by analysing a simulated power trace. The attack exploited key-dependent addressing of registers. & As an example of atomic pattern countermeasure & Yes & No & Giraud also proposes an atomic pattern which could result in a vulnerability due to address bit leakages & Yes & A similar atomicity principle by Rondepierre is implemented and examined. \\ \hline

\rowcolor{lightblue_table}
\cite{kabin_ec_2021-2} & EC point multiplication kP following the atomicity principle by Rondepierre. SCA resistance assessment through analysis of a single simulated power trace of kP execution & As an example of SCA atomicity countermeasures proposed to reduce the number of dummy operations in the Regularity countermeasure & {Yes} & No & - & No & - \\ \hline

\rowcolor{lightblue_table}
\cite{kabin_horizontal_2017} & Successful horizontal, i.e., single-trace, Address-Bit DPA attack against Montgomery kP implementation in an ECDSA implementation running on a spartan 6 FPGA Board. & As an example of an SCA atomicity countermeasure using only 8 field multiplications and 10 field additions per atom. & {Yes} & No & - & No & - \\ \hline

\rowcolor{lightblue_table}
\cite{kabin_randomized_2023} & Breaking well-known randomised addressing countermeasures by M. Izumi et al. and K. Itoh et al. with a single trace attack, shown only theoretically, No measurements/simulation. & Cited as an example of SCA atomicity countermeasure & Yes & No & - & No & - \\ \hline

\rowcolor{lightblue_table}
\cite{kabin_vulnerability_2023} & A simple SCA attack was successful by analysing a simulated power trace. The attack exploited key-dependent addressing of registers. & Cited as an example of atomic pattern countermeasure & Yes & No & - & Yes & A similar atomicity principle by Rondepierre is implemented and evaluated. \\ \hline

\rowcolor{lightblue_table}
\cite{sigourou_successful_2023} & Investigation of atomic patterns kP algorithm proposed by Rondepierre for elliptic curves over prime finite fields utilizing FLECC open-source cryptographic library. The scalar k processed during kP execution revealed through analysis of measured electromagnetic trace & Cited as an implementation focused on the atomicity principle for scalar multiplication in embedded systems & {Yes} & No & Giraud’s proposed countermeasure may also be susceptible to horizontal attacks due to the key-dependent addressing of registers. & Yes & Similar ECC Cryptographic library used in this thesis and interested in results. \\ \hline

\rowcolor{darkblue_table}
\cite{2013} & This paper, “A Study on Efficiency of Scalar Multiplication in Elliptic Curve Cryptography,” aims to improve the efficiency of scalar multiplication (kP) in elliptic curve cryptography. In particular, we propose a method to reduce the computational cost of scalar multiplication by using the Double-Base Number System (DBNS) representation and verify its effectiveness by comparing it with existing methods. & - & {-} & - & - & - & - \\ \hline

\rowcolor{darkblue_table}
\cite{lucas_support_2019} & This paper focuses on developing and evaluating software protections against both fault attacks (FA) and side-channel attacks (SCA) in the context of elliptic curve cryptography (ECC). The research proposes two main protections for scalar multiplication (SM): point verification (PV) and iteration counter (IC), which enhance security with minimal performance overhead. & - & {-} & - & - & - & - \\ \hline

\rowcolor{darkblue_table}
\cite{russon_probleme_2022} & The author focuses on the security of elliptic curve cryptography, particularly the difficulty of solving the discrete logarithm problem on elliptic curves. He examines various algorithms and implementation techniques, highlighting potential vulnerabilities and proposing new fault injection attacks. & - & {-} & - & - & - & - \\ \hline

\rowcolor{darkblue_table}
\cite{venelli_contribution_2011} & The author focuses on the study of side-channel attacks and their impact on the secure implementation of cryptographic algorithms. The author proposes improvements to mutual information analysis attacks and introduces a secure scalar multiplication algorithm for elliptic curve cryptography. & - & {-} & - & - & - & - \\ \hline

\end{longtable}
}

    \end{shiftedchapter}
    
    \begin{shiftedappendixB}
    \refstepcounter{chapter}                                
    \chapter*{Appendix \thechapter}                         
    \label{appendix:B}
    \addcontentsline{toc}{chapter}{Appendix \thechapter}    
    \markboth{Appendix \thechapter}{Appendix \thechapter}   

\section{Execution times of each field operation within each atomic block for all ten key bits.}
\renewcommand{\arraystretch}{1.8} 

{\scriptsize
\begin{longtable}[H]{|>{\raggedright\arraybackslash}m{1.3cm}|>{\raggedright\arraybackslash}m{1.6cm}|*{10}{>    {\raggedright\arraybackslash}m{0.85cm}|}>{\raggedright\arraybackslash}m{0.6cm}|>{\raggedright\arraybackslash}m{0.6cm}|> {\raggedright\arraybackslash}m{0.7cm}|}
\hline
\rowcolor{green!40}
\multicolumn{15}{|c|}{\textbf{Addition 1 Atomic Block}} \\ \hline
\rowcolor{orange!20}
\textbf{Operation Sequence} & \textbf{Field Operation} & \textbf{Key Bit Index 1} & \textbf{Key Bit Index 2} & \textbf{Key Bit Index 3} & \textbf{Key Bit Index 4}     & \textbf{Key Bit Index 5} & \textbf{Key Bit Index 6} & \textbf{Key Bit Index 7} & \textbf{Key Bit Index 8} & \textbf{Key Bit Index 9} &\textbf{Key Bit Index 10}  & \textbf{Max} & \textbf{Min} & \textbf{Range} \\ \hline
\endfirsthead
OP 1 & Squaring & 33158 & 33158 & 33153 & 33153 & 33153 & 33153 & 33158 & 33153 & 33158 & 33153 & 33158 & 33153 & 5 \\ \hline
OP 2 & Addition & 1355 & 1355 & 1360 & 1360 & 1355 & 1355 & 1360 & 1360 & 1360 & 1355 & 1360 & 1355 & 5 \\ \hline
OP 3 & Multiplication & 33158 & 33153 & 33153 & 33153 & 33153 & 33153 & 33158 & 33158 & 33153 & 33158 & 33158 & 33153 & 5 \\ \hline
OP 4 & Addition & 1355 & 1360 & 1360 & 1355 & 1355 & 1360 & 1360 & 1355 & 1360 & 1360 & 1360 & 1355 & 5 \\ \hline
OP 5 & Multiplication & 33153 & 33153 & 33153 & 33153 & 33153 & 33153 & 33158 & 33158 & 33158 & 33153 & 33158 & 33153 & 5 \\ \hline
OP 6 & Addition & 1355 & 1360 & 1360 & 1355 & 1355 & 1360 & 1360 & 1355 & 1360 & 1360 & 1360 & 1355 & 5 \\ \hline
OP 7 & Multiplication & 33152 & 33157 & 33157 & 33152 & 33152 & 33152 & 33152 & 33152 & 33157 & 33152 & 33157 & 33152 & 5 \\ \hline
OP 8 & Addition & 1355 & 1360 & 1360 & 1360 & 1355 & 1355 & 1360 & 1355 & 1360 & 1355 & 1360 & 1355 & 5 \\ \hline
OP 9 & Addition & 1355 & 1355 & 1355 & 1355 & 1355 & 1360 & 1360 & 1360 & 1360 & 1360 & 1360 & 1355 & 5 \\ \hline
OP 10 & Squaring & 33153 & 33158 & 33153 & 33158 & 33158 & 33158 & 33158 & 33158 & 33158 & 33153 & 33158 & 33153 & 5 \\ \hline
OP 11 & Multiplication & 33152 & 33152 & 33152 & 33152 & 33152 & 33152 & 33152 & 33152 & 33152 & 33152 & 33152 & 33152 & 0 \\ \hline
OP 12 & Addition & 1355 & 1355 & 1355 & 1360 & 1355 & 1355 & 1355 & 1355 & 1355 & 1360 & 1360 & 1355 & 5 \\ \hline
OP 13 & Subtraction & 1351 & 1356 & 1356 & 1351 & 1356 & 1356 & 1356 & 1356 & 1356 & 1351 & 1356 & 1351 & 5 \\ \hline
OP 14 & Multiplication & 33152 & 33152 & 33157 & 33157 & 33152 & 33152 & 33152 & 33152 & 33157 & 33152 & 33157 & 33152 & 5 \\ \hline
OP 15 & Subtraction & 1351 & 1356 & 1351 & 1356 & 1356 & 1351 & 1351 & 1356 & 1356 & 1351 & 1356 & 1351 & 5 \\ \hline
OP 16 & Subtraction & 1356 & 1356 & 1356 & 1351 & 1351 & 1356 & 1356 & 1356 & 1356 & 1351 & 1356 & 1351 & 5 \\ \hline
OP 17 & Multiplication & 33153 & 33153 & 33153 & 33153 & 33153 & 33153 & 33153 & 33153 & 33153 & 33153 & 33153 & 33153 & 0 \\ \hline
OP 18 & Subtraction & 1356 & 1356 & 1361 & 1361 & 1361 & 1361 & 1356 & 1356 & 1361 & 1361 & 1361 & 1356 & 5 \\ \hline
\rowcolor{gray!10}
& \textbf{Total CC} & \textbf{278775} & \textbf{278805} & \textbf{278805} & \textbf{278795} & \textbf{278780} & \textbf{278795} & \textbf{278815} & \textbf{278800} & \textbf{278830} & \textbf{278790} & \textbf{-} &\textbf{-} & \textbf{-} \\ \hline

\multicolumn{15}{c}{} \\ 
\hline
\rowcolor{green!40}
\multicolumn{15}{|c|}{\textbf{Addition 2 Atomic Block}} \\ \hline
\rowcolor{orange!20}
\textbf{Operation Sequence} & \textbf{Field Operation} & \textbf{Key Bit Index 1} & \textbf{Key Bit Index 2} & \textbf{Key Bit Index 3} & \textbf{Key Bit Index 4} & \textbf{Key Bit Index 5} & \textbf{Key Bit Index 6} & \textbf{Key Bit Index 7} & \textbf{Key Bit Index 8} & \textbf{Key Bit Index 9} & \textbf{Key Bit Index 10} & \textbf{Max} & \textbf{Min} & \textbf{Range} \\ \hline
OP 1 & Squaring & 33151 & 33151 & 33151 & 33151 & 33151 & 33156 & 33151 & 33156 & 33156 & 33151 & 33156 & 33151 & 5 \\ \hline
OP 2 & Addition & 1355 & 1360 & 1360 & 1355 & 1360 & 1360 & 1360 & 1360 & 1360 & 1355 & 1360 & 1355 & 5 \\ \hline
OP 3 & Multiplication & 33152 & 33152 & 33152 & 33152 & 33152 & 33152 & 33152 & 33152 & 33152 & 33157 & 33157 & 33152 & 5 \\ \hline
OP 4 & Addition & 1355 & 1360 & 1360 & 1360 & 1355 & 1360 & 1355 & 1355 & 1355 & 1360 & 1360 & 1355 & 5 \\ \hline
OP 5 & Multiplication & 33152 & 33157 & 33152 & 33152 & 33152 & 33157 & 33152 & 33152 & 33157 & 33157 & 33157 & 33152 & 5 \\ \hline
OP 6 & Addition & 1355 & 1355 & 1355 & 1355 & 1360 & 1360 & 1355 & 1355 & 1360 & 1355 & 1360 & 1355 & 5 \\ \hline
OP 7 & Multiplication & 33153 & 33153 & 33153 & 33153 & 33153 & 33153 & 33158 & 33158 & 33158 & 33153 & 33158 & 33153 & 5 \\ \hline
OP 8 & Addition & 1355 & 1355 & 1355 & 1360 & 1360 & 1360 & 1360 & 1360 & 1360 & 1355 & 1360 & 1355 & 5 \\ \hline
OP 9 & Addition & 1355 & 1355 & 1360 & 1355 & 1360 & 1360 & 1360 & 1360 & 1360 & 1355 & 1360 & 1355 & 5 \\ \hline
OP 10 & Squaring & 33151 & 33151 & 33156 & 33151 & 33156 & 33156 & 33156 & 33156 & 33156 & 33151 & 33156 & 33151 & 5 \\ \hline
OP 11 & Multiplication & 33153 & 33158 & 33158 & 33153 & 33153 & 33153 & 33153 & 33153 & 33158 & 33153 & 33158 & 33153 & 5 \\ \hline
OP 12 & Addition & 1355 & 1360 & 1360 & 1360 & 1360 & 1360 & 1355 & 1360 & 1355 & 1360 & 1360 & 1355 & 5 \\ \hline
OP 13 & Subtraction & 1351 & 1356 & 1351 & 1351 & 1351 & 1356 & 1351 & 1351 & 1351 & 1356 & 1356 & 1351 & 5 \\ \hline
OP 14 & Multiplication & 33152 & 33152 & 33152 & 33152 & 33152 & 33152 & 33157 & 33152 & 33152 & 33157 & 33157 & 33152 & 5 \\ \hline
OP 15 & Subtraction & 1352 & 1352 & 1357 & 1357 & 1357 & 1352 & 1357 & 1357 & 1357 & 1352 & 1357 & 1352 & 5 \\ \hline
OP 16 & Subtraction & 1351 & 1351 & 1356 & 1356 & 1351 & 1351 & 1356 & 1356 & 1356 & 1351 & 1356 & 1351 & 5 \\ \hline
OP 17 & Multiplication & 33152 & 33152 & 33157 & 33152 & 33152 & 33152 & 33152 & 33157 & 33152 & 33152 & 33157 & 33152 & 5 \\ \hline
OP 18 & Subtraction & 1357 & 1357 & 1357 & 1357 & 1357 & 1362 & 1362 & 1357 & 1362 & 1357 & 1362 & 1357 & 5 \\ \hline
\rowcolor{gray!10}
& \textbf{Total CC} & \textbf{278757} & \textbf{278787} & \textbf{278802} & \textbf{278782} & \textbf{278792} & \textbf{278812} & \textbf{278802} & \textbf{278807} & \textbf{278817} & \textbf{278787} & \textbf{-} & \textbf{-} & \textbf{-} \\ \hline

\multicolumn{15}{c}{} \\ 
\hline
\rowcolor{blue!40}
\multicolumn{15}{|c|}{\textbf{Doubling Atomic Block}} \\ \hline
\rowcolor{orange!20}
\textbf{Operation Sequence} & \textbf{Field Operation} & \textbf{Key Bit Index 1} & \textbf{Key Bit Index 2} & \textbf{Key Bit Index 3} & \textbf{Key Bit Index 4} & \textbf{Key Bit Index 5} & \textbf{Key Bit Index 6} & \textbf{Key Bit Index 7} & \textbf{Key Bit Index 8} & \textbf{Key Bit Index 9} & \textbf{Key Bit Index 10} & \textbf{Max} & \textbf{Min} & \textbf{Range} \\ \hline
OP 1 & Squaring & 33153 & 33153 & 33153 & 33153 & 33153 & 33153 & 33158 & 33158 & 33158 & 33153 & 33158 & 33153 & 5 \\ \hline
OP 2 & Addition & 1355 & 1355 & 1360 & 1355 & 1360 & 1355 & 1360 & 1360 & 1360 & 1355 & 1360 & 1355 & 5 \\ \hline
OP 3 & Multiplication & 33153 & 33153 & 33153 & 33153 & 33158 & 33158 & 33153 & 33153 & 33158 & 33153 & 33158 & 33153 & 5 \\ \hline
OP 4 & Addition & 1360 & 1360 & 1355 & 1360 & 1355 & 1355 & 1360 & 1360 & 1360 & 1355 & 1360 & 1355 & 5 \\ \hline
OP 5 & Multiplication & 33152 & 33157 & 33157 & 33152 & 33157 & 33157 & 33152 & 33152 & 33152 & 33152 & 33157 & 33152 & 5 \\ \hline
OP 6 & Addition & 1360 & 1360 & 1355 & 1360 & 1355 & 1355 & 1360 & 1360 & 1355 & 1360 & 1360 & 1355 & 5 \\ \hline
OP 7 & Multiplication & 33157 & 33157 & 33152 & 33152 & 33152 & 33152 & 33152 & 33152 & 33152 & 33157 & 33157 & 33152 & 5 \\ \hline
OP 8 & Addition & 1354 & 1359 & 1354 & 1354 & 1354 & 1354 & 1354 & 1359 & 1359 & 1354 & 1359 & 1354 & 5 \\ \hline
OP 9 & Addition & 1360 & 1355 & 1360 & 1360 & 1360 & 1360 & 1355 & 1360 & 1360 & 1360 & 1360 & 1355 & 5 \\ \hline
OP 10 & Squaring & 33156 & 33151 & 33156 & 33151 & 33151 & 33151 & 33156 & 33151 & 33151 & 33151 & 33156 & 33151 & 5 \\ \hline
OP 11 & Multiplication & 33152 & 33152 & 33157 & 33152 & 33152 & 33152 & 33152 & 33152 & 33157 & 33157 & 33157 & 33152 & 5 \\ \hline
OP 12 & Addition & 1360 & 1355 & 1360 & 1360 & 1360 & 1360 & 1355 & 1360 & 1355 & 1360 & 1360 & 1355 & 5 \\ \hline
OP 13 & Subtraction & 1351 & 1356 & 1356 & 1351 & 1356 & 1351 & 1351 & 1356 & 1351 & 1351 & 1356 & 1351 & 5 \\ \hline
OP 14 & Multiplication & 33153 & 33153 & 33153 & 33153 & 33153 & 33153 & 33153 & 33158 & 33158 & 33158 & 33158 & 33153 & 5 \\ \hline
OP 15 & Subtraction & 1352 & 1357 & 1357 & 1352 & 1357 & 1352 & 1357 & 1357 & 1352 & 1357 & 1357 & 1352 & 5 \\ \hline
OP 16 & Subtraction & 1356 & 1351 & 1351 & 1351 & 1351 & 1356 & 1351 & 1356 & 1356 & 1351 & 1356 & 1351 & 5 \\ \hline
OP 17 & Multiplication & 33152 & 33152 & 33152 & 33152 & 33152 & 33152 & 33152 & 33157 & 33152 & 33152 & 33157 & 33152 & 5 \\ \hline
OP 18 & Subtraction & 1362 & 1362 & 1357 & 1362 & 1362 & 1362 & 1357 & 1362 & 1357 & 1357 & 1362 & 1357 & 5 \\ \hline
\rowcolor{gray!10}
& \textbf{Total CC} & \textbf{278798} & \textbf{278798} & \textbf{278798} & \textbf{278783} & \textbf{278798} & \textbf{278788} & \textbf{278788} & \textbf{278823} & \textbf{278803} & \textbf{278793} & \textbf{-} & \textbf{-} & \textbf{-} \\ \hline

\caption{Execution times for each field operation in each atomic block across all 10 key bits, showing maximum, minimum, and range values.}
\label{tab:full_kP_execution_times}
\end{longtable}
}

    \end{shiftedappendixB}
    
    \refstepcounter{chapter}                                
    \chapter*{Appendix \thechapter}                         
    \label{appendix:C}
    \addcontentsline{toc}{chapter}{Appendix \thechapter}    
    \markboth{Appendix \thechapter}{Appendix \thechapter}   
    \section{Domain Parameters for NIST Curve P-256 used in the work}
\label{appendixC:DomainParams}

\medskip

\( p = 2^{256} - 2^{224} + 2^{192} + 2^{96} - 1 \)

\medskip

\( h = 1 \)

\medskip

\( n = \text{0x FFFFFFFF 00000000 FFFFFFFF FFFFFFFF BCE6FAAD A7179E84 F3B9CAC2 FC632551} \)

\medskip

\text{Seed} \( = \text{0x C49D3608 86E70493 6A6678E1 139D26B7 819F7E90} \)

\medskip

\( a = -3 \)

\medskip

\( \text{  } = \text{0x FFFFFFFF 00000001 00000000 00000000 00000000 FFFFFFFF FFFFFFFF FFFFFFFC} \)

\medskip

\( b = \text{0x 5AC635D8 AAA3A93E7 B3EBBD55 769886BC 651D06B0 CC53B0F6 3BCE3C3E 27D2604B} \)

\medskip

\( x = \text{0x 6B17D1F2 E12C4247 F8BCE6E5 63A440F2 77037D81 2DEB33A0 F4A13945 D898C296} \)

\medskip

\( y = \text{0x 4FE342E2 FE1A7F9B 8EE7EB4A 7C0F9E16 2BCE3357 6B315ECE CBB64068 37BF51F5} \)

    \begin{shiftedappendixB}
    \refstepcounter{chapter}                                
    \chapter*{Appendix \thechapter}                         
    \label{appendix:D}
    \addcontentsline{toc}{chapter}{Appendix \thechapter}    
    \markboth{Appendix \thechapter}{Appendix \thechapter}   

\section{Start of the Desynchronisation Region Where 5-Clock and 1-Clock Cycle Processes Were Identified}
\renewcommand{\arraystretch}{1.8}

{\scriptsize
\begin{longtable}{|>{\centering\arraybackslash}m{1.5cm}|>{\raggedright\arraybackslash}m{1.7cm}|>{\raggedright\arraybackslash}m{2.5cm}|>{\raggedright\arraybackslash}m{2.7cm}|>{\raggedright\arraybackslash}m{2.7cm}|>{\raggedright\arraybackslash}m{6cm}|}
\hline
\textbf{Key Bit Index} & \textbf{Main Loop Iterations} & \textbf{Index of Desync Start - Doubling} & \textbf{Index of Desync Start - Addition 1} & \textbf{Index of Desync Start - Addition 2} & \textbf{Number of Clock Cycles Removed} \\ \hline
\endfirsthead
\hline


\rowcolor{orange!20} & 1 &  & 1,634,900 &  & 250 samples removed from (Addition 1)\textsubscript{1} \\
\rowcolor{orange!20} & 2 & 3,428,800 &  &  & 250 samples removed from (Doubling)\textsubscript{1} \\
\rowcolor{orange!20} \multirow{-3}{*}{1st Key Bit} & 3 &  &  & 1,653,400 & 250 samples removed from (Addition 2)\textsubscript{1} \\
\hline

\rowcolor{blue!20} & 1 &  & 806,950 &  & 250 samples removed from (Addition 1)\textsubscript{2} \\
\rowcolor{blue!20} & 2 & 4,257,700 &  & 4,257,700 & 250 samples removed from (Doubling)\textsubscript{2} and (Addition 2)\textsubscript{2} \\
\rowcolor{blue!20} & 3 &  &  & 1,653,000 & 250 samples removed from (Addition 2)\textsubscript{2} \\
\rowcolor{blue!20} & \cellcolor{green!20} 4 & \cellcolor{green!20} & \cellcolor{green!20} & \cellcolor{green!20} 1,654,250 & \cellcolor{green!20} 50 samples removed from (Addition 2)\textsubscript{2} \\
\rowcolor{blue!20} &  & 1,655,300 &  & 1,655,300 & 250 samples removed from (Doubling)\textsubscript{2} \\
\rowcolor{blue!20} &  &  &  & 1,702,650 & 250 samples removed from (Doubling)\textsubscript{2} \\ 
\rowcolor{blue!20} \multirow{-7}{*}{2nd Key Bit} &  & 1,702,800 &  & 1,702,800 & 250 samples removed from (Doubling)\textsubscript{2} \\
\hline

\rowcolor{orange!20} & 1 &  & 3,429,100 & 3,429,100 & 250 samples removed from (Addition 1)\textsubscript{3} and (Addition 2)\textsubscript{3} \\
\rowcolor{orange!20} & 2 &  &  & 1,652,950 & 250 samples removed from (Addition 2)\textsubscript{3} \\
\rowcolor{orange!20} & \cellcolor{green!20} & \cellcolor{green!20} & \cellcolor{green!20} & \cellcolor{green!20} 1,654,150 & \cellcolor{green!20} 50 samples removed from (Addition 2)\textsubscript{3} \\
\rowcolor{orange!20} &  & 1,654,300 &  & 1,654,300 & 250 samples removed from (Doubling)\textsubscript{3} and (Addition 2)\textsubscript{3} \\
\rowcolor{orange!20} \multirow{-5}{*}{3rd Key Bit} &  & 5,153,400 &  & 5,153,400 & 250 samples removed from (Doubling)\textsubscript{3} and (Addition 2)\textsubscript{3} \\
\hline

\rowcolor{blue!20} & 1 &  &  & 1,653,600 & 250 samples removed from (Addition 2)\textsubscript{4} \\
\rowcolor{blue!20} & 2 &  & 1,703,350 & 1,703,350 & 250 samples removed from (Addition 1)\textsubscript{4} and (Addition 2)\textsubscript{4} \\
\rowcolor{blue!20} \multirow{-3}{*}{4th Key Bit} & 3 & 3,428,200 &  & 3,428,200 & 250 samples removed from (Doubling)\textsubscript{4} and (Addition 2)\textsubscript{4} \\
\hline

\rowcolor{orange!20} & 1 &  &  & 1,653,100 & 250 samples removed from Addition 5 b \\
\rowcolor{orange!20} & 2 & 2,532,500 &  &  & 250 samples removed from Doubling 5 \\ \hline \hline
\rowcolor{orange!20} & 3 & 1,702,850 &  & 1,702,850 & 250 samples removed from (Doubling)\textsubscript{5} and (Addition 2)\textsubscript{5} \\ 
\rowcolor{orange!20} \multirow{-4}{*}{5th Key Bit} & \cellcolor{green!20} & \cellcolor{green!20} & \cellcolor{green!20} & \cellcolor{green!20} 1,654,150 & \cellcolor{green!20} 50 samples removed from (Addition 2)\textsubscript{5} \\ 
\rowcolor{orange!20} &  & 1,657,000 &  & 1,657,000 & 250 samples removed from (Doubling)\textsubscript{5} and (Addition 2)\textsubscript{5} \\ 
\hline

\rowcolor{blue!20} & 1 &  &  & 806,500 & 250 samples removed from (Addition 2)\textsubscript{6} \\
\rowcolor{blue!20} & 2 & 2,532,000 &  & 2,532,000 & 250 samples removed from (Doubling)\textsubscript{6} and (Addition 2)\textsubscript{6} \\
\rowcolor{blue!20} & 3 &  & 3,427,600 & 3,427,600 & 250 samples removed from (Addition 1)\textsubscript{6} and (Addition 2)\textsubscript{6} \\
\rowcolor{blue!20} \multirow{-4}{*}{6th Key Bit} & 4 &  &  & 1,652,900 & 250 samples removed from (Addition 2)\textsubscript{6} \\
\hline

\rowcolor{orange!20} & 1 & 2,532,500 & 2,532,500 &  & 250 samples removed from (Doubling)\textsubscript{7} and (Addition 1)\textsubscript{7} \\
\rowcolor{orange!20} & 2 & 4,258,600 & 4,258,600 &  & 250 samples removed from (Doubling)\textsubscript{7} and (Addition 1)\textsubscript{7} \\
\rowcolor{orange!20} & 3 & 807,150 & 807,150 &  & 250 samples removed from (Doubling)\textsubscript{7} and (Addition 1)\textsubscript{7} \\
\rowcolor{orange!20} \multirow{-4}{*}{7th Key Bit} & 4 &  &  & 1,653,300 & 250 samples removed from (Addition 2)\textsubscript{7} \\
\rowcolor{orange!20} & \cellcolor{green!20} & \cellcolor{green!20} & \cellcolor{green!20} & \cellcolor{green!20} 1,654,200 & \cellcolor{green!20} 50 samples removed from (Addition 2)\textsubscript{7} \\
\rowcolor{orange!20} &  & 1,657,800 &  & 1,657,800 & 250 samples removed from (Doubling)\textsubscript{7} and (Addition 2)\textsubscript{7} \\
\hline

\rowcolor{blue!20} & 1 &  &  &  &  \\
\rowcolor{blue!20} \multirow{-2}{*}{8th Key Bit} & 2 &  &  &  &  \\
\rowcolor{blue!20} & 3 &  &  &  &  \\
\hline

\rowcolor{orange!20} & 1 & 2,532,550 &  &  & 250 samples removed from (Doubling)\textsubscript{9} \\
\rowcolor{orange!20} & 2 &  & 4,258,150 &  & 250 samples removed from (Addition 1)\textsubscript{9} \\
\rowcolor{orange!20} & 3 &  &  & 1,653,850 & 250 samples removed from (Addition 2)\textsubscript{9} \\
\rowcolor{orange!20} \multirow{-4}{*}{9th Key Bit} & 4 & 5,171,900 &  &  & 250 samples removed from (Doubling)\textsubscript{9} \\
\hline

\rowcolor{blue!20} & 1 &  & 2,531,900 & 2,531,900 & 250 samples removed from (Addition 1)\textsubscript{10} and (Addition 2)\textsubscript{10} \\
\rowcolor{blue!20} & 2 &  & 3,428,800 & 3,428,800 & 250 samples removed from (Addition 1)\textsubscript{10} and (Addition 2)\textsubscript{10} \\
\rowcolor{blue!20} & 3 &  &  & 1,653,000 & 250 samples removed from (Addition 2)\textsubscript{10} \\
\rowcolor{blue!20} & \cellcolor{green!20} & \cellcolor{green!20} & \cellcolor{green!20} & \cellcolor{green!20} 1,654,000 & \cellcolor{green!20} 50 samples removed from (Addition 2)\textsubscript{10} \\
\rowcolor{blue!20} \multirow{-6}{*}{10th Key Bit} &  & 1,654,750 &  & 1,654,750 & 250 samples removed from (Doubling)\textsubscript{10} and (Addition 2)\textsubscript{10} \\
\hline

\caption{Sample index numbers for the start of the desynchronised regions where both 5-clock and 1-clock cycle processes were identified and removed in the sub-trace triplets} 
\label{tab:desync_regions} 
\end{longtable}
}
    \end{shiftedappendixB}

\end{document}